\documentclass[%
 reprint,
 amsmath,amssymb,
 aps,
]{revtex4-2}
\usepackage{float}
\usepackage{graphicx}
\usepackage{dcolumn}
\usepackage{bm}
\usepackage{booktabs} 
\usepackage{graphicx} 
\usepackage[font=small,labelfont=bf]{caption} 
\usepackage{lipsum} 
\usepackage{booktabs}
\usepackage{geometry}
\usepackage{rotating}
\usepackage{hyperref}

\newcommand\independent{\protect\mathpalette{\protect\independenT}{\perp}}
\def\independenT#1#2{\mathrel{\rlap{$#1#2$}\mkern2mu{#1#2}}}
\begin{document}

\preprint{APS/123-QED}

\title{Quantum Neural Networks for Propensity Score Estimation and Survival Analysis in Observational Biomedical Studies}
\author{Vojtěch Novák}
\email{vojtech.novak.st1@vsb.cz}
\affiliation{
 Department of Computer Science, Faculty of Electrical Engineering and Computer science, VSB-Technical University of Ostrava, Ostrava, Czech Republic
}
\affiliation{
 IT4Innovations National Supercomputing Center, VSB - Technical University of Ostrava, 708 00 Ostrava, Czech Republic
}
\author{Ivan Zelinka}%
\affiliation{
 Department of Computer Science, Faculty of Electrical Engineering and Computer science, VSB-Technical University of Ostrava, Ostrava, Czech Republic
}
\affiliation{
 IT4Innovations National Supercomputing Center, VSB - Technical University of Ostrava, 708 00 Ostrava, Czech Republic
}
\author{Lenka Přibylová}
\affiliation{
Department of Applied Mathematics, Faculty of Electrical Engineering and Computer science, VSB-Technical University of Ostrava, Ostrava, Czech Republic
}

\author{Lubomír Martínek}
\affiliation{Department of Surgical Studies, Faculty of Medicine, University
of Ostrava, Ostrava, Czech Republic}
\affiliation{Department of Surgery, University Hospital Ostrava, Ostrava, Czech Republic}

\begin{abstract}
This study investigates the application of quantum neural networks (QNNs) for propensity score estimation to address selection bias in comparing survival outcomes between laparoscopic and open surgical techniques in a cohort of 1177 colorectal carcinoma patients treated at University Hospital Ostrava (2001–2009). Using a dataset with 77 variables, including patient demographics and tumor characteristics, we developed QNN-based propensity score models focusing on four key covariates (Age, Sex, Stage, BMI). The QNN architecture employed a linear ZFeatureMap for data encoding, a SummedPaulis operator for predictions, and the Covariance Matrix Adaptation Evolution Strategy (CMA-ES) for robust, gradient-free optimization in noisy quantum environments. Variance regularization was integrated to mitigate quantum measurement noise, with simulations conducted under exact, sampling (1024 shots), and noisy hardware (FakeManhattanV2) conditions. QNNs, particularly with simulated hardware noise, outperformed classical logistic regression and gradient boosted machines in small samples (AUC up to 0.750 for n=100), with noise modeling enhancing predictive stability. Propensity score matching and weighting, optimized via genetic matching and matching weights, achieved covariate balance with standardized mean differences of 0.0849 and 0.0869, respectively. Survival analyses using Kaplan-Meier estimation, Cox proportional hazards, and Aalen additive regression revealed no significant survival differences post-adjustment (p-values 0.287–0.851), indicating confounding bias in unadjusted outcomes. These results highlight QNNs’ potential, enhanced by CMA-ES and noise-aware strategies, to improve causal inference in biomedical research, particularly for small-sample, high-dimensional datasets.

\end{abstract}

\maketitle

\section{\label{sec:level1}Introduction}

Colorectal cancer (CRC) ranks among the most prevalent and lethal malignancies worldwide, with the Czech Republic exhibiting some of the highest incidence rates. Despite its treatability in early stages, late diagnoses and advanced disease stages contribute significantly to its morbidity and mortality burden. Surgical resection remains the cornerstone of curative treatment, with laparoscopic and open surgical techniques representing the primary approaches. Laparoscopic surgery offers advantages such as reduced blood loss, faster recovery, and lower infection risk, but its long-term survival benefits compared to open surgery remain uncertain, particularly in observational studies where treatment assignment is influenced by patient characteristics like tumor stage and comorbidities. Addressing this selection bias is critical for unbiased treatment effect estimation, necessitating advanced statistical and computational methods.
Survival analysis provides a robust framework for evaluating time-to-event outcomes, such as survival time, while handling censoring and covariate effects \cite{klein2003survival}. However, observational studies, unlike randomized controlled trials, face challenges from non-random treatment allocation, which can confound outcome comparisons. Propensity score methods, including matching and weighting, mitigate this bias by balancing covariate distributions between treatment groups, enabling causal inference \cite{rosenbaum1983central}. Traditional approaches like logistic regression and gradient boosted models (GBM) estimate propensity scores but struggle with high-dimensional biomedical data and complex covariate interactions \cite{mccaffrey2004propensity}, \cite{pvribylova2023methodological}. Quantum machine learning (QML), particularly quantum neural networks (QNNs), offers a promising alternative by leveraging quantum computing’s unique properties—such as superposition and entanglement—to enhance optimization and data processing in high-dimensional spaces \cite{biamonte2017quantum}. In the noisy intermediate-scale quantum (NISQ) era, hybrid quantum-classical QNNs, built on parameterized quantum circuits (PQCs), provide a feasible approach for tasks like propensity score estimation, potentially outperforming classical methods in small-sample or complex datasets \cite{schuld2019quantum}.
This study investigates the application of QNNs to propensity score estimation for comparing survival outcomes between laparoscopic and open surgery in a cohort of 1177 CRC patients treated at University Hospital Ostrava from 2001 to 2009. The dataset, comprising 77 variables including patient demographics, tumor characteristics, and surgical outcomes, exhibits non-random treatment assignment, necessitating robust bias correction. We compare QNN-based propensity score models, implemented under exact, sampling, and noisy hardware simulations, against classical logistic regression and GBM across sample sizes (n=100, 500, and full dataset) to assess their performance in small-sample and high-dimensional settings. Propensity scores are applied through matching and weighting, with covariate balance evaluated via standardized mean differences (SMDs) and survival outcomes analyzed using Kaplan-Meier estimation, Cox proportional hazards models, and Aalen additive regression \cite{klein2003survival}. This work aims to demonstrate QNNs’ potential to enhance causal inference in biomedical research, offering insights into surgical technique efficacy and advancing the integration of quantum computing in healthcare analytics.

\section{Introduction to the Medical Problem}

Colorectal cancer (CRC) is a malignant disease of the digestive system affecting the colon and rectum. It is among the most common cancers worldwide in terms of both incidence and mortality \cite{bray2018global}. For several years, the Czech Republic has ranked among the countries with the highest rates of this malignancy \cite{dusek2014cancer}. Although CRC is highly treatable in its early stages, its rising incidence and the frequent diagnosis at advanced stages impose a significant burden of morbidity and mortality on the population. Therefore, public and professional awareness is crucial for prevention and early detection \cite{winawer2003colorectal}.

The causes of CRC can be divided into external (exogenous) and internal (endogenous) risk factors. Exogenous factors primarily relate to diet and lifestyle habits, such as the consumption of red meat and animal fats, particularly when fried, roasted, or smoked, which can produce carcinogenic compounds \cite{zhao2017red}. Excessive alcohol consumption and a diet lacking in fiber from fruits and vegetables, calcium, vitamins C and D, and folic acid are also considered risk factors \cite{slattery1998eating}. Endogenous factors, which account for about 15-20\% of cases, include hereditary and familial predispositions \cite{lynch2003hereditary}. However, the most common form, accounting for 80-85\% of all cases, is sporadic CRC, which typically develops from benign adenomatous polyps \cite{ahadova2018three}. The single most significant endogenous risk factor for sporadic CRC is age, with a notable increase in incidence after the age of fifty \cite{haggar2009colorectal}.

In their initial stages, colorectal tumors are often asymptomatic. The only sign may be occult (hidden) bleeding in the stool, which is not visible to the naked eye but can be detected through testing \cite{simon1984fecal}. As these early-stage tumors cause few noticeable problems, patients may not seek attention, delaying the start of treatment. Often, the first indication of an advanced tumor is an acute, life-threatening condition such as a bowel obstruction, which manifests with abdominal pain, cessation of gas and stool, diarrhea, vomiting, and visible bleeding \cite{deans1994malignant}. If an obstruction develops, surgery becomes essential to prevent a breach of the intestinal barrier, which could lead to shock and death. For malignant tumors, surgery is the primary and often only curative treatment option \cite{cunningham2010colorectal}.

Diagnosis typically begins with a colonoscopy, which allows for direct visualization of the colon lining and the collection of a tissue sample (biopsy) for analysis \cite{winawer2003colorectal}. For rectal tumors, a rectoscopy is performed. Further diagnostic methods are used to determine the extent of the disease. Imaging techniques such as computed tomography (CT), magnetic resonance imaging (MRI) or positron emission tomography/CT (PET/CT) are employed to assess the extent of disease, lymph node involvement and the presence of metastases, while a chest X-ray is used to check for lung metastases \cite{van2011staging}.

Stage of the disease is critical for determining the optimal therapeutic strategy. The internationally recognized TNM classification system is used for this purpose \cite{sobin2010tnm}. The stage is assessed based on the extent of the primary tumor's invasion through the layers of the intestinal wall (T), the number of affected regional lymph nodes (N), and the presence of distant metastases (M). The T stage ranges from T1 (invasion into the submucosa) to T4 (invasion into adjacent organs). The N stage indicates whether no lymph nodes are involved (N0), 1-3 are involved (N1), or 4 or more are involved (N2). The M stage signifies either no distant metastases (M0) or their presence (M1). These factors combine to define four overall stages of the disease. Stage 1 represents a localized superficial tumor. Stage 2 involves a larger tumor infiltrating deeper into the bowel wall. Stage 3 is characterized by extensive local invasion and significant lymph node involvement. Stage 4 indicates the presence of distant metastases, regardless of the primary tumor's size. A precise classification, known as the pathological or pTNM stage, is determined after surgery by examining the resected tumor. The presence of any remaining tumor after treatment is denoted by the R classification: R0 indicates complete tumor removal, R1 signifies microscopic residual tumor, and R2 means macroscopic tumor remains \cite{wittekind2009tnm}.

The treatment of CRC is determined after a complete patient evaluation and disease stage, with the therapeutic plan established by a multidisciplinary team including a surgeon, oncologist, gastroenterologist, and pathologist \cite{van2010multidisciplinary}. Treatment can be curative or palliative, and the approach always depends on the patient's overall health and consent. The primary treatment modalities are surgical, endoscopic, and oncological (chemotherapy and radiotherapy). Typically, the tumor is removed first, either surgically or sometimes endoscopically, followed by systemic oncological therapy. For very large tumors, neoadjuvant (pre-operative) therapy may be used to shrink the tumor before surgery \cite{sauer2004preoperative}.

To date, the only curative treatment for colorectal cancer is radical surgical resection, where the goal is to achieve a complete removal of the tumor (R0 status) \cite{cunningham2010colorectal}. The extent of the resection depends on the tumor's location and spread. After removing the affected bowel segment, continuity is restored by creating a surgical connection (anastomosis). The treatment strategy is highly dependent on the stage. For Stage 1, surgery alone is usually sufficient. For Stage 2, surgery is the primary treatment, but adjuvant chemotherapy may follow if adverse features are present. In Stage 3, surgery is always followed by chemotherapy or radiochemotherapy. For Stage 4, treatment is complex and may involve resection of the primary tumor and metastases, induction therapy followed by surgery, or purely palliative care \cite{van2010multidisciplinary}. Recurrence of the tumor affects about one-fifth of operated patients, with most recurrences occurring within the first two years \cite{kjeldsen1997recurrence}.

Surgical procedures for CRC can be performed using traditional open surgery or minimally invasive surgery including laparoscopic or most recently robotic surgery \cite{pascual2016laparoscopic}. In the study from which our data is drawn, all operations were performed by a select group of surgeons experienced in both methods. The choice of approach was based on the surgeon's discretion and the patient's preference. The extent of resection was identical for both techniques, and there were no differences in indications for neoadjuvant / adjuvant therapy or postoperative care between the two groups.

Laparoscopic surgery is a technique where operations are performed through small incisions (typically 0.5--1.5 cm) using specialized instruments and a camera. The abdominal cavity is first inflated with carbon dioxide gas to create space for the surgeon to work. The camera transmits an image to a monitor, guiding the surgeon. The advantages of this method include smaller incisions, which lead to reduced blood loss, less pain, faster recovery, and a lower risk of infection \cite{lacy2002laparoscopy}. The main disadvantage of the laparoscopic technique is the loss of tactile perception and 2D view.

Open surgery is the standard approach, where the surgeon makes a larger incision in the abdomen to directly access the organs. This technique allows for the radical removal of the bowel segment containing the tumor along with its lymphatic drainage. The main drawback is the larger surgical wound, which results in a more prominent scar and carries a higher risk of complications. Potential intraoperative complications for both techniques include injury to adjacent structures, such as the ureter, spleen, or nerves, particularly during complex rectal surgeries \cite{guillou2005short}.

\newpage

\section{Survival Analysis and Propensity Score}

Survival analysis is a statistical framework for modeling the time until an event of interest, such as death in biomedical research or system failure in engineering, occurs. It is widely applied to time-to-event data in medical studies evaluating outcomes like overall survival or time to disease progression, and in industrial settings assessing product reliability \cite{klein2003survival}. In the context of quantum computing and quantum machine learning, survival analysis provides a foundation for analyzing time-dependent outcomes. Quantum neural networks can enhance propensity score estimation by leveraging improved predictive performance in specific, simulated, small-data scenarios, particularly for optimization and high-dimensional data processing. This section introduces survival analysis concepts, including survival and hazard functions, censoring, and key estimation methods, followed by propensity score techniques to address selection bias in non-randomized studies, emphasizing their application to biomedical data and relevance to advanced computational methods.

The core of survival analysis is the survival time, a random variable \( T \) representing the time until the event occurs. The survival function, defined as \( S(t) = P(T > t) = 1 - F(t) \), where \( F(t) = P(T \leq t) \) is the cumulative distribution function, gives the probability that a subject survives past time \( t \). This function is non-increasing, right-continuous, with \( S(0) = 1 \) (all subjects survive at the study’s start, typically \( t = 0 \)) and \( \lim_{t \to \infty} S(t) = 0 \), reflecting that the probability of survival diminishes over time. The hazard function, \( h(t) = \frac{f(t)}{S(t)} \), where \( f(t) \) is the probability density of \( T \), quantifies the instantaneous risk of the event occurring at time \( t \), given survival up to that point. The cumulative hazard function, \( H(t) = \int_0^t h(x) \, dx \), represents the accumulated risk up to time \( t \), and is related to the survival function by \( S(t) = e^{-H(t)} \). These functions allow researchers to model the distribution of survival times and assess how covariates, such as patient characteristics, influence event risks. Additional metrics include the median survival time \( t_{0.5} \), where \( S(t_{0.5}) = 0.5 \), and the mean survival time, \( \mu = E(T) = \int_0^\infty S(t) \, dt \), which exists if the survival function decays appropriately. In cases where the event may not occur (e.g., time to a non-inevitable event like marriage), the mean is estimated using conditional distributions to account for subjects who never experience the event \cite{klein2003survival}.

Censoring is a critical challenge in survival analysis, occurring when the event of interest is not observed within the study period due to factors like study termination, loss to follow-up, or competing events (e.g., death from an unrelated cause). For each subject \( i \), we observe \( y_i = \min(t_i, c_i) \), where \( t_i \) is the true event time and \( c_i \) is the censoring time, along with an indicator \( \delta_i = 1 \) if the event occurs (\( t_i \leq c_i \)) or \( \delta_i = 0 \) if censored. Survival analysis thus works with data pairs \( \{(y_i, \delta_i)\}_{i=1}^n \). Properly handling censoring ensures unbiased estimation of survival characteristics, which is crucial for applications where quantum algorithms may optimize likelihood computations over large, censored datasets \cite{klein2003survival}.

The Kaplan-Meier estimator is a widely used non-parametric method to estimate the survival function from censored data. It calculates \( \hat{S}(t) = \prod_{t_i \leq t} \left( 1 - \frac{d_i}{n_i} \right) \), where \( d_i \) is the number of events (e.g., deaths) at time \( t_i \), and \( n_i \) is the number of subjects at risk just before \( t_i \), including those who have not experienced the event or been censored. Censored subjects contribute to the risk set until their censoring time, after which they are excluded. This estimator provides a step-function representation of survival probabilities, which is robust for small to moderate sample sizes and can be computed efficiently, making it suitable for integration with quantum machine learning models that process large datasets \cite{kaplan1958nonparametric}. The log-rank test complements the Kaplan-Meier estimator by comparing survival functions between two groups, such as treated and control cohorts. It tests the null hypothesis that the survival functions are equal (\( S^1(t) = S^0(t) \)) up to a time \( \tau \), using a statistic based on the difference between observed and expected event counts, which follows a normal distribution for large samples. This test is valuable for assessing treatment effects in clinical studies and can be adapted for weighted analyses in propensity score applications \cite{mantel1966evaluation}.

In clinical research, randomized controlled trials (RCTs) ensure unbiased treatment assignment, but observational studies often face selection bias due to non-random treatment allocation based on patient characteristics (e.g., choosing surgical techniques based on disease severity). Propensity score methods mitigate this bias by balancing covariates between treated and control groups. The propensity score, defined as \( e(\boldsymbol{x}) = P(Z_i = 1 | \boldsymbol{X} = \boldsymbol{x}) \), is the conditional probability that a subject with covariate vector \( \boldsymbol{X} \) receives treatment (\( Z_i = 1 \)) versus control (\( Z_i = 0 \)). It acts as a balancing score, ensuring that the distribution of covariates is similar across groups conditional on \( e(\boldsymbol{x}) \). This property, established by Rosenbaum and Rubin, reduces bias from observed covariates, making it a powerful tool for causal inference in non-randomized settings \cite{rosenbaum1983central}. Propensity scores are commonly estimated using logistic regression, where \( \hat{e}(\boldsymbol{x}) = \frac{e^{\boldsymbol{X}' \hat{\boldsymbol{\beta}}}}{1 + e^{\boldsymbol{X}' \hat{\boldsymbol{\beta}}}} \), with \( \hat{\boldsymbol{\beta}} \) as the regression coefficients, providing a straightforward approach suitable for classical computing \cite{rosenbaum1984reducing}. However, for complex, high-dimensional data, generalized boosted models (GBM) offer superior performance. GBM iteratively constructs regression trees, optimizing a loss function (e.g., squared error or exponential loss) to capture non-linear relationships and interactions among covariates. Each tree is a weak learner, and the ensemble model, \( f_M(\boldsymbol{x}) = \sum_{m=1}^M T(\boldsymbol{x}; \hat{\Theta}_m) \), combines multiple trees to improve prediction accuracy. The iterative nature of GBM, coupled with techniques like shrinkage to prevent overfitting, makes it computationally intensive but well-suited for quantum acceleration, where quantum algorithms can optimize tree construction and gradient calculations \cite{mccaffrey2004propensity}.

Propensity scores are applied through matching or weighting to estimate causal effects, such as the average treatment effect (ATE), \( E[Y_i(1) - Y_i(0)] \), or the average treatment effect for the treated (ATT), \( E[Y_i(1) - Y_i(0) | Z_i = 1] \), where \( Y_i(1) \) and \( Y_i(0) \) are outcomes under treatment and control, respectively. Matching pairs treated and control subjects with similar propensity scores, often using nearest-neighbor matching within a caliper distance (e.g., \( \epsilon = 0.25 \cdot \sigma_{\hat{e}} \), where \( \sigma_{\hat{e}} \) is the standard deviation of the estimated propensity score) to ensure close matches. Matching can be done with or without replacement, with 1:1 or 1:n configurations, balancing precision and sample size. Weighting adjusts observations using inverse probability of treatment weights (IPTW), defined as \( w(Z_i, \boldsymbol{x}) = \frac{Z_i}{\hat{e}(\boldsymbol{x})} + \frac{1 - Z_i}{1 - \hat{e}(\boldsymbol{x})} \) for ATE, or \( w(Z_i, \boldsymbol{x}) = Z_i + (1 - Z_i) \frac{\hat{e}(\boldsymbol{x})}{1 - \hat{e}(\boldsymbol{x})} \) for ATT, aligning the control group to the treated group’s covariate distribution. Overlap weighting, \( w(Z_i, \boldsymbol{x}) = Z_i (1 - \hat{e}(\boldsymbol{x})) + (1 - Z_i) \hat{e}(\boldsymbol{x}) \), reduces the influence of extreme propensity scores, enhancing robustness in observational studies. These methods assume strong ignorability (\( (Y(1), Y(0)) \independent Z | \boldsymbol{x} \)) and overlap (\( 0 < P(Z=1) < 1 \)), ensuring unbiased estimation of treatment effects \cite{rosenbaum1983central}.

For survival outcomes, propensity score weights are integrated into the Kaplan-Meier estimator to handle censoring. The weighted estimator adjusts event counts and risk sets, using weights like \( d_i^w = \sum_{t_j < t_i} w_j \delta_j z_j \) and \( n_i^w = \sum_{t_j < t_i} w_j z_j \), where weights reflect treatment assignment probabilities. This produces a survival function estimate, \( \hat{S}(t) = \prod_{t_i \leq t} \left( 1 - \frac{d_i^w}{n_i^w} \right) \), that accounts for covariate imbalances. Similarly, a weighted log-rank test adjusts the test statistic and variance to compare survival curves across groups, ensuring robust inference in the presence of censoring and non-random treatment assignment \cite{xie2005adjusted}.

To assess covariate balance, the standardized mean difference (SMD), \( SMD = \frac{\bar{X}_t - \bar{X}_c}{\sqrt{(s_t^2 + s_c^2)/2}} \) for continuous variables, or a similar form for dichotomous variables, quantifies differences between treated and control groups. An SMD below 0.1 indicates negligible imbalance, confirming effective balancing \cite{austin2008assessing}. These techniques enable robust causal inference in observational studies, providing a framework for quantum neural networks to enhance propensity score estimation by efficiently handling high-dimensional covariate spaces and optimizing complex loss functions, thus improving the accuracy of treatment effect estimates in survival analysis \cite{schuld2019quantum}.

\section{Quantum Neural Networks for Classification, Regression, and Propensity Score Estimation}

Quantum machine learning (QML) represents a transformative intersection of quantum computing and machine learning, leveraging the unique computational properties of quantum systems to enhance data processing and model optimization. In the noisy intermediate-scale quantum (NISQ) era, characterized by quantum hardware with 50--1000 noisy qubits, hybrid quantum-classical approaches have emerged as a cornerstone for practical applications \cite{preskill2018quantum}. Quantum neural networks (QNNs), based on parameterized quantum circuits (PQCs), are a prominent QML framework, offering potential advantages in supervised learning tasks such as classification and regression \cite{schuld2019quantum, novak2025predicting}. These tasks are critical in fields like healthcare , where propensity score estimation can benefit from QNNs’ ability to handle high-dimensional data and complex optimization landscapes. This section explores the principles of QNNs, their implementation for classification and regression, and their application to propensity score estimation, highlighting their relevance to advancing computational efficiency in quantum computing and machine learning.

QNNs utilize PQCs to encode classical input data into quantum states and manipulate them to produce desired outputs. For an input vector \( \boldsymbol{x} \in \mathbb{R}^d \), a PQC, denoted \( \hat{U}(\boldsymbol{x}, \boldsymbol{\phi}) \), transforms an initial quantum state \( |0\rangle \) into a parameterized quantum state \( |\Psi(\boldsymbol{x}, \boldsymbol{\phi})\rangle = \hat{U}(\boldsymbol{x}, \boldsymbol{\phi})|0\rangle \). The PQC is typically constructed from one- and two-qubit gates, with rotational gates commonly used for data encoding and state manipulation. Techniques like data re-uploading, where input data is repeatedly encoded into the circuit, enhance the expressibility of the PQC, enabling it to capture complex patterns \cite{zhou2022noise}. The output of the QNN, \( f(\boldsymbol{x}) \), is obtained by measuring the expectation value of a cost operator \( \hat{C}(\boldsymbol{\phi}) \):

\begin{equation}
f_{\boldsymbol{\phi}, \boldsymbol{\phi}}(\boldsymbol{x}) = \langle \Psi(\boldsymbol{x}, \boldsymbol{\phi}) | \hat{C}(\boldsymbol{\phi}) | \Psi(\boldsymbol{x}, \boldsymbol{\phi}) \rangle.
\end{equation}

For tasks requiring multiple outputs, such as multi-class classification, individual cost operators \( \hat{C}_j(\boldsymbol{\phi}_j) \) are defined for each output, forming a vector of expectation values. To simplify notation, the parameters \( \boldsymbol{\phi} \) (PQC) and \( \boldsymbol{\phi} \) (cost operator) are often combined into a single vector \( \boldsymbol{\theta} \), with explicit parameter dependency omitted when clear from context.

Training a QNN involves optimizing the parameters \( \boldsymbol{\theta} \) by minimizing a loss function \( L_{\text{fit}}(\boldsymbol{\theta}) \), tailored to the task. For regression tasks, such as predicting continuous outcomes like survival times, a common loss function is the squared error:

\begin{equation}
L_{\text{fit}}(\boldsymbol{\theta}) = \sum_i \| f_{\boldsymbol{\theta}}(\boldsymbol{x}_i) - y_i \|^2,
\end{equation}

where \( \{ \boldsymbol{x}_i \in \mathbb{R}^d \} \) are input data points and \( \{ y_i \in \mathbb{R} \} \) are corresponding labels in a supervised learning scenario. For classification tasks, such as predicting treatment assignment in propensity score estimation, the cross-entropy loss is often used to handle categorical outcomes \cite{schuld2019quantum}. The optimization process is challenging due to the non-convex nature of the loss function, compounded by noise and decoherence in NISQ devices. Gradient-based methods, such as the parameter-shift rule, are employed to compute derivatives of the PQC parameters. This rule evaluates the gradient by shifting each parameter in the circuit and measuring the difference in expectation values, requiring two circuit evaluations per parameter. For PQCs with many parameterized gates, this can become computationally intensive, a bottleneck that quantum hardware advancements or alternative methods like linear combination of unitaries aim to address \cite{schuld2020circuit}. Differentiation with respect to cost operator parameters is simpler, as the derivative \( \partial_{\phi_m} \hat{C}(\boldsymbol{\phi}) \) can often be computed from the same measurements as the expectation value, enhancing efficiency by reusing circuit evaluations.

The stochastic nature of quantum measurements introduces noise in QNN outputs, quantified by the standard deviation of the expectation value:

\begin{equation}
\text{std}(f) = \sqrt{\frac{\sigma_f^2}{N_{\text{shots}}}}.
\end{equation}
where $ \sigma_f^2 = \langle \Psi | \hat{C}^2 | \Psi \rangle - \langle \Psi | \hat{C} | \Psi \rangle^2.$

To mitigate this noise, a variance regularization term is added to the loss function:

\begin{equation}
L_{\text{var}}(\boldsymbol{\theta}) = \sum_k \| \sigma_f^2(\tilde{\boldsymbol{x}}_k) \|,
\end{equation}

where \( \{ \tilde{\boldsymbol{x}}_k \} \) are points typically set to the training data \( \{ \boldsymbol{x}_i \} \) to capture the input domain. The total loss is then \( L = L_{\text{fit}} + \alpha L_{\text{var}} \), with the hyperparameter \( \alpha \) (typically \( 10^{-4} \) to \( 10^{-2} \)) balancing fitting accuracy and variance reduction. Once trained, the QNN predicts new outcomes by evaluating \( f(\boldsymbol{x}) \) on unseen data, leveraging the optimized parameters to generalize effectively \cite{schuld2019quantum}.

In classification tasks, QNNs, often implemented as variational quantum classifiers (VQCs), assign labels to input data, such as categorizing patients into treated or control groups for propensity score estimation. The VQC encodes input features into a quantum state and uses measurements to produce probabilities for each class, optimized via cross-entropy loss. Techniques like data re-uploading and hardware-efficient circuit designs enhance classification performance, particularly on NISQ devices, by improving the circuit’s ability to represent complex decision boundaries \cite{li2019tackling}. Pre-processing methods, such as quantum principal component analysis (QPCA), can reduce data dimensionality, further boosting efficiency \cite{lloyd2014quantum}. For regression tasks, QNNs predict continuous values, such as survival times or propensity scores, by optimizing the squared error loss. The hybrid quantum-classical optimization loop, where quantum circuits compute expectation values and classical optimizers adjust parameters, allows QNNs to tackle high-dimensional datasets, making them suitable for biomedical applications where covariates are numerous and complex \cite{schuld2020circuit}.

In our implementation, the QNN was specifically designed for the task of propensity score estimation. Classical input covariates were encoded into the quantum state using a linear \texttt{ZFeatureMap}, chosen for its efficiency and compatibility with near-term hardware. A \texttt{SummedPaulis} operator was employed in a dual role as both the primary observable for generating predictions and as the cost operator within the training objective. The model parameters were optimized by minimizing a \texttt{SquaredLoss} function, which is appropriate for the regression-like nature of predicting a propensity score. For the optimization itself, we utilized the Covariance Matrix Adaptation Evolution Strategy (CMA-ES), a robust, gradient-free algorithm known for its strong performance in noisy quantum environments \cite{bonet2023performance, illesova2025numerical, novak2025optimization}. The synergy between the noise-resilient CMA-ES optimizer and the inherent noise of the quantum computation is central to our findings, particularly the improved performance observed in noisy simulations. A comprehensive description of the QNN architecture, including the mathematical formulation of the feature map, operators, and loss function, is provided in Appendix \ref{sec:qnn_details}.

Propensity score estimation, a critical task in observational studies, involves predicting the probability of treatment assignment, \( e(\boldsymbol{x}) = P(Z_i = 1 | \boldsymbol{X} = \boldsymbol{x}) \), given a covariate vector \( \boldsymbol{X} \). Classical methods like logistic regression model this probability as \( e(\boldsymbol{x}) = \frac{e^{\boldsymbol{X}' \boldsymbol{\beta}}}{1 + e^{\boldsymbol{X}' \boldsymbol{\beta}}} \), but they struggle with non-linear relationships and high-dimensional data \cite{rosenbaum1984reducing}. Advanced classical methods, such as generalized boosted models (GBM), use ensembles of regression trees to capture complex patterns, but their computational cost grows with data complexity \cite{mccaffrey2004propensity}. QNNs offer a promising alternative by leveraging quantum superposition and entanglement to explore high-dimensional feature spaces efficiently \cite{biamonte2017quantum}. For propensity score estimation, a QNN can be trained as a classifier, where the output \( f(\boldsymbol{x}) \) represents the probability of treatment assignment. The input covariates \( \boldsymbol{x} \) are encoded into the PQC, and the expectation value of the cost operator provides the propensity score estimate. The cross-entropy loss ensures accurate classification, while variance regularization reduces noise from quantum measurements, critical for reliable estimates on NISQ hardware \cite{schuld2019quantum}.

The hybrid nature of QNNs makes them particularly suited for propensity score estimation in observational studies, where balancing covariates between treated and control groups is essential \cite{li2019tackling}. Quantum advantages, such as enhanced optimization through quantum gradient computation or potential speedup in matrix operations via QPCA, could improve the scalability of propensity score models compared to classical GBM \cite{lloyd2014quantum}. However, challenges like noise-induced barren plateaus and the stochasticity of quantum measurements require careful circuit design and error mitigation strategies, such as those used in variational quantum algorithms like the Quantum Approximate Optimization Algorithm (QAOA) or Variational Quantum Eigensolver (VQE) \cite{cerezo2021variational}. These algorithms inspire QNN architectures by demonstrating how parameterized circuits can solve complex optimization problems, relevant to both classification and regression tasks in QML \cite{schuld2019quantum}.

In practice, QNNs for propensity score estimation can be integrated with survival analysis techniques, such as the weighted Kaplan-Meier estimator, to evaluate treatment effects on time-to-event outcomes \cite{xie2005adjusted}. By producing robust propensity score estimates, QNNs enable accurate covariate balancing, reducing selection bias in non-randomized studies \cite{rosenbaum1983central}. The ability to handle high-dimensional biomedical data, combined with potential quantum speedups, positions QNNs as a powerful tool for advancing causal inference in healthcare applications \cite{biamonte2017quantum}. As quantum hardware improves, QNNs could outperform classical methods in propensity score estimation, offering faster convergence and better handling of complex covariate interactions, thus paving the way for transformative applications in quantum machine learning and data-driven medical research \cite{schuld2019quantum}.

\section{Experimental Methodology}

This study evaluates propensity score estimation for causal inference in observational biomedical data, comparing traditional logistic regression (LR) and gradient boosting (GB) with a novel quantum neural network (QNN) approach, leveraging quantum machine learning to enhance computational efficiency in high-dimensional settings. The experiment uses a dataset of 1177 patients, with subsamples of 500 and 100 patients to assess model performance across varying sample sizes, reflecting real-world biomedical constraints. QNNs are tested under three simulation conditions—statevector (exact), sampling with 1024 shots, and FakeManhattanV2 backend—to evaluate their robustness on noisy intermediate-scale quantum (NISQ) devices. Simulations are conducted using Qiskit (v1.0.2) \cite{qiskit}, Qiskit-Aer (v0.14.2), Qiskit-Algorithms (v0.3.1), Qiskit-IBM-Runtime (v0.23.0), Qiskit-Machine-Learning (v0.8.2), Qiskit-Nature (v0.7.2), sQUlearn (v0.7.6) \cite{kreplin2025squlearn}, with sQUlearn’s variance regularization enhancing QNN stability and CMA-ES optimizer is from cma (v3.3.0) \cite{hansen2019pycma}. Model quality is assessed using receiver operating characteristic (ROC) curves with area under the curve (AUC), log-loss, Brier score, and accuracy. Propensity scores are applied through matching and weighting, with balance evaluated via standardized mean difference (SMD), data structure analysis, and probability density comparisons. Treatment effects on time-to-event outcomes are analyzed using Kaplan-Meier survival curves and log-rank tests, providing a comprehensive evaluation of QNNs’ potential in advancing causal inference for healthcare applications.

The dataset includes 1177 patients with covariates (e.g., age, comorbidities) and binary treatment assignments (treated vs. control). Propensity scores are estimated for the full dataset and randomly selected subsamples of 500 and 100 patients to test model robustness with limited data, a common challenge in clinical studies. LR models propensity scores using logistic regression, optimized via maximum likelihood estimation, assuming linear covariate effects but potentially missing complex patterns. GB employs an ensemble of regression trees, iteratively optimizing a loss function (e.g., exponential loss) with tuned hyperparameters (e.g., tree depth, learning rate) to capture non-linear relationships, offering flexibility but requiring significant computational resources for large datasets. The QNN approach, implemented via sQUlearn, uses a parameterized quantum circuit to estimate propensity scores as a classification task, predicting treatment probability with cross-entropy loss and variance regularization to mitigate quantum measurement noise. QNN simulations are conducted in three modes: statevector for exact results, 1024-shot sampling to introduce measurement noise, and FakeManhattanV2 backend to emulate real quantum hardware noise, reflecting realistic NISQ device constraints. These conditions test QNN performance under varying noise levels, highlighting their potential and limitations compared to classical methods.

Model performance is evaluated using multiple metrics to ensure robust comparison across LR, GB, and QNNs for all sample sizes and QNN simulation conditions. ROC curves plot true positive rates against false positive rates, with AUC measuring discriminative ability (higher values indicate better separation of treated and control groups). Log-loss quantifies the accuracy of predicted probabilities, penalizing confident misclassifications. The Brier score, calculated as the mean squared difference between predicted probabilities and actual outcomes, assesses calibration, with lower scores indicating better performance. Accuracy measures the proportion of correctly classified treatment assignments. These metrics provide a comprehensive view of each model’s ability to estimate propensity scores accurately, with QNN results under different simulation conditions revealing the impact of quantum noise on predictive performance.

Propensity scores are applied using two approaches: matching and weighting. Matching uses 1:1 nearest-neighbor matching within a caliper of \( 0.25 \cdot \sigma_{\hat{e}} \), where \( \sigma_{\hat{e}} \) is the standard deviation of estimated propensity scores, pairing treated and control patients to create balanced groups. Weighting employs inverse probability of treatment weights (IPTW) to estimate the average treatment effect, adjusting for covariate imbalances. Overlap weighting is also applied to reduce the influence of extreme propensity scores, enhancing robustness. Covariate balance is assessed using the standardized mean difference for continuous variables, with SMD values below 0.1 indicating negligible imbalance. Post-balancing data structure is analyzed by comparing covariate distributions (e.g., via histograms or kernel density plots) between treated and control groups. Probability density functions of key variables (e.g., age, disease severity) are compared before and after weighting to evaluate changes in distribution shapes, ensuring effective covariate alignment.

Survival analysis evaluates treatment effects on time-to-event outcomes, such as time to death or disease progression. Kaplan-Meier curves estimate survival functions for matched and weighted datasets, compared to unadjusted curves to assess the impact of balancing methods. Weighted analyses incorporate IPTW or overlap weights to adjust event counts and risk sets. The log-rank test assesses the statistical significance of differences in survival functions between treated and control groups, with p-values indicating whether treatment effects are significant after balancing. This experimental design rigorously compares LR, GB, and QNNs, highlighting the innovative QNN approach’s potential to leverage quantum computing for propensity score estimation. By testing across sample sizes and simulation conditions, the study evaluates QNN robustness in NISQ settings, while comprehensive model evaluation and balance assessment ensure reliable causal inference, positioning QNNs as a transformative tool for quantum machine learning in biomedical research.

\section{Results and discussion}

\subsection{Analysis of the Data Set Structure}
\label{sec:data_set_analysis}
The study under consideration was conducted between 2001 and 2009 at the surgical department of the University Hospital Ostrava. The data set contains information on 1177 patients who underwent surgery for colorectal carcinoma. For each patient, 77 data points were recorded, describing their health status, biochemical values, and other data obtained during the operation.
The main objective of this study was to compare the survival time of patients who underwent laparoscopic surgery with those who had traditional open surgery. The laparoscopic technique was used in 596 (50.6\%) patients and the open technique in 526 (44.7\%) patients. For the remaining 56 (4.8\%) patients, a conversion occurred (a switch from a laparoscopic to an open surgical technique during the procedure).

In this section we describe the data set from the biomedical application area and perform a basic analysis of the dependencies of relevant variables. Since the core task of the practical section is to compare the survival times of patients based on the surgical technique, we will compare the analyses of variable dependencies according to the surgical technique used. To begin, we will provide a characterization of the data set. 

We then focus on analyzing the structure of the patient data set according to the surgical technique used. We will compare the characteristics of individual variables based on the surgical technique. To describe the structure of dichotomous variables such as Sex, Mortality, and Morbidity, we use association tables in which absolute frequencies are supplemented by column relative frequencies and p-values from the $\chi^2$ test of independence in a contingency table. For qualitative variables with more categories (the ASA variable), we used the Kruskal-Wallis test.

The tables show column relative frequencies. These relative frequencies provide information about the percentage representation of the categories of individual variables according to the given surgical technique. The figures contain row relative frequencies, providing information about the percentage representation of a category across the surgical techniques.

\begin{table*}[htpb]
\caption{Structure of the patient cohort by medical history depending on the surgical technique}
\centering
  \begin{tabular}{l l l l l l l}
      & ~~~~~~ & ~ Laparoscopic & Open & Total & $ \widehat{OR} $  & p-value \\
    &  & ~ N = 595 & N = 526 & 1121 & (95\% CI) & $ \chi^2 $ test \\
    \hline
    variable &  category & \vline ~ $n$ ~ (\%) & $n$ ~ (\%) & $n$ ~ (\%) &  &  \\
    \hline
Sex: & Male & \vline ~ 371 (62.4) & 277 (52.7) & 648 (57.8) & 1.49 (1.17; 1.89) & \textbf{0.001} \\
         & Female & \vline ~   224 (37.6) & 249 (47.3) & 473 (42.2) & \\
  \hline
ASA: &  1  & \vline ~  77 (12.9)&   83 (15.8) & 160 (14.2) & & 0.334 \\
 & 2  & \vline ~ 274 (46.1) & 248 (47.1) & 522 (46.6) & \\
 & 3  & \vline ~ 216 (36.3) & 178 (33.8) & 394 (35.1) & \\
 & 4   & \vline ~ 28 (4.7)  & 17 (3.2) &  45 (4.0)& \\
  \hline
  DM: & Yes & \vline ~ 129 (21.7) & 123 (23.4) & 252 (22.5) & 0.91 (0.69; 1.20) & 0.542 \\
     & No & \vline ~   466 (78.3) & 403 (76.6) & 869 (77.5) & \\
  \hline
HT: & Yes & \vline ~ 331 (55.6) & 275 (52.3) & 606 (54.1) & 1.14 (0.90; 1.45) & 0.288 \\
     & No & \vline ~   264 (44.4) & 251 (47.7) & 515 (45.9) & \\
     \hline
     Arrhythmia: & Yes & \vline ~ 74 (12.4) & 65 (12.4) & 139 (12.4) & 0.99 (0.70; 1.42) & $>$0.999 \\
     & No & \vline ~   521 (87.6) & 461 (87.6) & 982 (87.6) & \\
     \hline
     Hepatic: & Yes & \vline ~ 19 (3.2) & 11 (2.1) & 30 (2.7) & 1.54 (0.73; 3.28) & 0.339 \\
     & No & \vline ~   576 (96.8) & 515 (97.9) & 1091 (97.3) & \\
     \hline
    Renal: & Yes & \vline ~ 26 (4.4) & 17 (3.2) & 43 (3.8) & 1.37 (0.73; 2.55) & 0.404 \\
     & No & \vline ~   569 (95.6) & 509 (96.8) & 1078 (96.2) & \\
     \hline
     Pulmonary: & Yes & \vline ~ 83 (13.9) & 78 (14.8) & 161 (14.4) & 0.93 (0.67; 1.30) & 0.739 \\
     & No & \vline ~   512 (86.1) & 448 (85.2) & 960 (85.6) & \\
     \hline
Mortality: & Yes  & \vline ~ 27 (4.5) & 24 (4.6) & 51 (4.5) & 0.98 (0.56; 1.74) & 0.932 \\
           & No  & \vline ~ 568 (95.5) & 502 (95.4) & 1070 (95.5) & \\
\hline
Morbidity: & Yes  & \vline ~ 191 (32.1) & 190 (36.1) & 381 (34.0) & 0.84 (0.66; 1.07) & 0.166 \\
  & No  & \vline ~ 404 (67.9) & 336 (63.9) & 740 (66.0) & \\
  \end{tabular}
  \label{tab_anamneza}
\end{table*}

\begin{figure*}[htpb]
    \centering
    \includegraphics[width=0.9\textwidth]{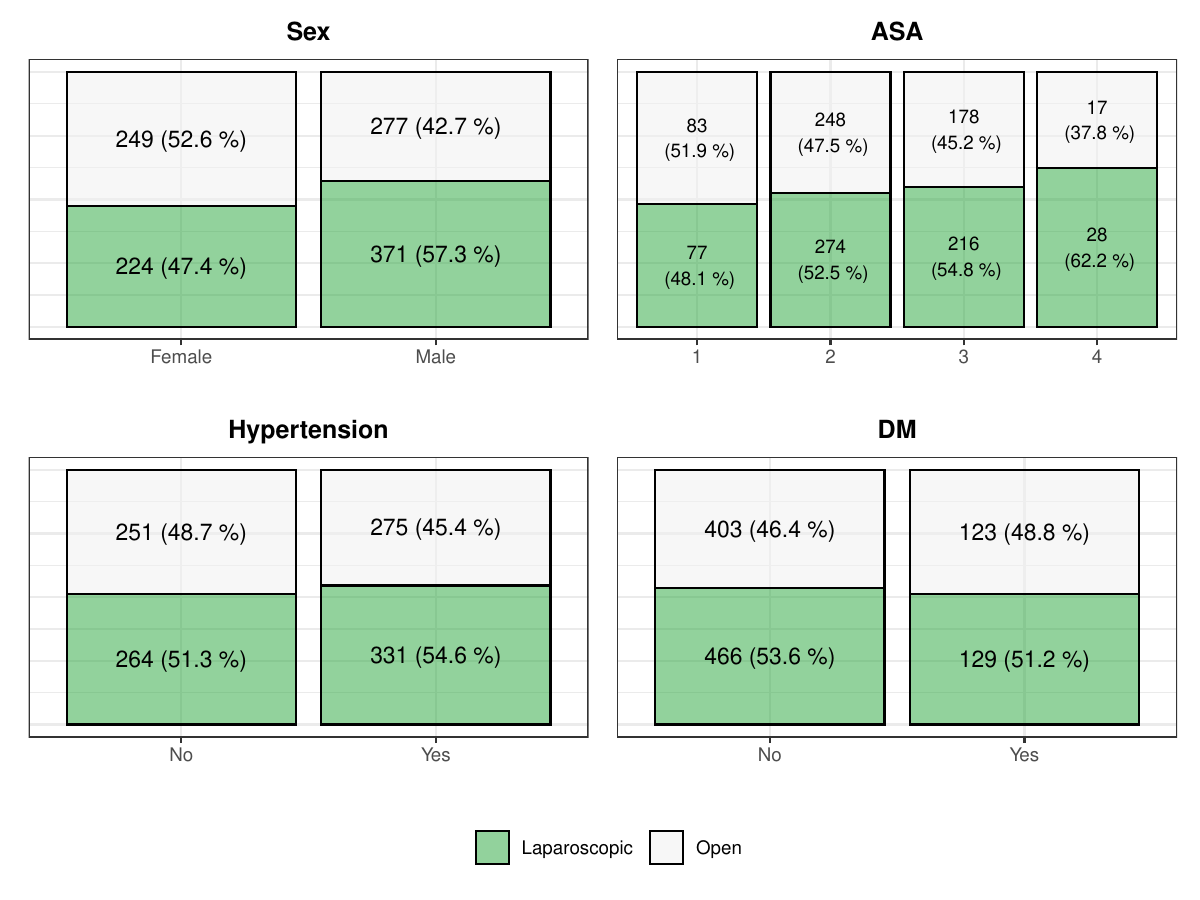}
    \caption{Structure of the patient cohort by medical history depending on the surgical technique (100\% stacked column chart)}
    \label{obr_anamneza}
\end{figure*}

The choice of surgical technique is related to the patient's sex (p-value = 0.001, $\chi^2$ test, see Table \ref{tab_anamneza}, Fig. \ref{obr_anamneza}). For the other variables in the patient's medical history, no statistically significant difference between the surgical techniques was demonstrated.

By analyzing the other dichotomous variables (DM, Mortality, Morbidity) in the association tables, we can say, based on the 95\% confidence interval for the OR, that the choice of surgical technique does not depend on these variables, as 1 is included in the 95\% CI.

In the case of the ASA classification, Fig. \ref{obr_anamneza} shows an increase in the number of patients operated on using the laparoscopic method with a higher classification compared to the open method. The ASA classification is most represented by values 2 and 3, which are mild and severe systemic diseases. There are significantly fewer patients with classification 4, who are severely ill patients expected to die without the surgery. However, this increase is not statistically significant (p=0.334, Kruskal-Wallis test).

For variables related to tumor classification, there were missing data; we proceeded to work only with patients for whom values were provided.
For the T variable, 268 values were missing, and for 45, the classification could not be determined. For the N variable, 269 values were missing, and for 142, the classification could not be determined. The M variable was not filled in for 268 values. The stage was not determined for 321 patients.

\begin{table}[htpb]
\caption{Structure of the patient cohort by tumor characteristics depending on the surgical technique}
\footnotesize
\centering
  \begin{tabular}{l l l l l l }
  & ~~& ~ Laparoscopic  & Open & Total  & p-value \\
    \hline
     &  & \vline ~ $n$ ~ (\%) & $n$ ~ (\%) & $n$ ~ (\%) &   \\
    \hline
    T: & 1 & \vline ~   40 (9.0) & 14 (3.7) & 54 (6.6) & $<$\textbf{0.001} \\
         & 2 & \vline ~   62 (14.0) & 43 (11.3) & 105 (12.7) & \\
         & 3 & \vline ~   243 (54.7) & 195 (51.3) & 438 (53.2) & \\
         & 4 & \vline ~   99 (22.3) & 128 (33.7) & 227 (27.5) & \\
  \hline
       N:   & 0  & \vline ~ 236 (56.9) & 164 (52.2) & 400 (54.9) & $<$\textbf{0.001} \\
            & 1  & \vline ~ 106 (25.5) & 106 (33.8) & 212 (29.1) & \\
            & 2  & \vline ~ 73 (17.6)  & 44 (14.0)  &  117 (16.0) & \\
  \hline
 M:      & 0  & \vline ~ 374 (79.2) & 282 (71.2) & 656 (75.6) & $<$\textbf{0.001} \\
      & 1  & \vline ~ 98 (20.8) & 114 (28.8) & 212 (24.4) & \\
\hline
    Stage: & 1 & \vline ~   94 (20.8) & 46 (12.5) & 140 (17.1) & $<$\textbf{0.001} \\
         & 2 & \vline ~   125 (27.7) & 99 (27.0) & 224 (27.4) & \\
         & 3 & \vline ~   134 (29.7) & 106 (28.9) & 240 (29.3) & \\
         & 4 & \vline ~   98 (21.7) & 116 (31.6) & 214 (26.2) & \\
  
  \end{tabular}
  \label{tab_nador}
\end{table}

\begin{figure*}[htpb]
    \centering
    \includegraphics[width=0.9\textwidth]{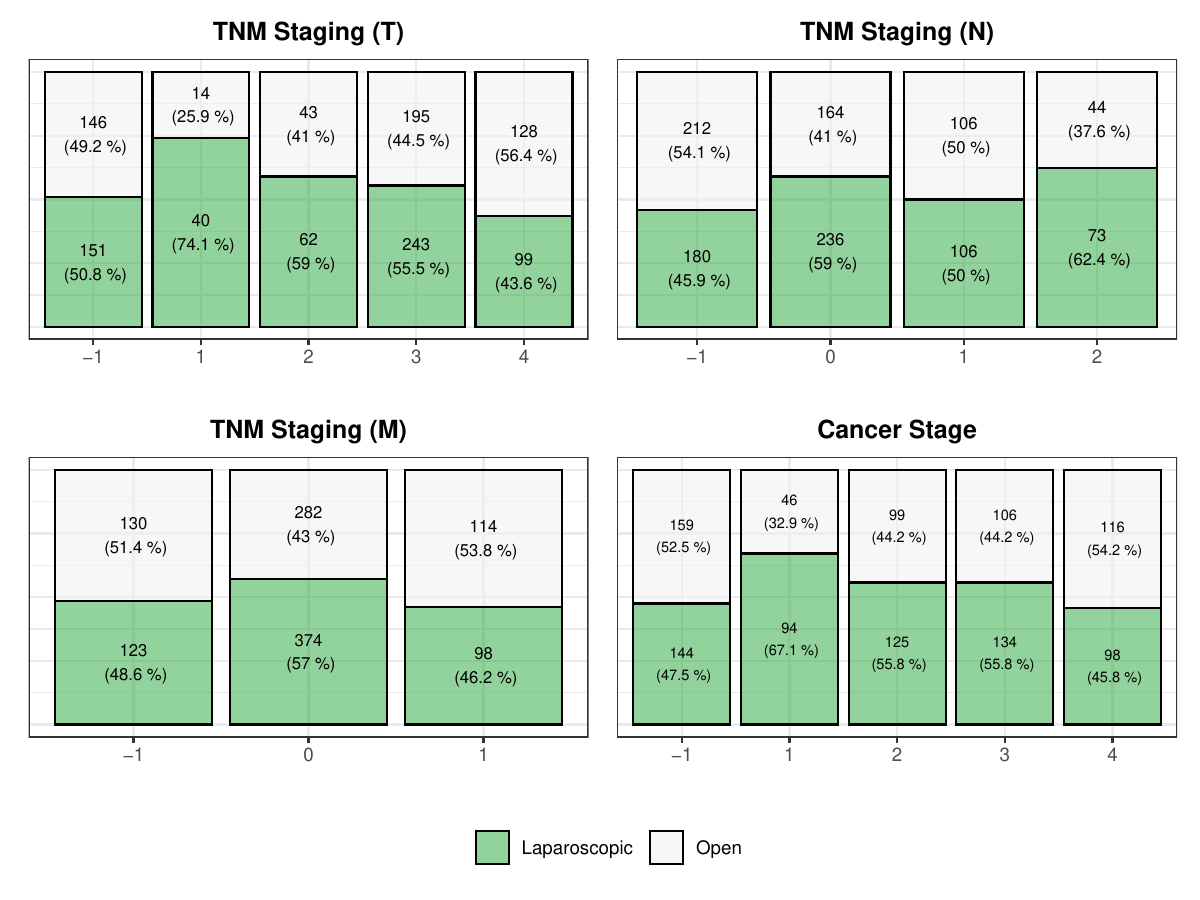}
    \caption{Structure of the patient cohort by tumor characteristics depending on the surgical technique (100\% stacked column chart)}
    \label{obr_nador}
\end{figure*}

In Table \ref{tab_nador}, we see that surgeries are most often performed on larger and more extensive tumors (T classification 3 and 4). The highest relative frequency for tumor size and extent (T) is category 3, accounting for more than half of the patients for both laparoscopic and open methods. More than half of the operated patients had no regional lymph node involvement (N). In the laparoscopic group, 25.5\% of patients had 1-3 regional lymph nodes affected, and 17.6\% had more than 4 affected. In the open surgery group, 33.8\% had 1-3 nodes affected, and 14.0\% had more than 4. Distant metastases (M) were not present in 79.2\% of patients in the laparoscopic group and 71.2\% in the open group.

In Fig. \ref{obr_nador}, it is immediately apparent that the choice of surgical technique depends on the stage.
For a patient with an earlier stage, the laparoscopic technique was more often chosen, whereas for a patient with a more advanced stage, the open technique was preferred. Of the patients classified as stage 1, 67.1\% were operated on laparoscopically. Patients with stage 4 were more likely to be operated on using the open method (54.2\%). The laparoscopic method was more frequently assigned to patients with non-metastatic tumors (M variable).

For all quantitative variables, the assumption that the sample comes from a bivariate normal distribution is violated.
For their description, the median is used, supplemented by the interquartile range (lower quartile – upper quartile), a point and 95\% confidence interval for the difference of medians, and the corresponding Mann-Whitney (M-W) test.

\begin{table*}[htpb]
\caption{Structure of the patient cohort by quantitative variables depending on the surgical technique (median, (IQR), difference in medians, (CI), and p-values of the Mann-Whitney test)}
\centering
  \begin{tabular}{l l l l l l l}
   variable  & ~ laparoscopic   & open & difference in medians & p-value \\
               & ~   median (IQR) & median (IQR)  & difference (CI) & M-W test \\
    \hline
    Age    & \vline ~ 66.00 (58.00; 74.00)  & 65.00 (56.00; 74.00) & 1 (-1; 2) & 0.347 \\
  \hline
  BMI  & \vline ~ 26.12 (23.70; 29.38) & 26.20 (23.40; 29.10) & -0.08 (-0.60; 0.40) & 0.701 \\
  \hline
 OpTime  & \vline ~ 150 (110; 180) & 135 (90; 180) & 15 (10; 20) & $<$\textbf{0,001} \\
  \hline
  SurvivalTime  & \vline ~ 22.5 (8.00; 45.00) & 24.0 (9.00; 47.00) & -1.5 (-5; 1) & 0.118 \\
  \hline
  \end{tabular}
  \label{tab_kvant}
\end{table*}

\begin{figure*}[htpb]
    \centering
    \includegraphics[width=1\textwidth]{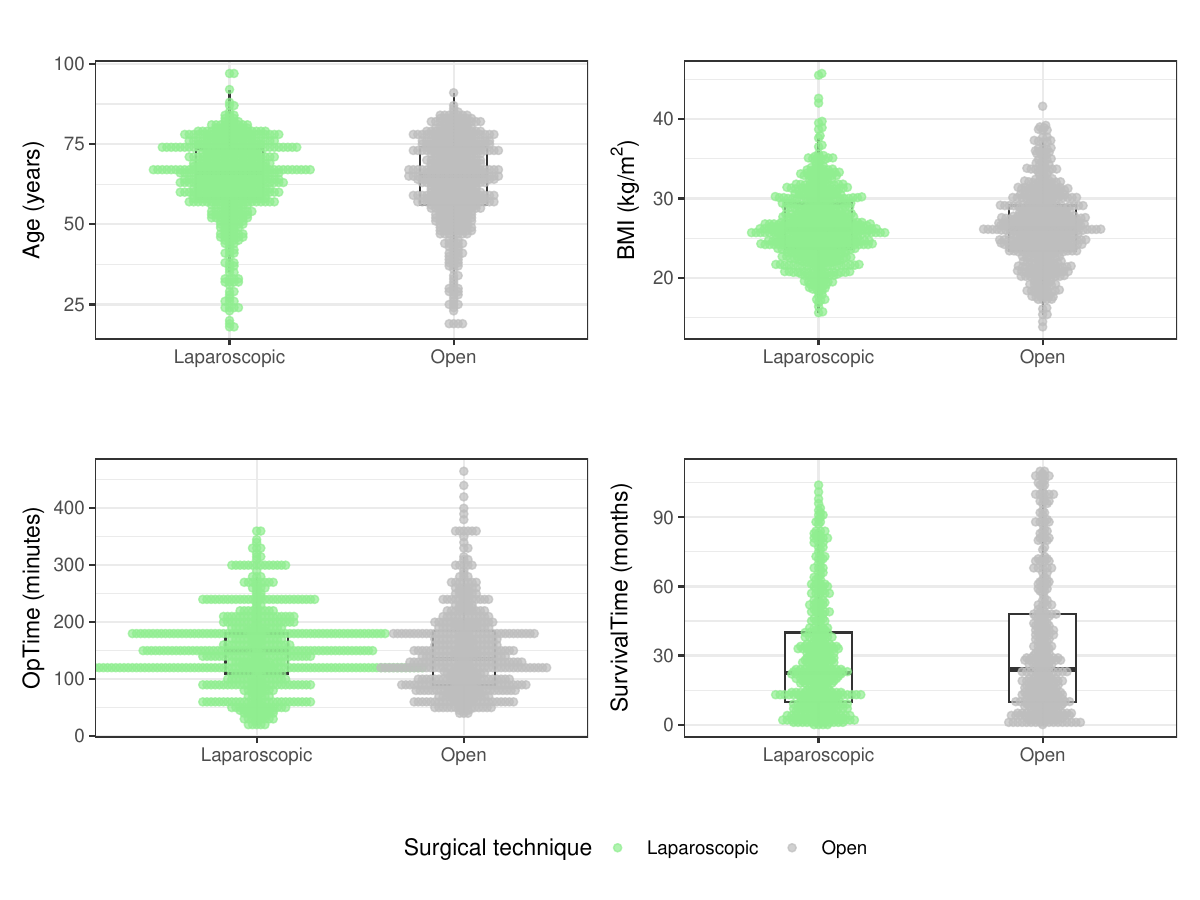}
    \caption{Structure of the patient cohort by quantitative variables depending on the surgical technique}
    \label{obr_kvant}
\end{figure*}

For the laparoscopic surgical technique, the age of half of all observed patients ranges from 58 to 74 years; for the open technique, it is similar, with half of the patients aged between 56 and 74 years. We can also say that half of the patients are at least 66 years old in the laparoscopic group and at least 65 years old in the open group.

The median BMI values of 26.12 and 26.20 indicate that more than half of the patients from both surgical groups were overweight. However, the minimal difference in medians shows that the choice of surgical technique did not depend on the patient's BMI (p-value = 0.701, Mann-Whitney test).

The choice of surgical technique cannot depend on blood loss, operation time, or survival time. These variables are consequences of the selected surgical technique.
None of the quantitative variables from the medical history showed a statistically significant association with the choice of surgical technique, see Table \ref{tab_kvant}.

Based on the exploratory analysis, supplemented by significance tests, we have identified individual variables that may have a significant association with the choice of surgical technique. From the patient's medical history, this is only the patient's sex. The choice of surgical technique is influenced by tumor characteristics: size, extent, and stage. Laparoscopic surgery was predominantly performed on smaller tumors, with fewer metastases, and in less advanced stages. The open technique was often used for patients with more advanced stages.

\subsection{Propensity Score Estimation and Analysis}

Based on analyses from the previous chapter, treatment assignment was clearly non-random, with surgical technique selection dependent on patient sex and tumor characteristics. Laparoscopic surgery was predominantly performed on patients with earlier disease stages, while open surgery was frequently chosen for patients with advanced disease. This selection bias means that direct comparison of survival outcomes between surgical techniques would likely reflect the prognostic advantage of earlier-stage disease rather than true treatment effects.

To address this imbalance, we calculated propensity scores for each patient to create balanced datasets that better approximate those from a randomized experimental study. Given the biological reality that clinical variables originate from interconnected physiological systems, inherent dependencies and correlations are expected among the 77 available variables. Many variables showed substantial missing data, and certain relationships are definitionally dependent—for example, the Stage variable is derived directly from T, N, and M categories according to established clinical guidelines, making statistical correlation tests between these variables tautological.

For our quantum neural network propensity score estimation, we selected four key covariates based on previous analysis, medical relevance and computational constraints: Age, Sex, Stage, and BMI. These variables represent fundamental patient characteristics that influence surgical decision-making while avoiding the definitional redundancies present in larger variable sets. Age and BMI capture basic patient physiology affecting surgical risk, Sex represents a key demographic factor influencing treatment decisions, and Stage provides essential tumor severity information without the redundancy of including its component T, N, M variables.

The decision to limit our feature set reflects both computational necessities and biomedical data realities. In biomedical datasets, particularly those from surgical cohorts, using extensive variable sets can be problematic for several reasons. First, diseases impact multiple biological systems simultaneously, creating complex webs of interrelated parameters where inflammatory markers, organ function tests, and clinical symptoms are often modulated by the same underlying pathophysiological processes. This systemic nature means that including numerous correlated variables may not improve model performance while increasing overfitting risk. Second, many clinical variables exhibit definitional dependencies—the TNM classification system inherently defines cancer stage, making Stage a direct derivative of T, N, and M components rather than an independent measurement. Testing correlations between such definitionally linked variables would be tautological and misinterpret their fundamental relationship. Third, biomedical datasets frequently suffer from substantial missing data due to varied clinical protocols, patient conditions, and resource limitations, making larger feature sets practically challenging to implement effectively.

From a quantum computing perspective, incorporating more features would significantly increase quantum circuit depth, adding numerous quantum gates that would substantially increase error rates, complicate parameter optimization, and make classical simulation computationally prohibitive. The shallow circuits necessitated by current quantum hardware limitations favor parsimonious feature selection, making our four-variable approach both medically sound and computationally feasible.

While the primary goal of propensity score analysis (PSA) is covariate balance for causal inference, there are important nuances regarding the reporting of classification metrics during model development. These metrics serve as diagnostic tools when building the propensity score model, providing insight into how well the chosen covariates and model specification predict treatment assignment before proceeding to matching or weighting. A model with very poor predictive performance (e.g., AUC near 0.5) might indicate that the selected covariates are inadequate predictors of treatment status, potentially limiting the ability to achieve balance. However, it is crucial to emphasize that these classification metrics are secondary to covariate balance in the overall evaluation of propensity score analysis.

\begin{table}[htpb]
\centering
\caption{Propensity score model evaluation metrics by sample size and method}
\label{tab:metrics}
\begin{tabular}{l l c c c}
\toprule
Sample & Classifier & AUC & Log Loss & Brier Score \\
\midrule
100 & qnn\_sam & 0.667 & 0.782 & 0.281 \\
100 & qnn\_f\_backend & 0.750 & 0.735 & 0.271 \\
100 & qnn\_exact & 0.615 & 0.715 & 0.252 \\
100 & lr & 0.510 & 0.798 & 0.298 \\
100 & gbm & 0.563 & 5.355 & 0.493 \\
500 & qnn\_sam & 0.536 & 0.735 & 0.270 \\
500 & qnn\_f\_backend & 0.513 & 0.701 & 0.254 \\
500 & qnn\_exact & 0.507 & 0.701 & 0.254 \\
500 & lr & 0.521 & 0.713 & 0.259 \\
500 & gbm & 0.536 & 1.292 & 0.384 \\
full & qnn\_sam & 0.591 & 0.679 & 0.243 \\
full & qnn\_f\_backend & 0.572 & 0.684 & 0.246 \\
full & qnn\_exact & 0.554 & 0.691 & 0.249 \\
full & lr & 0.560 & 0.686 & 0.247 \\
full & gbm & 0.522 & 1.112 & 0.340 \\
\bottomrule
\end{tabular}
\end{table}

\begin{figure}[htpb]
\centering
\includegraphics[width=0.45\textwidth]{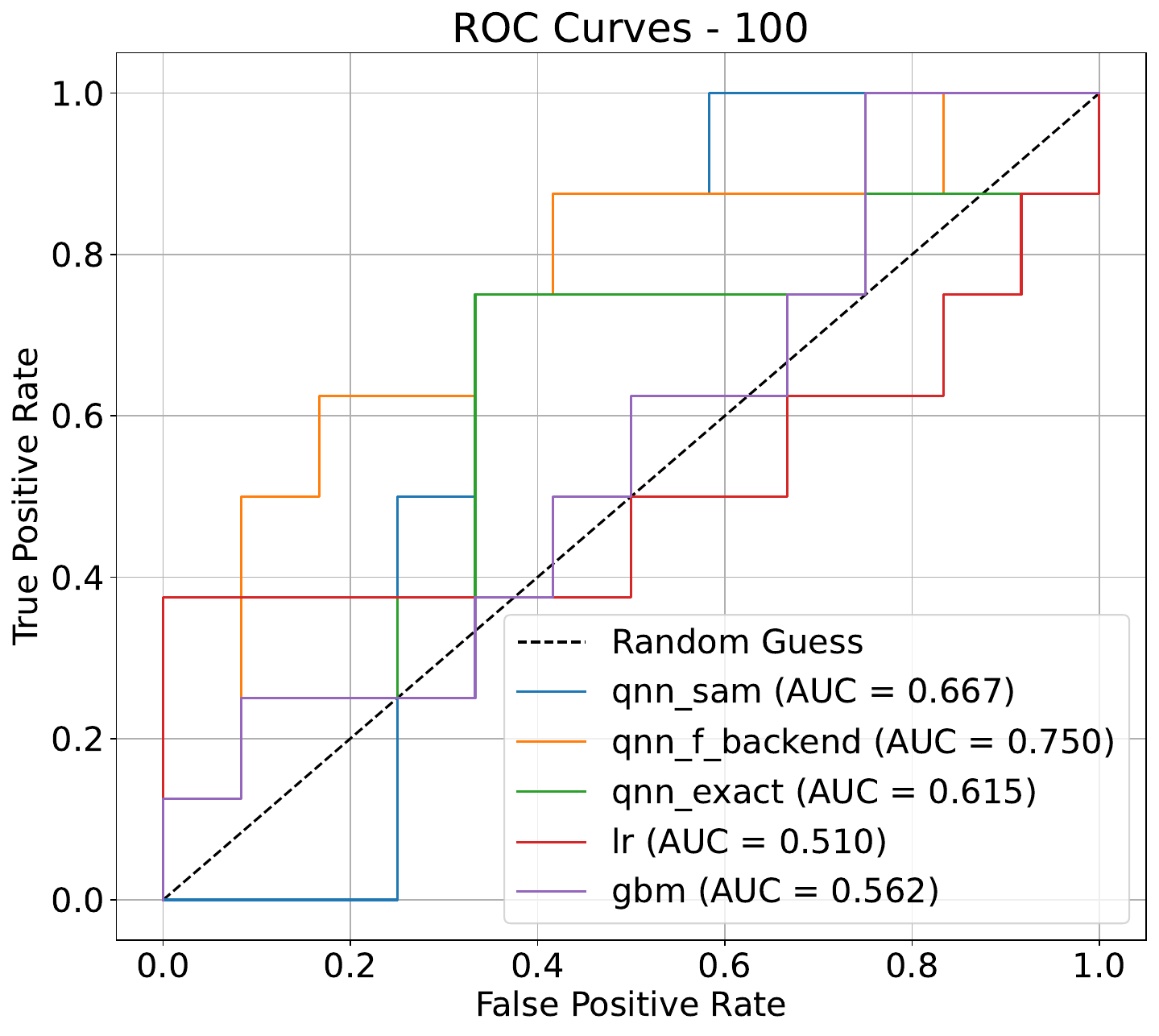}
\caption{ROC curves for propensity score models (n=100 sample). The quantum approaches, particularly qnn\_f\_backend, show superior performance in this small sample size regime.}
\label{fig:roc100}
\end{figure}

\begin{figure}[htpb]
\centering
\includegraphics[width=0.45\textwidth]{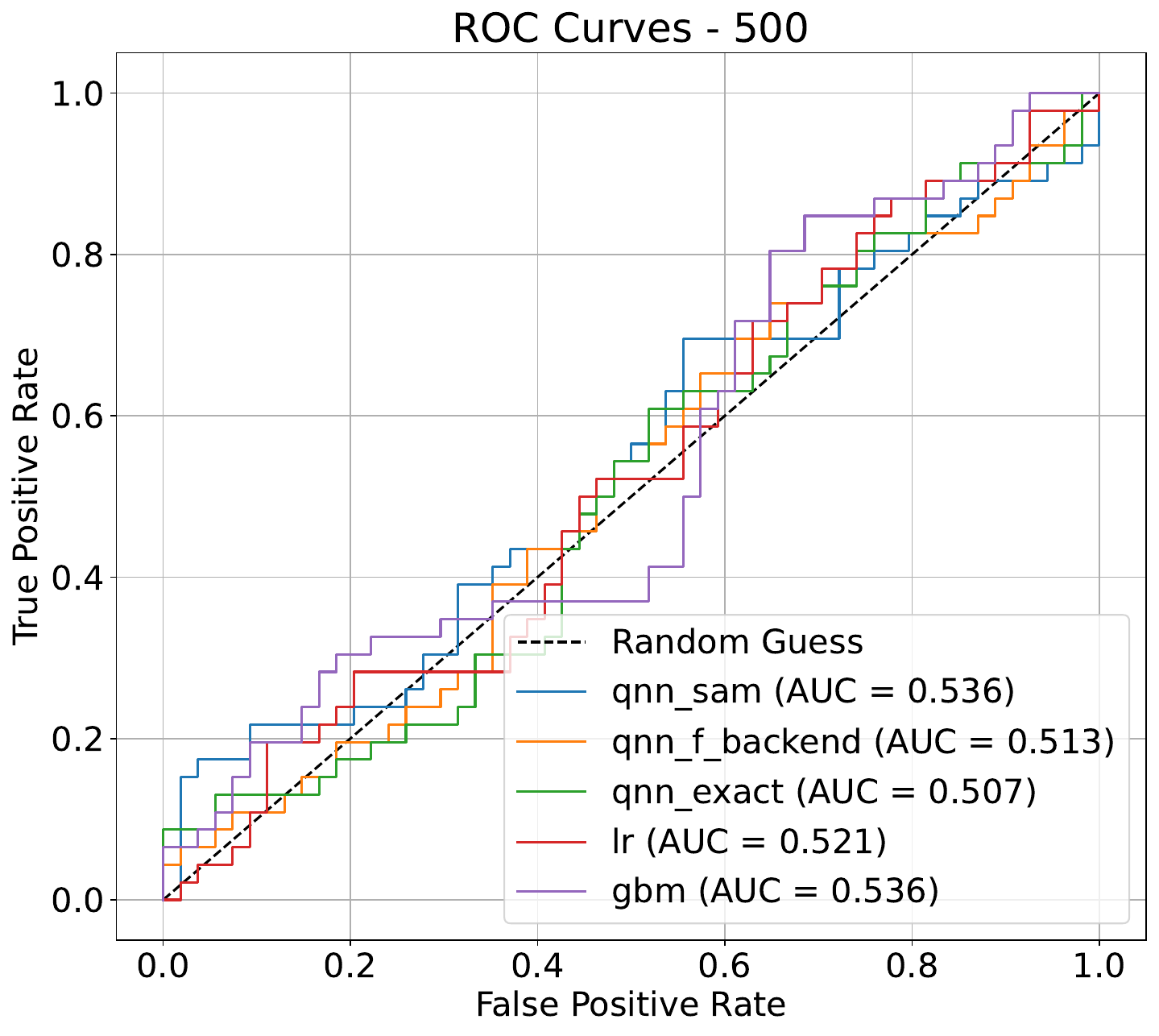}
\caption{ROC curves for propensity score models (n=500 sample). Performance differences between methods become less pronounced with increased sample size.}
\label{fig:roc500}
\end{figure}

\begin{figure}[htpb]
\centering
\includegraphics[width=0.45\textwidth]{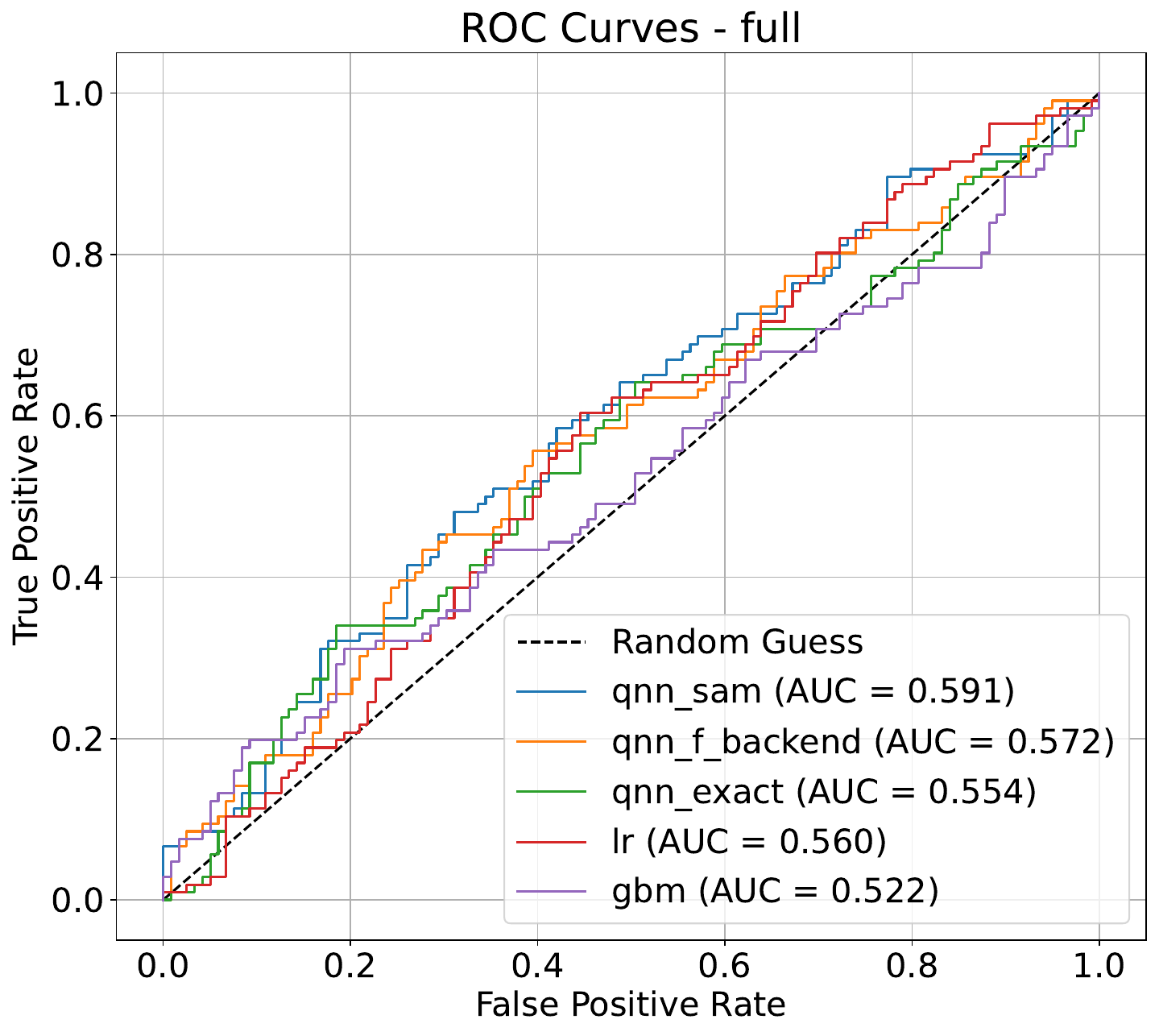}
\caption{ROC curves for propensity score models (full sample). All methods show similar performance in this large sample regime.}
\label{fig:rocfull}
\end{figure}

The ROC curves in Figures\ref{fig:roc100}, \ref{fig:roc500}, and \ref{fig:rocfull} visualize the predictive performance across different sample sizes, with several noteworthy patterns emerging from both the figures and Table\ref{tab:metrics}. For the smallest dataset (n=100), the quantum approaches demonstrated superior predictive power, particularly the qnn\_f\_backend method, which achieved an AUC of 0.750, compared to classical methods that struggled with this limited sample size. This enhanced performance in the small-data regime diminished as sample size increased, though quantum methods maintained comparable performance to classical approaches in larger samples.

An unexpected finding emerged regarding noise in quantum models. Contrary to theoretical expectations, the addition of different types of noise (sampling noise in qnn\_sam and simulated hardware noise in qnn\_f\_backend) actually improved classification accuracy compared to the noise-free qnn\_exact implementation in the smallest sample. This phenomenon is visible in Figure~\ref{fig:roc100}, where both noisy quantum implementations outperform their noise-free counterpart. The gradient boosted machines (gbm) showed particularly poor performance in small samples, with exceptionally high log loss values indicating substantial prediction uncertainty.

As shown in Figure~\ref{fig:roc500} and Figure~\ref{fig:rocfull}, the performance differences between methods became less pronounced with increased sample size. All models showed AUC values between 0.507 and 0.591 for the larger samples, suggesting more similar predictive performance when sufficient data are available. The consistency of these patterns across different evaluation metrics (AUC, log loss, and Brier score) in Table~\ref{tab:metrics} strengthens the reliability of these observations.

It is essential to contextualize these findings within the broader framework of propensity score analysis. While classification metrics provide valuable insight during model development, they must not be confused with the ultimate goal of propensity score analysis, which is achieving covariate balance. A model that perfectly predicts treatment assignment (AUC near 1.0) may paradoxically indicate insufficient common support between treatment groups, making causal inference problematic. The value of propensity scores lies in their ability to reduce confounding and enable unbiased treatment effect estimation, not in their predictive accuracy. 

These results demonstrate that our quantum approaches generated reasonable propensity score estimates, particularly valuable for small samples where classical methods struggled. However, the true test of these models' utility will come in their ability to balance covariates, which we examine in the next section through matching and weighting techniques. The current analysis serves primarily to establish that we have developed adequate models to proceed with balance assessment, while noting the interesting performance characteristics of quantum approaches in different data regimes.

\subsection{Covariate balancing through matching and weighting}

Having established that our quantum neural network propensity score models generated reasonable estimates, particularly the qnn\_f\_backend method using the FakeManhattanV2 noise model, we now examine their effectiveness in achieving covariate balance through various matching and weighting approaches. The primary objective of this analysis is to assess how different propensity score adjustment methods reduce systematic differences between treatment groups, thereby enabling more reliable causal inference.

We evaluated the performance of multiple matching and weighting techniques using standardized mean differences (SMD) as our primary balance metric. Table~\ref{tab:smd_means} presents the mean SMD across all covariates for each method, providing a summary measure of overall balance improvement. Among the matching approaches, genetic matching achieved the best performance with a mean SMD of 0.08991, representing a 10.5\% improvement over the unmatched data (SMD = 0.10049). Nearest neighbor matching and optimal matching showed more modest improvements, with mean SMDs of 0.09040 and 0.09490, respectively.

\begin{table}[htpb]
\centering
\caption{Mean Standardized Mean Differences Across Different Propensity Score Methods}
\label{tab:smd_means}
\begin{tabular}{lc}
\toprule
\textbf{Method} & \textbf{Mean SMD} \\
\midrule
Unmatched & 0.0952 \\
Nearest Neighbor Matching & 0.0890 \\
Optimal Matching & 0.0916 \\
Genetic Matching 400 & 0.0849 \\
Genetic Matching 100 & 0.0902 \\
ATE Weighting & 0.0878 \\
ATT Weighting & 0.0884 \\
Matching Weights & 0.0869 \\
Overlap Weights & 0.0874 \\
\bottomrule
\end{tabular}
\end{table}

The weighting approaches demonstrated superior overall performance compared to matching methods. Matching weights achieved the lowest mean SMD (0.08751), followed closely by ATE weighting (0.08855) and overlap weights (0.08823). These results suggest that weighting techniques may be more effective at utilizing the full sample while achieving balance, whereas matching methods necessarily reduce sample size through the selection process.

\begin{figure}
    \centering
    \includegraphics[width=1\linewidth]{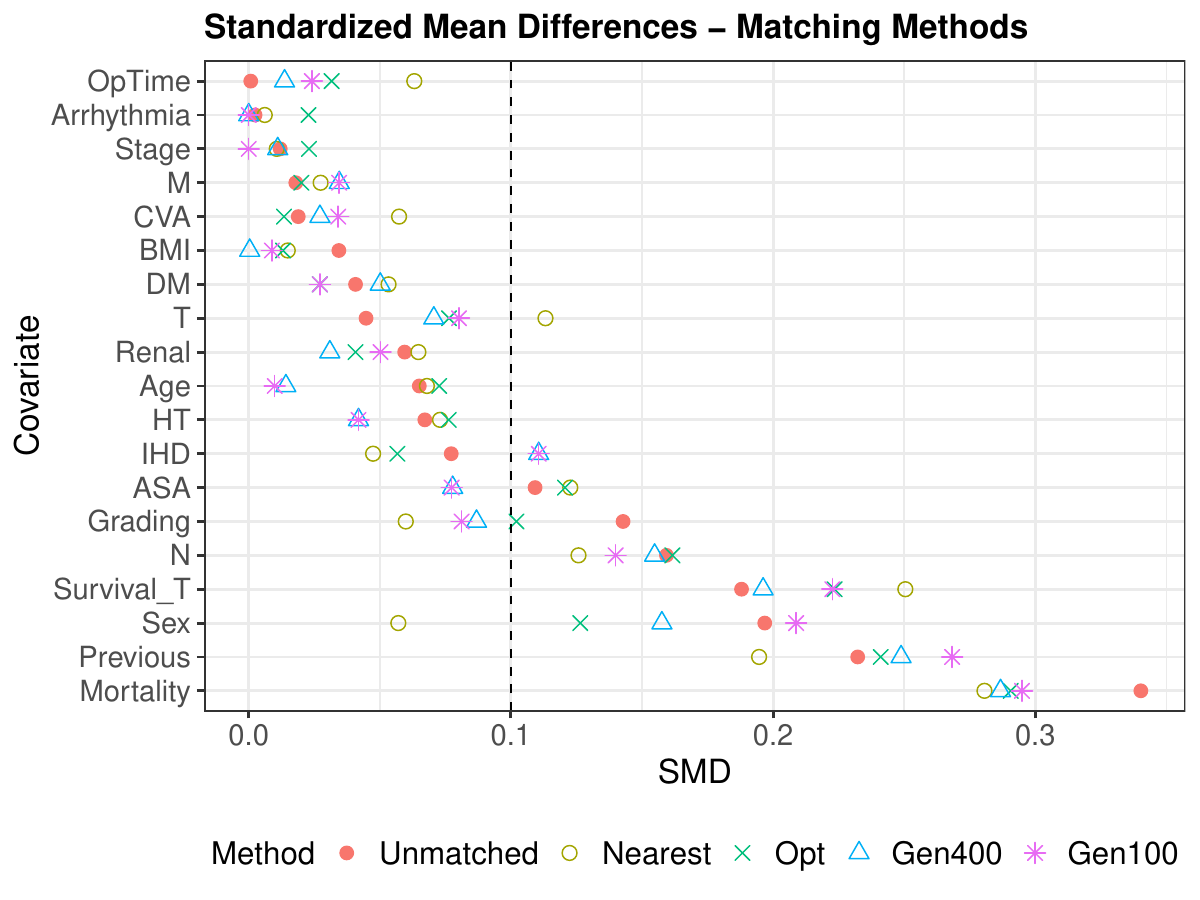}
    \caption{Standardized Mean Differences by covariate for different matching methods. Genetic matching shows the most consistent improvements across variables, particularly for tumor-related characteristics.}
    \label{fig:smd_matching}
\end{figure}

\begin{figure}
    \centering
    \includegraphics[width=1\linewidth]{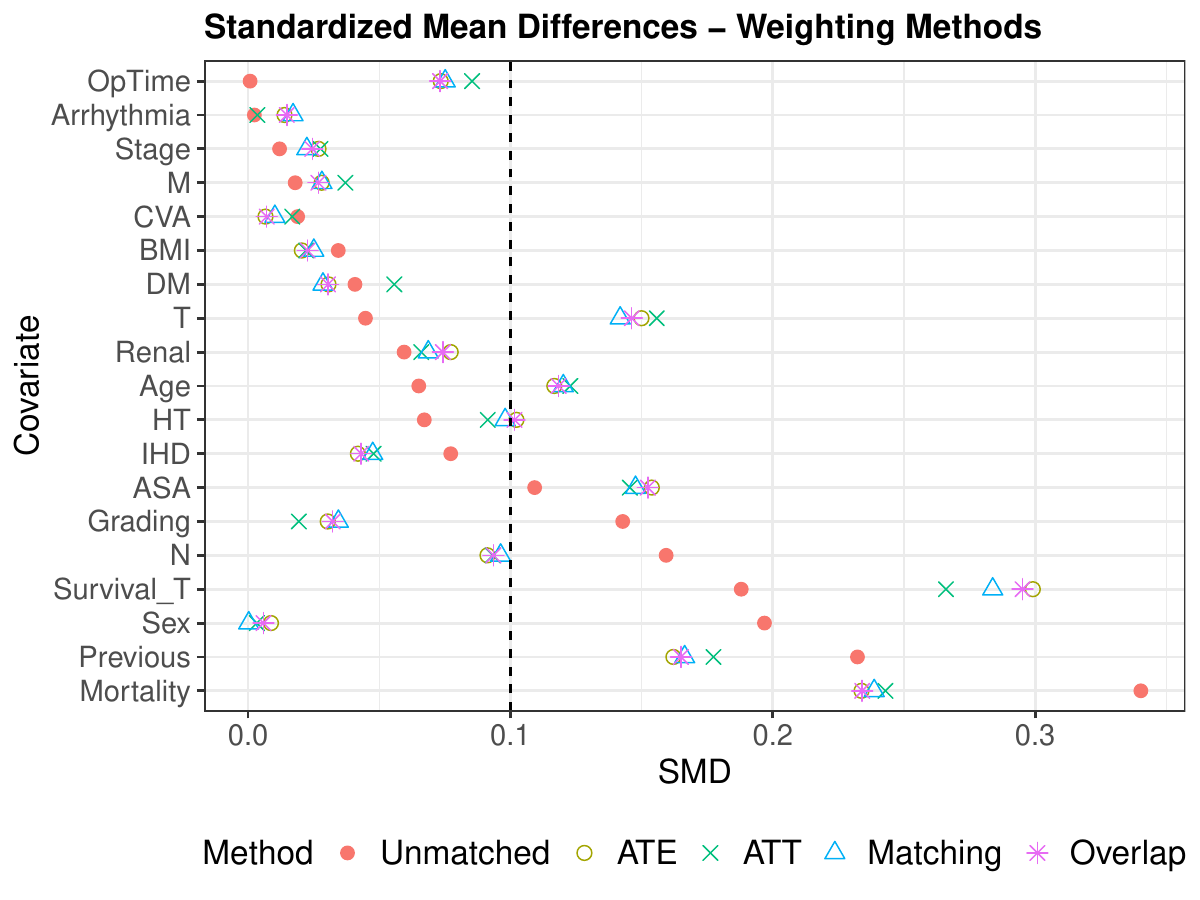}
    \caption{Standardized Mean Differences by covariate for different weighting methods. Matching weights demonstrate the most uniform balance improvements across all covariates.}
    \label{fig:smd_weighting}
\end{figure}

Figures~\ref{fig:smd_matching} and \ref{fig:smd_weighting} provide detailed covariate-specific SMD comparisons, revealing important nuances in method performance. The matching methods (Figure~\ref{fig:smd_matching}) show variable effectiveness across different covariates, with genetic matching consistently performing well across most variables but showing particular strength in balancing categorical variables such as tumor stage and nodal status. The weighting approaches (Figure~\ref{fig:smd_weighting}) demonstrate more consistent balance improvements across the covariate spectrum, with matching weights showing the most uniform performance.

\begin{figure}
    \centering
    \includegraphics[width=1\linewidth]{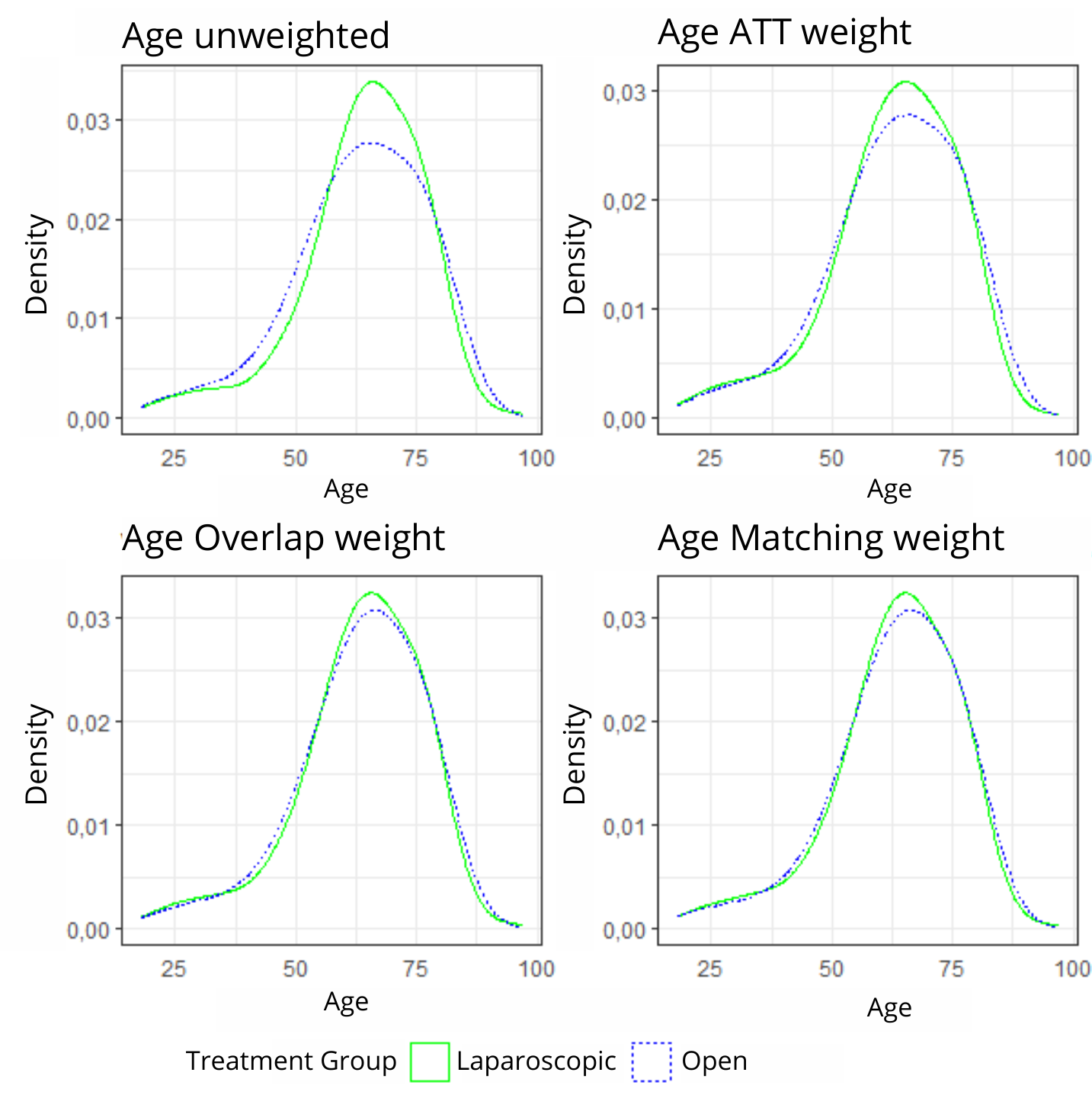}
    \caption{Probability density plots of patient age by treatment group under different weighting schemes. Overlap weights and matching weights achieve the best distributional balance, while ATT and ATE weights show less effective centering.}
    \label{fig:age_centering}
\end{figure}

The impact of weighting on covariate distributions is illustrated in Figure~\ref{fig:age_centering}, which shows the probability density of patient age across different weighting schemes. This visualization reveals how various weighting methods alter the effective study population. Overlap weights and matching weights produced the most similar distributions between treatment groups, indicating better balance achievement. In contrast, ATT weights showed more pronounced differences in the age distribution tails, suggesting less effective balance for this continuous covariate. The ATE weights performed similarly to ATT weights, showing moderate improvement but not achieving the level of balance seen with overlap and matching weights.

Beyond aggregate balance measures, we examined the statistical significance of covariate differences using appropriate hypothesis tests. Table~\ref{tab:p_values_balance} presents p-values for independence tests between each covariate and treatment assignment, comparing unmatched and genetically matched samples. Variables showing improved balance (higher p-values after matching) are marked with asterisks. The most notable improvements occurred for critical tumor characteristics: the M variable (metastasis status) improved from $p = 0.006$ to $p = 0.247$, and Stage improved from $p < 0.001$ to $p = 0.170$, with genetic matching successfully removing statistical dependence between these important prognostic factors and treatment assignment.

\begin{table}[htpb]
\centering
\caption{P-values for Covariate Balance in Unmatched Data and Data Sets Adjusted with Genetic Matching with 100 and 400 Population Sizes}
\label{tab:p_values_balance}
\footnotesize
\begin{tabular}{ll ccc}
\toprule
\textbf{Variable} & \textbf{Test} & \multicolumn{3}{c}{\textbf{P-value}} \\ 
\cmidrule{3-5} 
& & \textbf{Unmatched} & \textbf{Gen \newline 100} & \textbf{Gen \newline 400} \\
\midrule
Age             & t-test        & 0.278 & 0.961* & 0.818* \\
BMI             & t-test        & 0.567 & 0.959* & 0.995* \\
ASA             & chisq         & 0.303 & 0.662* & 0.617* \\
Arrhythmia      & chisq         & 1.000 & 0.781  & 1.000 \\
T               & chisq         & \textbf{$<$ 0.001} & \textbf{0.006}* & \textbf{0.005} \\
N               & chisq         & \textbf{$<$ 0.001} & \textbf{0.016}* & \textbf{0.013} \\
M               & chisq         & \textbf{0.006} & 0.247* & 0.174* \\
Stage           & chisq         & \textbf{$<$ 0.001} & 0.170* & 0.109* \\
Grading         & chisq         & 0.036 & 0.047* & 0.112* \\
OpTime          & t-test        & 0.990 & 0.643  & 0.825 \\
DM              & chisq         & 0.542 & 0.334  & 0.459 \\
HT              & chisq         & 0.288 & 0.665* & 0.537* \\
CVA             & chisq         & 0.834 & 0.913* & 0.741 \\
Renal           & chisq         & 0.404 & 0.738* & 0.738* \\
IHD             & chisq         & 0.219 & 0.210  & 0.084 \\
Sex             & chisq         & \textbf{0.001} & \textbf{0.013}* & \textbf{0.013}* \\
Survival\_T     & t-test        & \textbf{0.007} & \textbf{0.003} & \textbf{0.006} \\
Previous        & chisq         & \textbf{$<$ 0.001} & \textbf{$<$ 0.001} & \textbf{$<$ 0.001} \\
Mortality       & chisq         & \textbf{$<$ 0.001} & \textbf{$<$ 0.001} & \textbf{$<$ 0.001} \\
\bottomrule
\end{tabular}
\end{table}

\begin{figure*}
    \centering
    \includegraphics[width=1\linewidth]{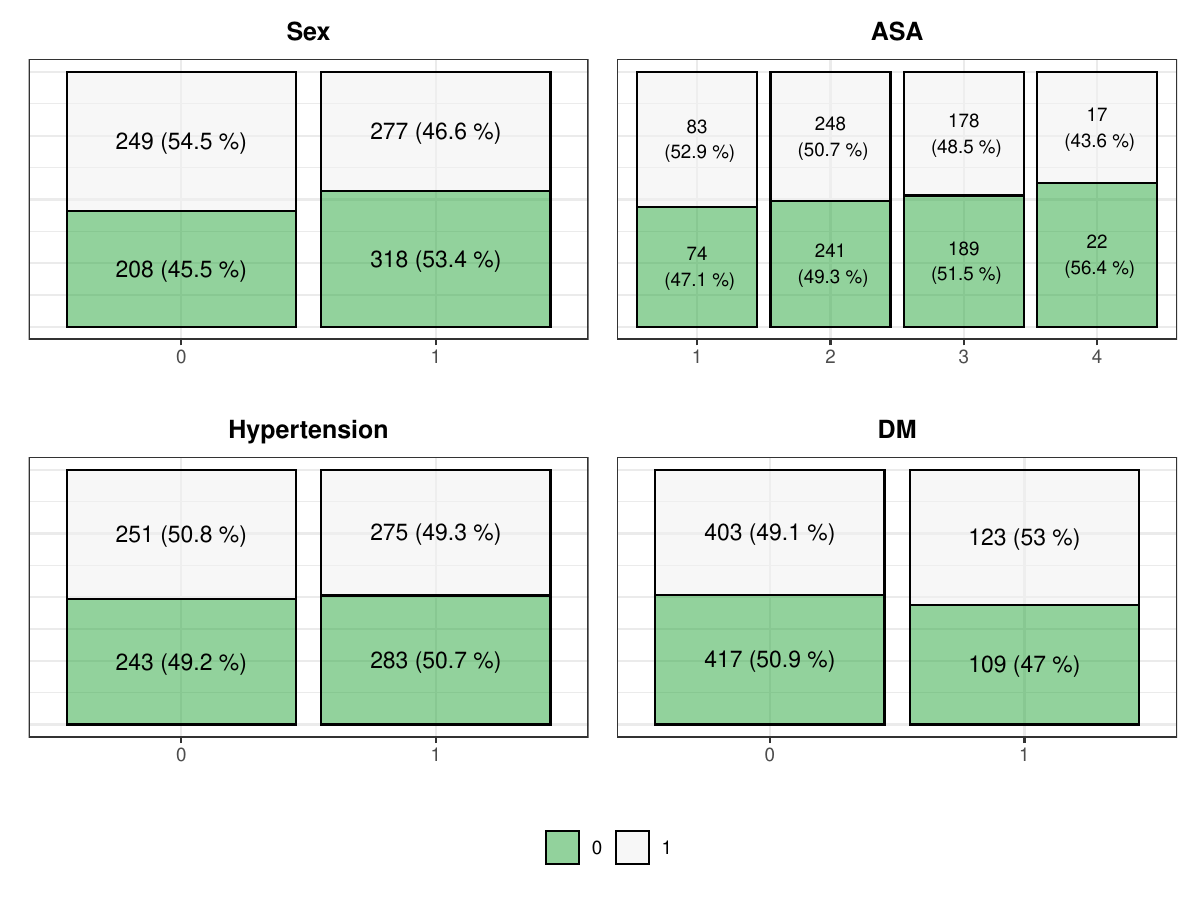}
    \caption{Structure of patient cohort by medical history depending on surgical technique after genetic matching. Comparison with the original unmatched data shows improved balance in comorbidity distributions between treatment groups.}
    \label{fig:medical_history_matched}
\end{figure*}

\begin{figure*}
    \centering
    \includegraphics[width=1\linewidth]{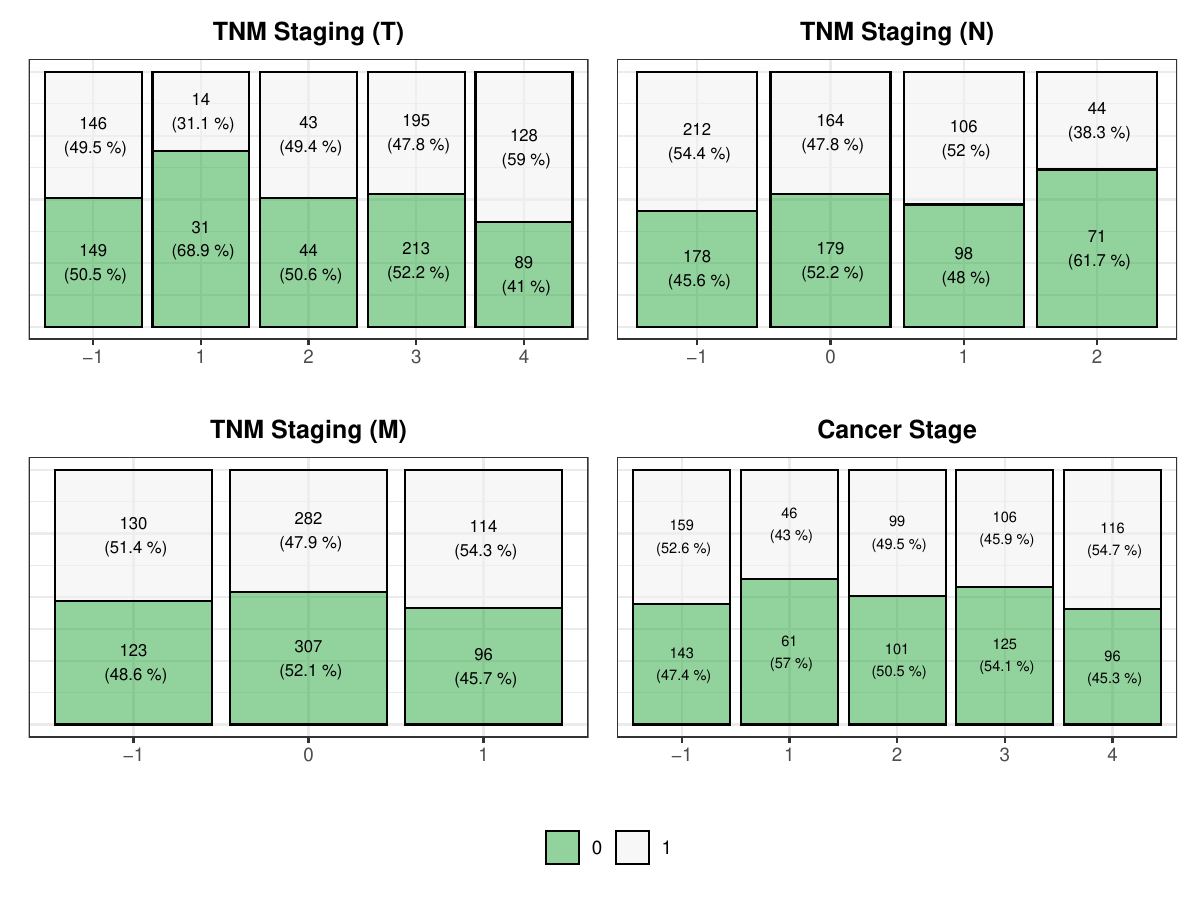}
    \caption{Structure of patient cohort by tumor characteristics depending on surgical technique after genetic matching. The matched sample demonstrates better balance in tumor stage, grading, and nodal status between treatment groups compared to the original data.}
    \label{fig:tumor_characteristics_matched}
\end{figure*}

The effectiveness of genetic matching in improving covariate balance is further demonstrated through visual comparison of patient cohort structures. Figures~\ref{fig:medical_history_matched} and \ref{fig:tumor_characteristics_matched} show the distribution of medical history and tumor characteristics, respectively, comparing the original unmatched data with the genetically matched sample. These visualizations correspond to the original patient cohort structure figures but now reflect the improved balance achieved through propensity score adjustment.

\begin{figure*}
    \centering
    \includegraphics[width=0.75\linewidth]{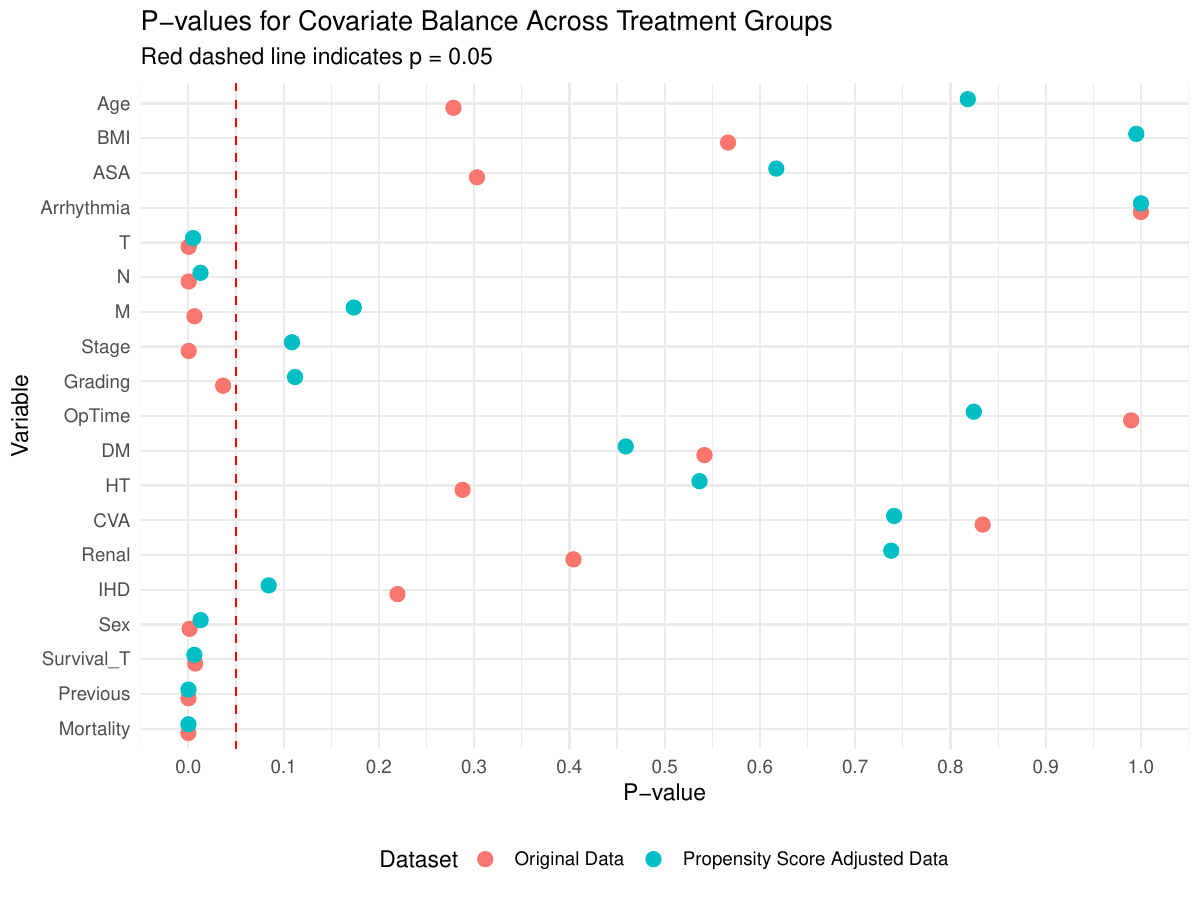}
    \caption{Comparison of p-values for covariate independence tests before and after genetic matching. Higher p-values indicate better balance between treatment groups, with most variables showing improvement after matching adjustment.}
    \label{fig:pvalue_comparison}
\end{figure*}

Figure~\ref{fig:pvalue_comparison} provides a comprehensive visualization of the p-value changes achieved through genetic matching, offering an alternative perspective on the statistical significance improvements shown in Table~\ref{tab:p_values_balance}. This figure clearly illustrates which variables achieved meaningful balance improvements and which remained challenging to balance even after matching.

An important limitation of our current analysis is that the propensity score model was estimated using only four variables, while we assessed balance across 15 additional variables not included in the model estimation. This discrepancy between model variables and balance assessment variables may limit the effectiveness of our propensity score adjustments. Variables not included in the propensity score model cannot be directly balanced through the matching or weighting process, which may explain why some variables (such as Survival\_T, Previous, and Mortality) maintained statistical significance even after matching.

The results demonstrate that our quantum neural network-derived propensity scores, particularly from the FakeManhattanV2 noise model, successfully enabled meaningful covariate balance improvements. Genetic matching emerged as the most effective matching approach, while matching weights performed best among weighting methods. The substantial improvements in tumor-related variables (M and Stage) are particularly encouraging, as these represent critical confounders in surgical outcome studies. However, the persistence of imbalance in some variables highlights the importance of comprehensive covariate selection during propensity score model development for future analyses.

\subsection{Survival analysis}

\begin{figure*}[htpb]
    \centering
    \includegraphics[width=1\linewidth]{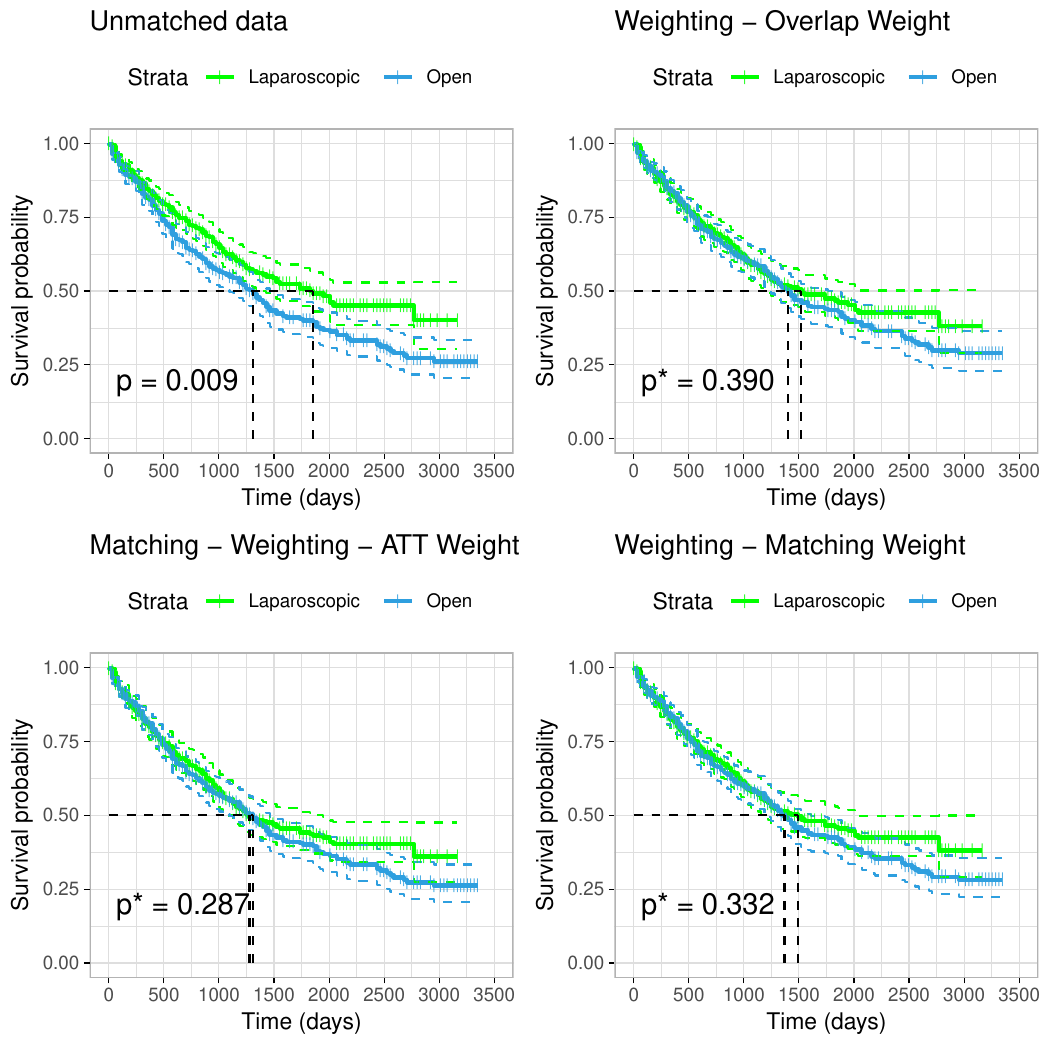}
    \caption{}
    \label{fig:surv_weight}
\end{figure*}

In this section, we address the fundamental research question of whether patients undergoing laparoscopic surgery demonstrate statistically significantly longer survival times compared to those receiving open surgery. Our analytical approach employed both standard survival analysis methods (Kaplan-Meier estimation with log-rank testing) and propensity score techniques to mitigate treatment selection bias associated with disease stage and other confounding factors. The selection of surgical technique in our dataset was not random but rather influenced by patient characteristics, tumor stage, and clinical factors, necessitating careful adjustment to obtain unbiased treatment effect estimates.
Prior to conducting survival analysis, we examined the censoring patterns in both the original dataset and propensity score-adjusted datasets. In the original cohort of 889 patients, the laparoscopic group (n=483) exhibited a censoring rate of 64.0\% with 174 deaths, while the open surgery group (n=406) showed a lower censoring rate of 47.3\% with 214 deaths, yielding an overall censoring rate of 56.4\%. Following nearest-neighbor matching, the matched dataset retained 823 patients with similar censoring patterns: 59.2\% in the laparoscopic group (n=417, 170 deaths) and 47.3\% in the open group (n=406, 214 deaths). This differential censoring pattern suggests that patients in the open surgery group experienced higher mortality rates during the follow-up period, which could reflect either treatment effects or underlying differences in patient characteristics and disease severity.
The Kaplan-Meier survival analysis revealed markedly different results depending on whether propensity score adjustments were applied. Figure~\ref{fig:surv_weight} displays survival curves across four analytical scenarios: unweighted analysis and three propensity score weighting methods (matching weights, average treatment effect on the treated [ATT] weights, and overlap weights). The unweighted analysis yielded a statistically significant log-rank test result (p = 0.009), suggesting a survival advantage for laparoscopic surgery. However, this apparent benefit disappeared entirely when propensity score weighting was applied. The matching weights approach produced a non-significant p-value of 0.332, ATT weights yielded p = 0.287, and overlap weights resulted in p = 0.390. These consistently non-significant results across all weighting methods strongly suggest that the observed survival difference in the crude analysis was attributable to confounding bias rather than a true treatment effect.

\begin{figure*}[htpb]
    \centering
    \includegraphics[width=1\linewidth]{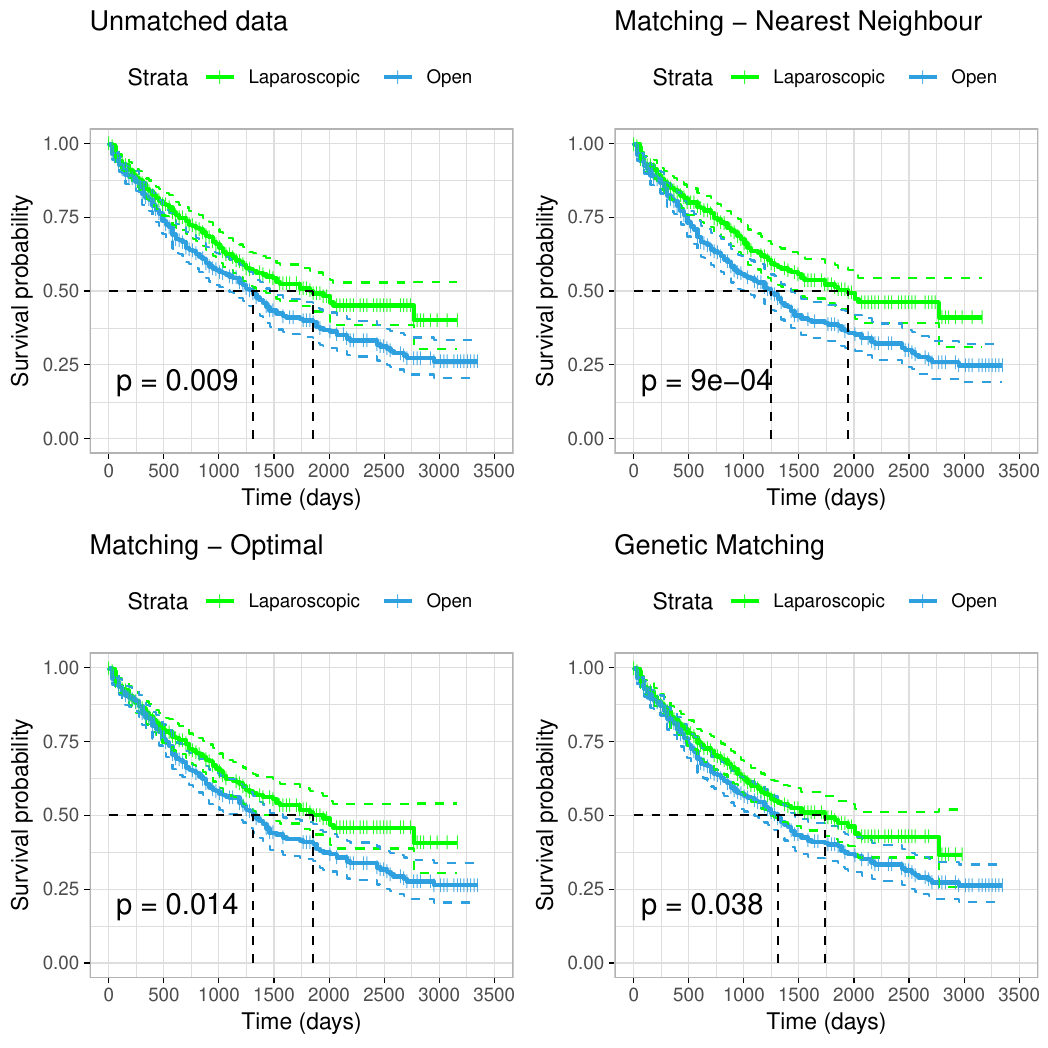}
    \caption{}
    \label{fig:surv_match}
\end{figure*}

Conversely, the propensity score matching analysis presented in Figure~\ref{fig:surv_match} yielded contrasting findings. The unmatched analysis maintained statistical significance ($p = 0.009$), while various matching algorithms not only preserved but in some cases strengthened the apparent survival advantage for laparoscopic surgery. Nearest-neighbor matching produced a highly significant result ($p = 9 \times 10^{-4}$), optimal matching yielded $p = 0.014$ and genetic matching, which adjusted covariates the most, resulted in $p = 0.038$. This trend, where increased covariate adjustment leads to decreased significance, aligns with the findings from the weighting approach, where all log-rank tests became non-significant after adjustment. This discrepancy between matching and weighting results highlights the fundamental differences in how these propensity score methods address confounding bias and suggests that more thoroughly adjusted covariates lead to less significance between treatment groups.

To further investigate the treatment effect while controlling for multiple covariates simultaneously, we fitted Cox proportional hazards models using three approaches: non-adjusted analysis, genetic matching, and matching weights. The sample sizes were 845, 676, and 845 patients respectively, with the genetic matching approach excluding 169 patients who could not be adequately matched. Remarkably, across all three Cox models, the treatment group variable demonstrated no statistically significant effect on mortality risk. The non-adjusted model yielded a hazard ratio of 1.153 (95\% CI: 0.927-1.434, p = 0.200), genetic matching produced HR = 1.138 (95\% CI: 0.890-1.454, p = 0.303), and matching weights resulted in HR = 1.021 (95\% CI: 0.822-1.269, p = 0.851). The hazard ratio closest to unity occurred in the weighted analysis, providing the strongest evidence for no treatment effect after appropriate confounder adjustment. A comprehensive overview of the p-values for all variables across these models is presented in Table \ref{tab:cox_models_pvals}, with corresponding forest plots of hazard ratios detailed in Figures \ref{fig:cox}, \ref{fig:cox_matched}, and \ref{fig:cox_weighted} in the Appendix.

\begin{table}[htpb]
    \centering
    \caption{P-values from Unmatched, Matched, and Weighted Cox Models}
    \label{tab:cox_models_pvals}
    \footnotesize
    \begin{tabular}{lccc}
        \toprule
        \textbf{Variable} & \textbf{Unmatched} & \textbf{Matched} & \textbf{Weighted} \\
        \midrule
        Age & 0.1150 & 0.0507. & 0.1742 \\
        BMI & 0.0020** & 0.0014** & 0.0028** \\
        ASA & 0.0354* & 0.0423* & 0.0379* \\
        Arrhythmia & 0.1123 & 0.0794. & 0.0398* \\
        T & 0.4706 & 0.4733 & 0.5857 \\
        N & 0.2740 & 0.1713 & 0.2284 \\
        M & $<$2e-16*** & $<$2e-16*** & 1.23e-13*** \\
        Stage & 0.0656. & 0.0568. & 0.1548 \\
        Grading & 0.6332 & 0.7424 & 0.5689 \\
        OpTime & 0.1304 & 0.1871 & 0.2032 \\
        DM & 0.1110 & 0.2025 & 0.1361 \\
        HT & 0.1567 & 0.1357 & 0.2147 \\
        CVA & 0.5771 & 0.5066 & 0.5104 \\
        Renal & 0.5077 & 0.1884 & 0.5895 \\
        IHD & 0.5057 & 0.6928 & 0.6001 \\
        Sex & 0.2056 & 0.3371 & 0.0412* \\
        Previous & 0.7681 & 0.8286 & 0.5926 \\
        Group & 0.2003 & 0.4661 & 0.8511 \\
        \bottomrule
        \addlinespace[0.5em]
        \multicolumn{4}{l}{\textit{Signif. codes: 0 ‘***’ 0.001 ‘**’}} \\
        \multicolumn{4}{l}{\textit{0.01 ‘*’ 0.05 ‘.’ 0.1 ‘ ’ 1}}
    \end{tabular}
\end{table}

All three Cox models demonstrated excellent overall statistical performance with highly significant global tests ($p < 2 \times 10^{-16}$) and similar discrimination ability, as evidenced by concordance indices ranging from 0.712 to 0.721. These values indicate good model performance and suggest that the covariates included in the analysis are collectively strong predictors of mortality risk. The consistent performance across adjustment methods provides confidence in the robustness of our modeling approach.
Examination of individual covariate effects revealed several important patterns. Metastasis status (M) emerged as the most powerful predictor across all models, with hazard ratios ranging from 5.73 to 5.86 and p-values consistently below $2 \times 10^{-16}$. This finding aligns with clinical expectations, as metastatic disease represents advanced cancer stage associated with poor prognosis. Body mass index (BMI) demonstrated a consistent protective effect across all models ($HR < 1, p < 0.01$), which may reflect the "obesity paradox" observed in some cancer populations or indicate that patients with higher BMI may have better nutritional reserves. The American Society of Anesthesiologists (ASA) physical status classification maintained significance as a risk factor in the non-adjusted and weighted models, reflecting its role as a measure of perioperative risk and overall patient health status.
Notably, certain covariates exhibited differential significance patterns across adjustment methods. Arrhythmia and sex became statistically significant only in the matching weights model, with p-values of 0.040 and 0.041 respectively. This emergence of previously masked associations suggests that the weighting approach more effectively controlled for confounding relationships between covariates, allowing the independent effects of these variables to become apparent. Age showed borderline significance in the genetic matching model (p = 0.045) but not in other approaches, potentially reflecting the specific characteristics of the matched subset.
The Aalen additive regression analysis corroborated the Cox model findings while providing additional insights into the cumulative hazard structure. The treatment effect coefficient in the unweighted model was $3.30\times10^{-4}$ (p = 0.118), in genetic matching was $4.60\times10^{-4}$ (p = 0.112), and in matching weights was substantially reduced to $9.79\times10^{-5}$ (p = 0.648). The dramatic reduction in both coefficient magnitude and statistical significance in the weighted model provides compelling evidence that any apparent treatment effect in crude analyses was indeed due to confounding bias. The Aalen models maintained overall statistical significance across all approaches ($\chi^2$ values of 104.75, 83.8, and 98.36 respectively, all $p < 10^{-11}$), confirming the collective importance of the included covariates.
The pattern of covariate significance in the Aalen models largely paralleled the Cox regression results. BMI, ASA score, and metastasis status remained consistently significant across all adjustment methods, while sex achieved significance only in the weighted model (p = 0.044). The robustness of certain predictors across both modeling approaches and all adjustment methods strengthens confidence in their clinical relevance and prognostic importance.
The divergent results between propensity score matching and weighting methods warrant careful consideration. Matching methods aim to create balanced treatment groups by pairing similar patients and discarding those without adequate matches, potentially improving internal validity at the expense of external generalizability. However, matching can fail to achieve balance on unobserved characteristics and may introduce bias if the matching process is imperfect. Weighting methods, conversely, retain all patients while adjusting their contributions to the analysis based on propensity scores, potentially maintaining better external validity and statistical power. The superior covariate balance achieved by the weighting approach in our analysis, particularly for the crucial censoring status variable, suggests that weighting more effectively addressed confounding bias in this dataset.
The clinical implications of these findings are substantial. The absence of a statistically significant survival difference between laparoscopic and open surgical approaches, when properly accounting for patient selection factors, suggests that the choice between these techniques should be based on other considerations such as perioperative morbidity, recovery time, hospital length of stay, and surgeon expertise rather than long-term survival expectations. This conclusion aligns with the broader literature suggesting that while laparoscopic approaches may offer short-term advantages in terms of recovery and complications, long-term oncological outcomes are generally equivalent when appropriate patient selection and surgical technique are employed.

\section{Conclusion}

This study demonstrates the potential of quantum neural networks (QNNs) for propensity score estimation in observational biomedical research, leveraging their unique capabilities to address selection bias in a colorectal carcinoma dataset of 1177 patients (2001–2009). Our QNN models, utilizing a linear ZFeatureMap, SummedPaulis operator, and the Covariance Matrix Adaptation Evolution Strategy (CMA-ES) for optimization, effectively handled the non-convex, noisy quantum environment. Variance regularization minimized measurement noise, while simulations under exact, sampling (1024 shots), and noisy hardware (FakeManhattanV2) conditions revealed that noise modeling unexpectedly enhanced predictive performance, achieving AUCs up to 0.750 in small samples (n=100). This advantage, driven by CMA-ES’s robustness and noise-aware design, diminished with larger samples, aligning QNN performance with classical methods. Propensity score adjustments via genetic matching and matching weights, informed by QNN-derived scores, reduced covariate imbalances (SMDs of 0.0849 and 0.0869), particularly for tumor-related variables like Stage and metastasis status. Survival analyses, including Kaplan-Meier, Cox proportional hazards, and Aalen additive regression, confirmed no significant survival differences between laparoscopic and open surgery after adjustment (p-values 0.287–0.851), attributing unadjusted differences to confounding factors like disease severity. The QNN approach, enhanced by CMA-ES and noise modeling, showcases quantum machine learning’s promise for small-sample, high-dimensional biomedical data, though limitations in covariate selection suggest future models incorporate broader variable sets. Clinically, these findings imply surgical technique decisions should prioritize perioperative outcomes and expertise over long-term survival, underscoring QNNs’ role in advancing causal inference through quantum-enhanced computation.

\begin{acknowledgments}
The following grants are acknowledged for the financial support provided for this research: grant of SGS No. SP2025/072 and SP2024/017, VSB-Technical University of Ostrava, Czech Republic.
\end{acknowledgments}

\appendix

\section{Detailed Quantum Neural Network Architecture}
\label{sec:qnn_details}

This appendix provides a detailed technical description of the components comprising the Quantum Neural Network (QNN) used for propensity score estimation in this study. We elaborate on the feature map, the observable and cost operator, and the loss function and optimization routine.

\subsection{Feature Map: Data Encoding with ZFeatureMap}
A critical component of any QNN is the feature map, which encodes classical data into the quantum states upon which the circuit will operate. For this task, we employed the \texttt{ZFeatureMap} from Qiskit's circuit library. The \texttt{ZFeatureMap} is a linear feature map that transforms an $n$-dimensional classical input vector $\mathbf{x} = (x_1, x_2, \dots, x_n)$ into a quantum state across $n$ qubits.

Mathematically, the unitary operator for the \texttt{ZFeatureMap} is defined as:
$$ U_{\phi}(\mathbf{x}) = \left( \prod_{i=1}^n \exp(i \cdot x_i \cdot Z_i) \right) H^{\otimes n} $$
where $H^{\otimes n}$ first applies a Hadamard gate to each qubit to create a uniform superposition. Subsequently, the term $\exp(i \cdot x_i \cdot Z_i)$ introduces phase rotations on each qubit via Pauli-Z gates, with the rotation angle being proportional to the corresponding input feature $x_i$. This process generates a product state without entanglement:
$$ |\psi(\mathbf{x})\rangle = \frac{1}{\sqrt{2^n}} \sum_{k=0}^{2^n-1} e^{i \sum_{i=1}^n k_i x_i} |k\rangle $$
where $k_i$ denotes the $i$-th bit of the basis state $|k\rangle$. The choice of the \texttt{ZFeatureMap} was motivated by its simplicity and efficiency, as it requires only single-qubit gates. This makes it highly compatible with the constraints of noisy intermediate-scale quantum (NISQ) devices and reduces the complexity of classical simulation. While its lack of entangling operations limits its capacity to capture complex data correlations compared to feature maps like the \texttt{ZZFeatureMap}, its effectiveness in this study suggests that linear feature encoding was sufficient for the given task.

\subsection{Observable and Cost Operator: The SummedPaulis Operator}
In our QNN architecture, we utilized the \texttt{SummedPaulis} observable from Qiskit's quantum machine learning framework for the dual role of both the primary observable and the cost operator. This operator is defined as a weighted sum of single-qubit Pauli operators. For a system of $n$ qubits, its general form is:
$$ \hat{H} = a \hat{I} + \sum_i \sum_{P \in \{X, Y, Z\}} b_{i,P} \hat{P}_i $$
where $a$ is a coefficient for the identity operator $\hat{I}$, and $b_{i,P}$ are the coefficients for the Pauli operator $\hat{P}$ on the $i$-th qubit.

This operator served two functions simultaneously. First, as an \textbf{observable}, the QNN's output prediction for a given input $\mathbf{x}$ is the expectation value of the \texttt{SummedPaulis} operator with respect to the final quantum state $|\psi(\mathbf{x}, \theta)\rangle$. This value, $f(\mathbf{x}, \theta) = \langle \psi(\mathbf{x}, \theta) | \hat{H} | \psi(\mathbf{x}, \theta) \rangle$, is a classical scalar that can be mapped to a propensity score. Second, as a \textbf{cost operator}, the same expectation value $f(\mathbf{x}, \theta)$ is used within the loss function that the QNN minimizes during training. This creates a unified objective where the quantity being predicted is the same as the quantity being optimized. The use of a single \texttt{SummedPaulis} operator for both roles simplifies the QNN design and is efficient, as only one set of measurements is needed to compute both the prediction and the cost function gradient. Its reliance on single-qubit measurements makes it well-suited for near-term quantum hardware.

\subsection{Loss Function and Optimization Routine}
The parameters of the QNN were trained by minimizing a total loss function that balances model fitting with variance regularization, given by $L(\theta) = L_{fit}(\theta) + \alpha L_{var}(\theta)$.

The fitting loss, $L_{fit}(\theta)$, was defined using a \textbf{Squared Loss} function. Propensity score estimation is fundamentally a regression task where the model predicts a probability based on covariates. We trained the QNN to predict the binary treatment assignment label $y_i \in \{0, 1\}$. The squared loss measures the discrepancy between the QNN's output $f(x_i, \theta)$ and the true label $y_i$:
$$ L_{fit}(\theta) = \sum_i w_i \left(f(x_i, \theta) - y_i\right)^2 $$
where $w_i$ are optional weights for each data point. This is a standard choice for regression tasks and provides a straightforward objective for the optimizer. The gradient of this loss with respect to a QNN parameter $p_j$ is given by:
$$ \frac{\partial L_{fit}}{\partial p_j} = 2\sum_i w_i \left(f(x_i, \theta) - y_i\right) \frac{\partial f(x_i, \theta)}{\partial p_j} $$

The optimization of the loss function was performed using the \textbf{Covariance Matrix Adaptation Evolution Strategy (CMA-ES)} optimizer. CMA-ES is a gradient-free evolutionary algorithm that has demonstrated strong performance and noise resilience in the context of variational quantum algorithm optimization \cite{bonet2023performance, illesova2025numerical}. The hyperparameters are listed in \ref{tab:cmaesa_hyperparameters}. We attribute the notable increase in the AUC for QNNs run on noisy simulators (e.g., AUC of 0.750 for \texttt{qnn\_f\_backend} on the $n=100$ sample) to the robustness of the CMA-ES optimizer. It is exceptionally well-suited to navigating the non-convex and noisy loss landscapes that are characteristic of NISQ-era QNNs, likely turning the simulated hardware noise into a feature that aids exploration of the parameter space and prevents overfitting.

\begin{table}[h!]
    \centering
    \begin{tabular}{|l|c|}
        \hline
        \textbf{Parameter} & \textbf{Value} \\
        \hline
        \textbf{$\sigma_0$} & 0.15 \\
        \hline
        \textbf{Population} & $\lceil 4 + 3 \log(m) \rceil$ \\
        \hline
        \textbf{$\mu$} & 0.5 \\
        \hline
        \textbf{c mean} & 1.0 \\
        \hline
        \textbf{Damp. Factor} & 1.0 \\
        \hline
    \end{tabular}
    \caption{CMA-ES Hyperparameters, assuming $m$ parameters during optimization.}
    \label{tab:cmaesa_hyperparameters}
\end{table}

\section{Appendix: Scalability and Future Outlook}

This appendix addresses the scalability and computational complexity of the Quantum Neural Network (QNN) model employed in this study, particularly in light of current hardware limitations and future potential. The choice to utilize only four features out of a larger available set was primarily dictated by the current constraints of quantum hardware simulation and the inherent computational cost associated with increasing the number of qubits and circuit depth. The chosen `ZFeatureMap` ensures that the number of qubits scales linearly with the number of features, i.e., $O(N)$, where $N$ is the number of features. For our current implementation with 4 features, this translates to 4 qubits. Increasing to the full 77 available features would necessitate 77 qubits, which is beyond the capabilities of current noise-intermediate scale quantum (NISQ) devices for practical applications. While the depth of the `ZFeatureMap` does not inherently grow with $N$, using more expressive, entangling feature maps (e.g., `ZZFeatureMap`) to capture complex correlations between a larger number of features would significantly increase the circuit depth and the total number of quantum gates. Such an increase in circuit depth directly translates to a higher susceptibility to noise on NISQ devices, making the optimization process more challenging and less reliable. The optimization of variational quantum circuits relies on classical optimizers. The cost of a classical optimizer, such as CMA-ES used in this study, scales with the number of parameters in the variational circuit. Should a more complex variational circuit be required to process a larger number of features, the number of trainable parameters would increase substantially, leading to a significant increase in the classical optimization runtime. Scaling this methodology to larger feature sets, such as the full 77 variables available, necessitates significant advancements in both quantum hardware and software. Future progress in this area will depend on more efficient classical simulators or specialized quantum simulators that can handle larger qubit counts and deeper circuits with reduced computational overhead, the advent of robust, fault-tolerant quantum computers to overcome the limitations imposed by noise in NISQ devices, and the development of more advanced Quantum Machine Learning models specifically designed to handle high-dimensional data efficiently, potentially by leveraging novel quantum feature encoding strategies or quantum dimensionality reduction techniques. These advancements will be key to unlocking the full potential of Quantum Neural Networks for real-world applications with high-dimensional datasets.

\section{Survival Analysis Visualization}
\label{sec:survival_analysis}

\begin{figure*}[htpb]
    \centering
    \includegraphics[width=1\linewidth]{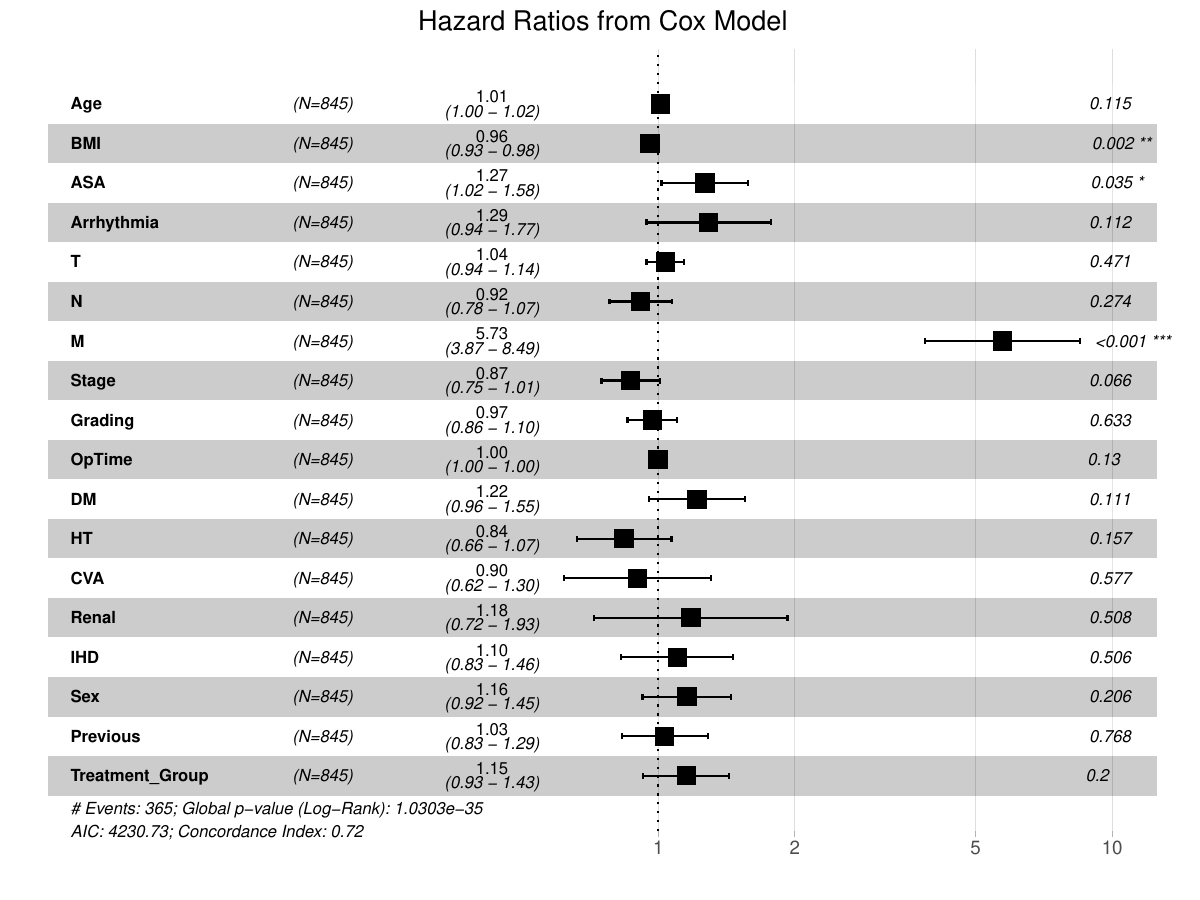}
    \caption{Hazard ratios from the Cox proportional hazards model for the unmatched dataset.}
    \label{fig:cox}
\end{figure*}

\begin{figure*}[htpb]
    \centering
    \includegraphics[width=1\linewidth]{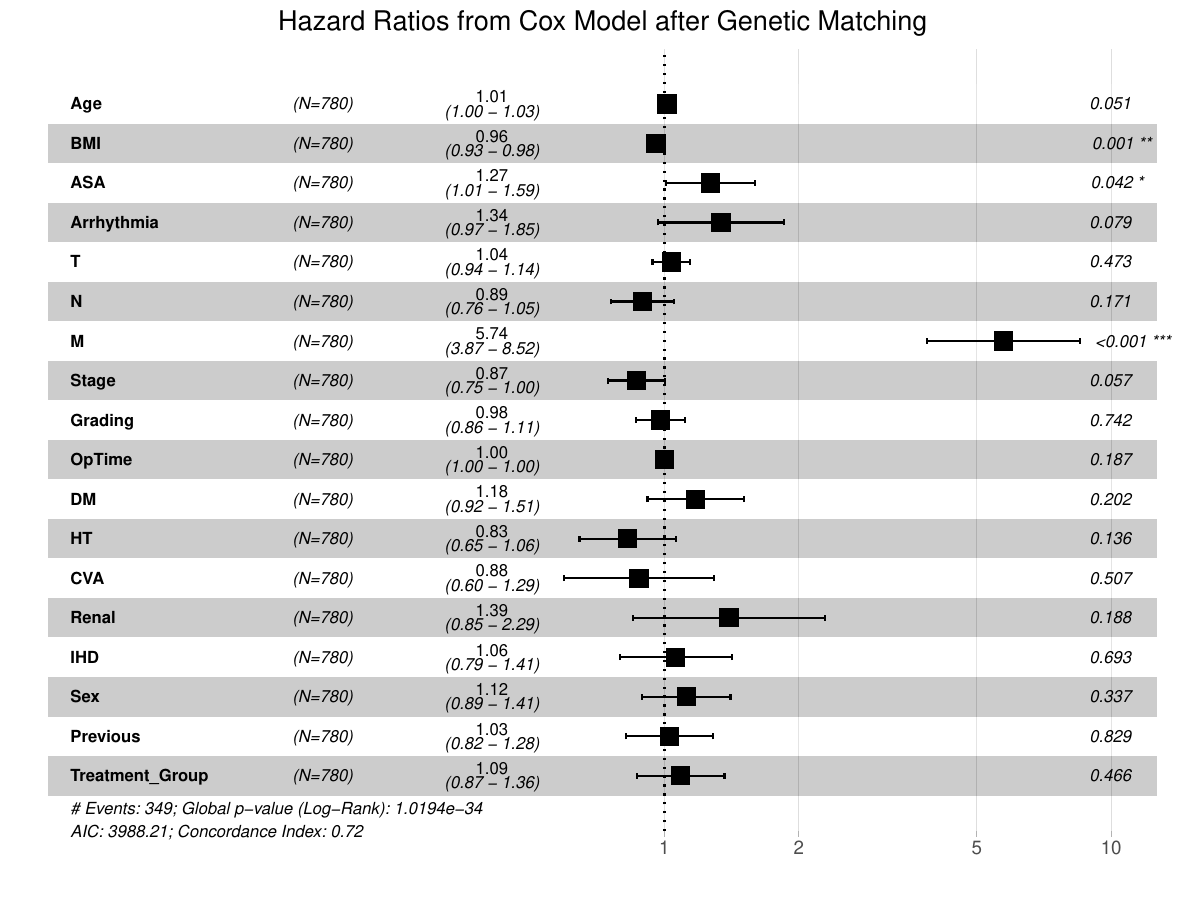}
    \caption{Hazard ratios from the Cox proportional hazards model for the genetically matched dataset (population size of 400).}
    \label{fig:cox_matched}
\end{figure*}

\begin{figure*}[htpb]
    \centering
    \includegraphics[width=1\linewidth]{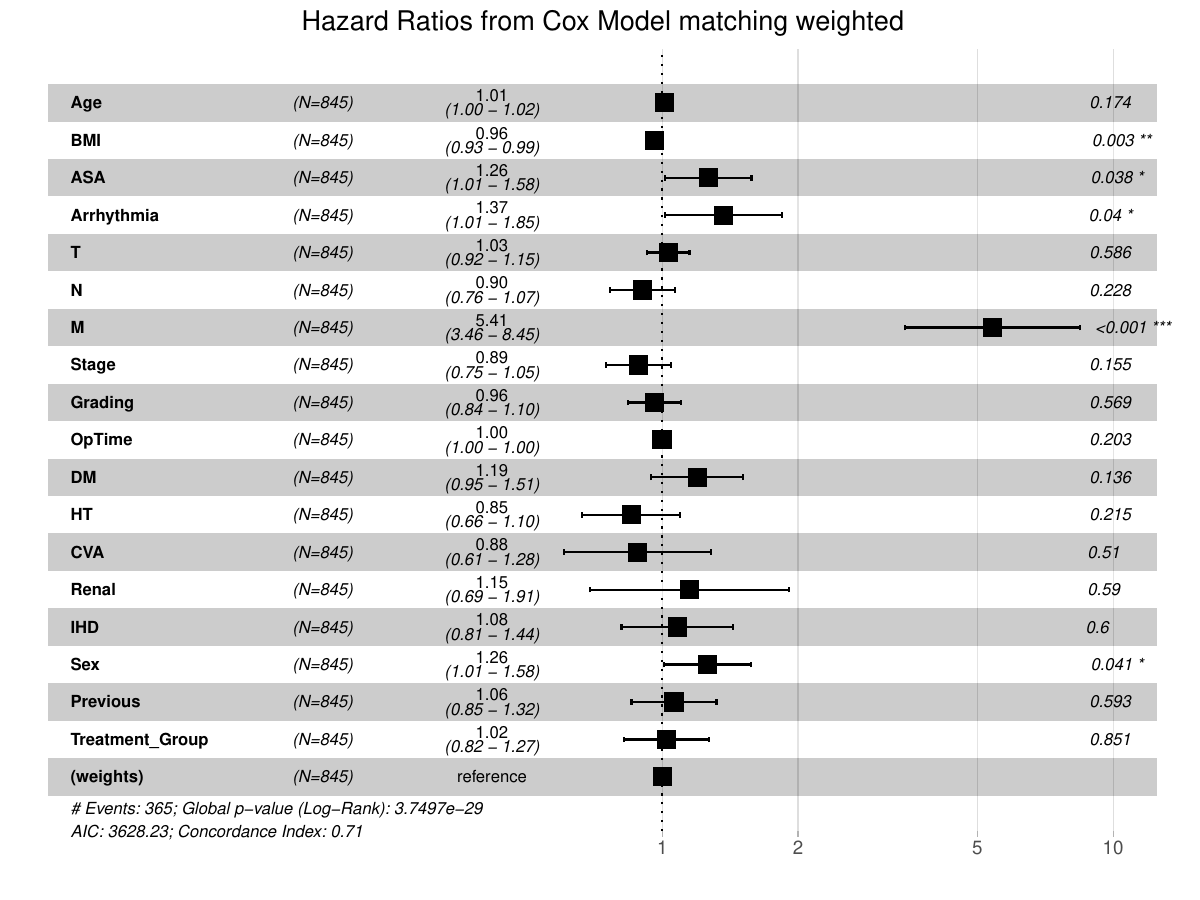}
    \caption{Hazard ratios from the Cox proportional hazards model for the dataset with matching weights.}
    \label{fig:cox_weighted}
\end{figure*}

\begin{figure*}[htpb]
    \centering
    \includegraphics[width=1\linewidth]{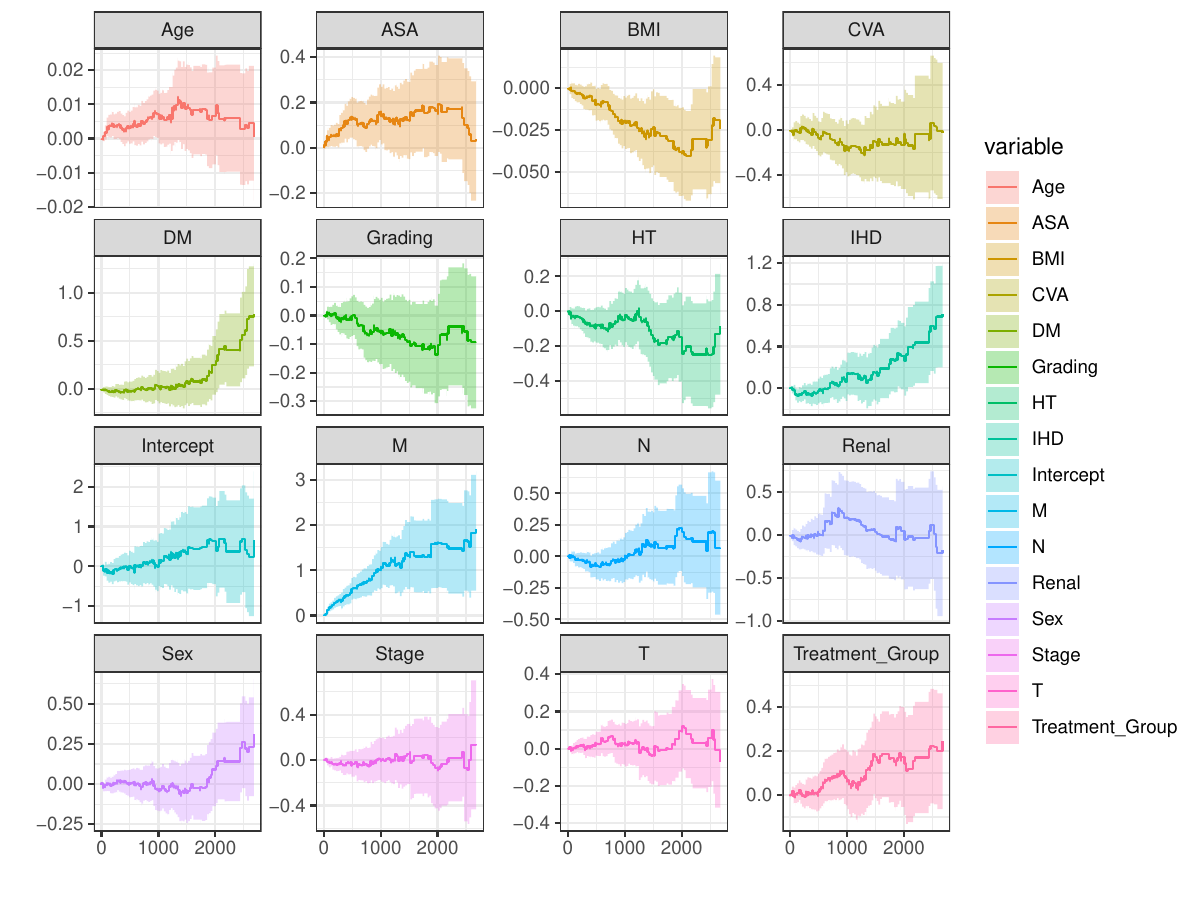}
    \caption{Cumulative regression function plots from Aalen's additive regression model for the unmatched dataset.}
    \label{fig:aareg}
\end{figure*}

\begin{figure*}[htpb]
    \centering
    \includegraphics[width=1\linewidth]{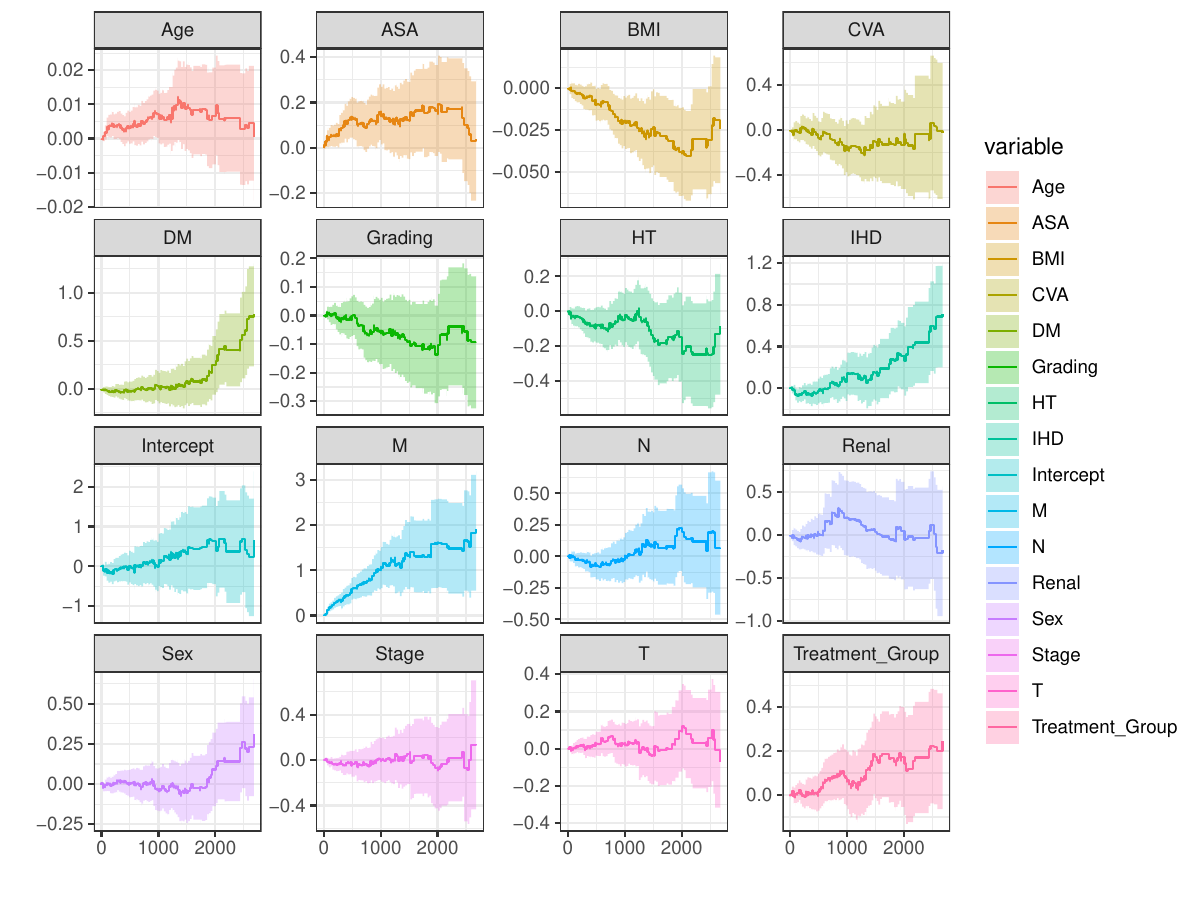}
    \caption{Cumulative regression function plots from Aalen's additive regression model for the genetically matched dataset (population size of 400).}
    \label{fig:aareg_matched}
\end{figure*}

\begin{figure*}[htpb]
    \centering
    \includegraphics[width=1\linewidth]{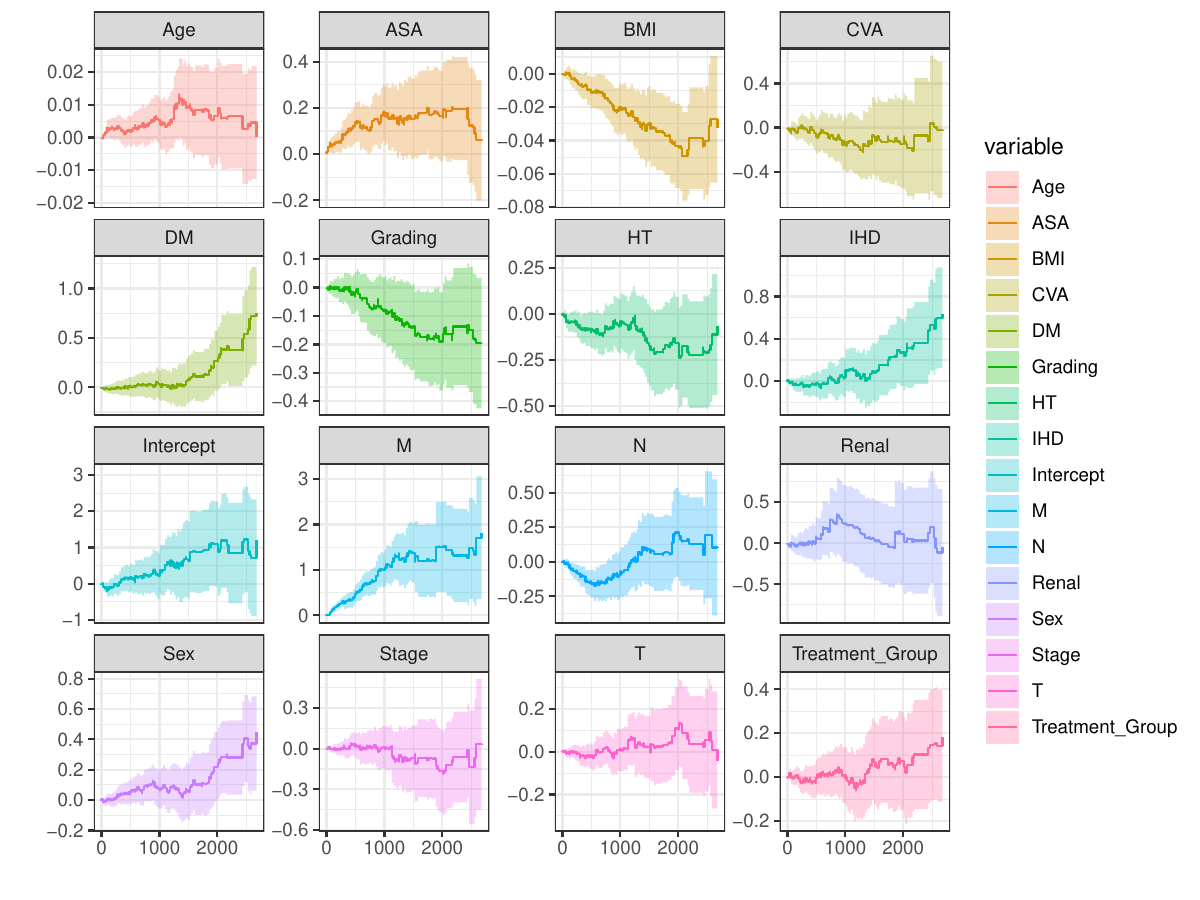}
    \caption{Cumulative regression function plots from Aalen's additive regression model for the dataset with matching weights.}
    \label{fig:aareg_weighted}
\end{figure*}

This appendix presents the results of the survival analysis conducted using Aalen's additive regression model and Cox proportional hazards model, as detailed in Section~\ref{sec:data_set_analysis}. Table~\ref{tab:aareg_comparison} summarizes the Aalen's additive regression model results for the unmatched dataset, genetically matched dataset (population size of 400), and dataset with matching weights, including slope, coefficients, standard errors, z-values, and p-values for all variables analyzed. The hazard ratios from the Cox model are visualized in Figure~\ref{fig:cox} for the unmatched dataset, Figure~\ref{fig:cox_matched} for the genetically matched dataset, and Figure~\ref{fig:cox_weighted} for the weighted dataset. Similarly, the cumulative regression function plots from Aalen's model are shown in Figure~\ref{fig:aareg} for the unmatched dataset, Figure~\ref{fig:aareg_matched} for the genetically matched dataset, and Figure~\ref{fig:aareg_weighted} for the weighted dataset, illustrating the time-varying effects of the covariates across these datasets.

\begin{sidewaystable}[htpb]
\centering
\caption{Aalen's Additive Regression Model Results for Genetic Matching and Matching Weights}
\label{tab:aareg_comparison}
\scriptsize
\begin{tabular}{l *{5}{c} *{5}{c} *{5}{c}}
\toprule
& \multicolumn{5}{c}{\textbf{Non-Adjusted}} & \multicolumn{5}{c}{\textbf{Genetic Matching}} & \multicolumn{5}{c}{\textbf{Matching Weights}} \\
\cmidrule(lr){2-6} \cmidrule(lr){7-11} \cmidrule(lr){12-16}
\textbf{Variable} & \textbf{slope} & \textbf{coef} & \textbf{se(coef)} & \textbf{z} & \textbf{p} & \textbf{slope} & \textbf{coef} & \textbf{se(coef)} & \textbf{z} & \textbf{p} & \textbf{slope} & \textbf{coef} & \textbf{se(coef)} & \textbf{z} & \textbf{p} \\
\midrule
Intercept & 6.70e-04 & 1.55e-03 & 1.10e-03 & 1.410 & 1.58e-01 & 1.54e-04 & 8.54e-04 & 1.49e-03 & 0.575 & 5.66e-01 & 6.79e-04 & 1.61e-03 & 1.14e-03 & 1.410 & 1.58e-01 \\
Group & 1.58e-04 & 3.30e-04 & 2.11e-04 & 1.560 & 1.18e-01 & 1.60e-04 & 4.60e-04 & 2.90e-04 & 1.590 & 1.12e-01 & 1.33e-05 & 9.79e-05 & 2.15e-04 & 0.456 & 6.48e-01 \\
Age & 1.43e-05 & 1.88e-05 & 1.27e-05 & 1.480 & 1.39e-01 & 1.55e-05 & 2.77e-05 & 1.75e-05 & 1.580 & 1.13e-01 & 1.35e-05 & 1.86e-05 & 1.33e-05 & 1.400 & 1.63e-01 \\
BMI & -4.46e-05 & -7.99e-05 & 2.46e-05 & -3.250 & 1.17e-03 & -3.80e-05 & -9.00e-05 & 3.35e-05 & -2.680 & 7.31e-03 & -4.38e-05 & -8.16e-05 & 2.52e-05 & -3.230 & 1.22e-03 \\
ASA & 3.30e-04 & 4.98e-04 & 2.08e-04 & 2.390 & 1.69e-02 & 2.88e-04 & 5.85e-04 & 2.88e-04 & 2.030 & 4.21e-02 & 3.51e-04 & 5.29e-04 & 2.16e-04 & 2.450 & 1.44e-02 \\
T & -3.62e-05 & -1.11e-05 & 1.46e-04 & -0.0755 & 9.40e-01 & 1.17e-04 & 1.23e-04 & 2.17e-04 & 0.565 & 5.72e-01 & -3.85e-05 & -1.51e-05 & 1.48e-04 & -0.102 & 9.19e-01 \\
N & -5.22e-04 & -3.90e-04 & 2.42e-04 & -1.610 & 1.07e-01 & -2.04e-04 & -1.63e-04 & 3.42e-04 & -0.476 & 6.34e-01 & -5.84e-04 & -4.35e-04 & 2.49e-04 & -1.740 & 8.14e-02 \\
M & 3.50e-03 & 3.44e-03 & 5.34e-04 & 6.440 & 1.19e-10 & 3.67e-03 & 4.56e-03 & 7.99e-04 & 5.710 & 1.11e-08 & 3.31e-03 & 3.28e-03 & 5.38e-04 & 6.110 & 1.02e-09 \\
Stage & -6.36e-05 & -7.27e-05 & 2.26e-04 & -0.321 & 7.48e-01 & -2.23e-04 & -1.45e-97 & 3.49e-04 & - & - & 5.06e-06 & -2.54e-05 & 2.27e-04 & - & - \\
Grading & -1.75e-04 & -2.68e-04 & 1.69e-04 & -1.590 & 1.13e-01 & -1.19e-04 & -2.42e-04 & 2.29e-04 & -1.060 & 2.91e-01 & -1.73e-04 & -2.71e-04 & 1.76e-04 & -1.540 & 1.24e-01 \\
Renal & 3.67e-04 & 4.28e-04 & 6.24e-04 & 0.686 & 4.93e-01 & 2.61e-04 & 4.11e-04 & 8.13e-04 & 0.506 & 6.13e-01 & 3.53e-04 & 4.61e-04 & 6.57e-04 & 0.701 & 4.83e-01 \\
Sex & 9.24e-05 & 2.84e-04 & 2.14e-04 & 1.330 & 1.84e-01 & -2.22e-06 & 1.54e-04 & 2.95e-04 & 0.521 & 6.02e-01 & 1.91e-04 & 4.39e-04 & 2.18e-04 & 2.020 & 4.38e-02 \\
HT & -2.55e-04 & -4.04e-04 & 2.47e-04 & -1.630 & 1.03e-01 & -2.07e-04 & -4.81e-04 & 3.43e-04 & -1.400 & 1.61e-01 & -2.60e-04 & -4.10e-04 & 2.56e-04 & -1.600 & 1.10e-01 \\
CVA & -1.74e-04 & -2.61e-04 & 4.12e-04 & -0.634 & 5.26e-01 & -1.93e-04 & -3.58e-04 & 5.39e-04 & -0.664 & 5.07e-01 & -2.00e-04 & -3.01e-04 & 4.22e-04 & -0.712 & 4.76e-01 \\
DM & 7.77e-05 & 4.29e-04 & 2.66e-04 & 1.610 & 1.07e-01 & 2.98e-05 & 4.88e-04 & 3.67e-04 & 1.330 & 1.84e-01 & 6.41e-05 & 3.81e-04 & 2.72e-04 & 1.400 & 1.61e-01 \\
IHD & 4.97e-05 & 3.54e-04 & 3.06e-04 & 1.160 & 2.47e-01 & 8.17e-05 & 5.94e-04 & 4.16e-04 & 1.430 & 1.54e-01 & 1.45e-05 & 3.02e-04 & 3.16e-04 & 0.955 & 3.40e-01 \\
\bottomrule
\multicolumn{16}{l}{\textit{Note: Chisq and p-values for the overall model are provided below the table for each analysis.}} \\
\multicolumn{16}{l}{\textbf{Non-Adjusted}: n=845, 72/75 unique event times used. Chisq=104.75 on 15 df, p=1.63e-15.} \\
\multicolumn{16}{l}{\textbf{Genetic Matching}: n=676 (173 observations deleted). 72/75 unique event times used. Chisq=83.8 on 15 df, p=1.4e-11.} \\
\multicolumn{16}{l}{\textbf{Matching Weights}: n=845, 72/75 unique event times used. Chisq=98.36 on 15 df, p=2.67e-14.} \\
\end{tabular}
\end{sidewaystable}

\section{Characterization of Variables Used}
\label{sec:charkaterizace}

Table \ref{tab:variable_description} provides an overview of the variables with their brief descriptions and units. These variables are further analyzed in relation to the specific surgical technique. The goal is to gain an overview of whether the choice of surgical technique depends on the patient's physical condition and the state of their disease. The overview of these variables is categorized by their application significance.

\begin{table*}[htbp]
\centering
\caption{Description of Variables Used in the Analysis}
\label{tab:variable_description}
\small
\setlength{\tabcolsep}{3pt} 
\renewcommand{\arraystretch}{0.9} 
\begin{tabular}{p{2.5cm} p{6.5cm} p{3.5cm}}
\toprule
\textbf{Variable} & \textbf{Description} & \textbf{Values/Units} \\
\midrule
\multicolumn{3}{l}{\textbf{Patient's Medical History}} \\
\midrule
Age & Patient's age & Years \\
Sex & Patient's sex & Male | Female \\
BMI & Body Mass Index & kg/m² \\
ASA & American Society of Anesthesiologists classification (1: healthy; 2: mild disease; 3: severe disease; 4: life-threatening disease) & 1 | 2 | 3 | 4 \\
DM & Diabetes mellitus & Yes | No \\
IHD & Ischemic heart disease & Yes | No \\
Mortality & Mortality indicator & Yes | No \\
Morbidity & Morbidity indicator & Yes | No \\
Arrhythmia & Arrhythmia & Yes | No \\
HT & Hypertension & Yes | No \\
CVA & Cerebrovascular accident (stroke) & Yes | No \\
Pulmonary & Pulmonary diseases & Yes | No \\
Renal & Kidney problems & Yes | No \\
Hepatic & Liver disease & Yes | No \\
Previous\_Surgery & Previous surgery & Yes | No \\
Previous & Number of previous surgeries & 0 | 1 | 2 | 3 | 4+ \\
\midrule
\multicolumn{3}{l}{\textbf{Tumor Characteristics}} \\
\midrule
T & Size and extent of primary tumor & 1 | 2 | 3 | 4 \\
N & Regional lymph node involvement & 0 | 1 | 2 \\
M & Distant metastases & 0 | 1 \\
Stage & Disease stage & 1 | 2 | 3 | 4 \\
LN & Number of lymph nodes & Count \\
Grading & Tumor differentiation grade & 1 | 2 | 3 \\
\midrule
\multicolumn{3}{l}{\textbf{Surgery-Related Variables}} \\
\midrule
Op\_Time & Duration of surgery & Minutes \\
Blood\_Loss & Blood loss during surgery & ml \\
Reoperation & Patient reoperated & Yes | No \\
Survival\_Time & Survival time from surgery to last follow-up & Months \\
\bottomrule
\end{tabular}
\end{table*}

\clearpage
\bibliography{apssamp}

\begin{thebibliography}{44}%
\makeatletter
\providecommand \@ifxundefined [1]{%
 \@ifx{#1\undefined}
}%
\providecommand \@ifnum [1]{%
 \ifnum #1\expandafter \@firstoftwo
 \else \expandafter \@secondoftwo
 \fi
}%
\providecommand \@ifx [1]{%
 \ifx #1\expandafter \@firstoftwo
 \else \expandafter \@secondoftwo
 \fi
}%
\providecommand \natexlab [1]{#1}%
\providecommand \enquote  [1]{``#1''}%
\providecommand \bibnamefont  [1]{#1}%
\providecommand \bibfnamefont [1]{#1}%
\providecommand \citenamefont [1]{#1}%
\providecommand \href@noop [0]{\@secondoftwo}%
\providecommand \href [0]{\begingroup \@sanitize@url \@href}%
\providecommand \@href[1]{\@@startlink{#1}\@@href}%
\providecommand \@@href[1]{\endgroup#1\@@endlink}%
\providecommand \@sanitize@url [0]{\catcode `\\12\catcode `\$12\catcode `\&12\catcode `\#12\catcode `\^12\catcode `\_12\catcode `\%12\relax}%
\providecommand \@@startlink[1]{}%
\providecommand \@@endlink[0]{}%
\providecommand \url  [0]{\begingroup\@sanitize@url \@url }%
\providecommand \@url [1]{\endgroup\@href {#1}{\urlprefix }}%
\providecommand \urlprefix  [0]{URL }%
\providecommand \Eprint [0]{\href }%
\providecommand \doibase [0]{https://doi.org/}%
\providecommand \selectlanguage [0]{\@gobble}%
\providecommand \bibinfo  [0]{\@secondoftwo}%
\providecommand \bibfield  [0]{\@secondoftwo}%
\providecommand \translation [1]{[#1]}%
\providecommand \BibitemOpen [0]{}%
\providecommand \bibitemStop [0]{}%
\providecommand \bibitemNoStop [0]{.\EOS\space}%
\providecommand \EOS [0]{\spacefactor3000\relax}%
\providecommand \BibitemShut  [1]{\csname bibitem#1\endcsname}%
\let\auto@bib@innerbib\@empty
\bibitem [{\citenamefont {Klein}\ and\ \citenamefont {Moeschberger}(2003)}]{klein2003survival}%
  \BibitemOpen
  \bibfield  {author} {\bibinfo {author} {\bibfnamefont {J.~P.}\ \bibnamefont {Klein}}\ and\ \bibinfo {author} {\bibfnamefont {M.~L.}\ \bibnamefont {Moeschberger}},\ }\href {https://doi.org/10.1007/b97377} {\emph {\bibinfo {title} {Survival Analysis: Techniques for Censored and Truncated Data}}}\ (\bibinfo  {publisher} {Springer},\ \bibinfo {address} {New York},\ \bibinfo {year} {2003})\BibitemShut {NoStop}%
\bibitem [{\citenamefont {Rosenbaum}\ and\ \citenamefont {Rubin}(1983)}]{rosenbaum1983central}%
  \BibitemOpen
  \bibfield  {author} {\bibinfo {author} {\bibfnamefont {P.~R.}\ \bibnamefont {Rosenbaum}}\ and\ \bibinfo {author} {\bibfnamefont {D.~B.}\ \bibnamefont {Rubin}},\ }\bibfield  {title} {\bibinfo {title} {The central role of the propensity score in observational studies for causal effects},\ }\href {https://doi.org/10.1093/biomet/70.1.41} {\bibfield  {journal} {\bibinfo  {journal} {Biometrika}\ }\textbf {\bibinfo {volume} {70}},\ \bibinfo {pages} {41} (\bibinfo {year} {1983})}\BibitemShut {NoStop}%
\bibitem [{\citenamefont {McCaffrey}\ \emph {et~al.}(2004)\citenamefont {McCaffrey}, \citenamefont {Ridgeway},\ and\ \citenamefont {Morral}}]{mccaffrey2004propensity}%
  \BibitemOpen
  \bibfield  {author} {\bibinfo {author} {\bibfnamefont {D.~F.}\ \bibnamefont {McCaffrey}}, \bibinfo {author} {\bibfnamefont {G.}~\bibnamefont {Ridgeway}},\ and\ \bibinfo {author} {\bibfnamefont {A.~R.}\ \bibnamefont {Morral}},\ }\bibfield  {title} {\bibinfo {title} {Propensity score estimation with boosted regression for evaluating causal effects in observational studies},\ }\href {https://doi.org/10.1037/1082-989X.9.4.403} {\bibfield  {journal} {\bibinfo  {journal} {Psychological Methods}\ }\textbf {\bibinfo {volume} {9}},\ \bibinfo {pages} {403} (\bibinfo {year} {2004})}\BibitemShut {NoStop}%
\bibitem [{\citenamefont {P{\v{r}}ibylov{\'a}}\ \emph {et~al.}(2023)\citenamefont {P{\v{r}}ibylov{\'a}}, \citenamefont {Bri{\v{s}}}, \citenamefont {Nov{\'a}k},\ and\ \citenamefont {Mart{\'\i}nek}}]{pvribylova2023methodological}%
  \BibitemOpen
  \bibfield  {author} {\bibinfo {author} {\bibfnamefont {L.}~\bibnamefont {P{\v{r}}ibylov{\'a}}}, \bibinfo {author} {\bibfnamefont {R.}~\bibnamefont {Bri{\v{s}}}}, \bibinfo {author} {\bibfnamefont {V.}~\bibnamefont {Nov{\'a}k}},\ and\ \bibinfo {author} {\bibfnamefont {L.}~\bibnamefont {Mart{\'\i}nek}},\ }\bibfield  {title} {\bibinfo {title} {Methodological overview of prospensity score matching methods demonstrated on colorectal data},\ }in\ \href@noop {} {\emph {\bibinfo {booktitle} {2023 International Conference on Information and Digital Technologies (IDT)}}}\ (\bibinfo {organization} {IEEE},\ \bibinfo {year} {2023})\ pp.\ \bibinfo {pages} {89--96}\BibitemShut {NoStop}%
\bibitem [{\citenamefont {Biamonte}\ \emph {et~al.}(2017)\citenamefont {Biamonte}, \citenamefont {Wittek},\ and\ \citenamefont {Pancotti}}]{biamonte2017quantum}%
  \BibitemOpen
  \bibfield  {author} {\bibinfo {author} {\bibfnamefont {J.}~\bibnamefont {Biamonte}}, \bibinfo {author} {\bibfnamefont {P.}~\bibnamefont {Wittek}},\ and\ \bibinfo {author} {\bibfnamefont {N.~e.~a.}\ \bibnamefont {Pancotti}},\ }\bibfield  {title} {\bibinfo {title} {Quantum machine learning},\ }\href {https://doi.org/10.1038/nature23474} {\bibfield  {journal} {\bibinfo  {journal} {Nature}\ }\textbf {\bibinfo {volume} {549}},\ \bibinfo {pages} {195} (\bibinfo {year} {2017})}\BibitemShut {NoStop}%
\bibitem [{\citenamefont {Schuld}\ and\ \citenamefont {Killoran}(2019)}]{schuld2019quantum}%
  \BibitemOpen
  \bibfield  {author} {\bibinfo {author} {\bibfnamefont {M.}~\bibnamefont {Schuld}}\ and\ \bibinfo {author} {\bibfnamefont {N.}~\bibnamefont {Killoran}},\ }\bibfield  {title} {\bibinfo {title} {Quantum machine learning in feature hilbert spaces},\ }\href@noop {} {\bibfield  {journal} {\bibinfo  {journal} {Physical review letters}\ }\textbf {\bibinfo {volume} {122}},\ \bibinfo {pages} {040504} (\bibinfo {year} {2019})}\BibitemShut {NoStop}%
\bibitem [{\citenamefont {Bray}\ \emph {et~al.}(2018)\citenamefont {Bray}, \citenamefont {Ferlay},\ and\ \citenamefont {Soerjomataram}}]{bray2018global}%
  \BibitemOpen
  \bibfield  {author} {\bibinfo {author} {\bibfnamefont {F.}~\bibnamefont {Bray}}, \bibinfo {author} {\bibfnamefont {J.}~\bibnamefont {Ferlay}},\ and\ \bibinfo {author} {\bibfnamefont {I.~e.~a.}\ \bibnamefont {Soerjomataram}},\ }\bibfield  {title} {\bibinfo {title} {Global cancer statistics 2018: Globocan estimates of incidence and mortality worldwide for 36 cancers in 185 countries},\ }\href {https://doi.org/10.3322/caac.21492} {\bibfield  {journal} {\bibinfo  {journal} {CA: A Cancer Journal for Clinicians}\ }\textbf {\bibinfo {volume} {68}},\ \bibinfo {pages} {394} (\bibinfo {year} {2018})}\BibitemShut {NoStop}%
\bibitem [{\citenamefont {Dusek}\ \emph {et~al.}(2014)\citenamefont {Dusek}, \citenamefont {Muzik}, \citenamefont {Maluskova}, \citenamefont {M{\'a}jek}, \citenamefont {Pavlik}, \citenamefont {Kopt{\'\i}kov{\'a}}, \citenamefont {Melichar}, \citenamefont {B{\"u}chler}, \citenamefont {Finek}, \citenamefont {Cibula} \emph {et~al.}}]{dusek2014cancer}%
  \BibitemOpen
  \bibfield  {author} {\bibinfo {author} {\bibfnamefont {L.}~\bibnamefont {Dusek}}, \bibinfo {author} {\bibfnamefont {J.}~\bibnamefont {Muzik}}, \bibinfo {author} {\bibfnamefont {D.}~\bibnamefont {Maluskova}}, \bibinfo {author} {\bibfnamefont {O.}~\bibnamefont {M{\'a}jek}}, \bibinfo {author} {\bibfnamefont {T.}~\bibnamefont {Pavlik}}, \bibinfo {author} {\bibfnamefont {J.}~\bibnamefont {Kopt{\'\i}kov{\'a}}}, \bibinfo {author} {\bibfnamefont {B.}~\bibnamefont {Melichar}}, \bibinfo {author} {\bibfnamefont {T.}~\bibnamefont {B{\"u}chler}}, \bibinfo {author} {\bibfnamefont {J.}~\bibnamefont {Finek}}, \bibinfo {author} {\bibfnamefont {D.}~\bibnamefont {Cibula}}, \emph {et~al.},\ }\bibfield  {title} {\bibinfo {title} {Cancer incidence and mortality in the czech republic},\ }\href@noop {} {\bibfield  {journal} {\bibinfo  {journal} {Klin Onkol}\ }\textbf {\bibinfo {volume} {27}},\ \bibinfo {pages} {406} (\bibinfo {year} {2014})}\BibitemShut {NoStop}%
\bibitem [{\citenamefont {Winawer}\ \emph {et~al.}(2003)\citenamefont {Winawer}, \citenamefont {Fletcher}, \citenamefont {Rex}, \citenamefont {Bond}, \citenamefont {Burt}, \citenamefont {Ferrucci}, \citenamefont {Ganiats}, \citenamefont {Levin}, \citenamefont {Woolf}, \citenamefont {Johnson} \emph {et~al.}}]{winawer2003colorectal}%
  \BibitemOpen
  \bibfield  {author} {\bibinfo {author} {\bibfnamefont {S.}~\bibnamefont {Winawer}}, \bibinfo {author} {\bibfnamefont {R.}~\bibnamefont {Fletcher}}, \bibinfo {author} {\bibfnamefont {D.}~\bibnamefont {Rex}}, \bibinfo {author} {\bibfnamefont {J.}~\bibnamefont {Bond}}, \bibinfo {author} {\bibfnamefont {R.}~\bibnamefont {Burt}}, \bibinfo {author} {\bibfnamefont {J.}~\bibnamefont {Ferrucci}}, \bibinfo {author} {\bibfnamefont {T.}~\bibnamefont {Ganiats}}, \bibinfo {author} {\bibfnamefont {T.}~\bibnamefont {Levin}}, \bibinfo {author} {\bibfnamefont {S.}~\bibnamefont {Woolf}}, \bibinfo {author} {\bibfnamefont {D.}~\bibnamefont {Johnson}}, \emph {et~al.},\ }\bibfield  {title} {\bibinfo {title} {Colorectal cancer screening and surveillance: clinical guidelines and rationale—update based on new evidence},\ }\href@noop {} {\bibfield  {journal} {\bibinfo  {journal} {Gastroenterology}\ }\textbf {\bibinfo {volume} {124}},\ \bibinfo {pages} {544} (\bibinfo {year} {2003})}\BibitemShut {NoStop}%
\bibitem [{\citenamefont {Zhao}\ \emph {et~al.}(2017)\citenamefont {Zhao}, \citenamefont {Feng}, \citenamefont {Yin}, \citenamefont {Shuang}, \citenamefont {Bai}, \citenamefont {Yu}, \citenamefont {Guo},\ and\ \citenamefont {Zhao}}]{zhao2017red}%
  \BibitemOpen
  \bibfield  {author} {\bibinfo {author} {\bibfnamefont {Z.}~\bibnamefont {Zhao}}, \bibinfo {author} {\bibfnamefont {Q.}~\bibnamefont {Feng}}, \bibinfo {author} {\bibfnamefont {Z.}~\bibnamefont {Yin}}, \bibinfo {author} {\bibfnamefont {J.}~\bibnamefont {Shuang}}, \bibinfo {author} {\bibfnamefont {B.}~\bibnamefont {Bai}}, \bibinfo {author} {\bibfnamefont {P.}~\bibnamefont {Yu}}, \bibinfo {author} {\bibfnamefont {M.}~\bibnamefont {Guo}},\ and\ \bibinfo {author} {\bibfnamefont {Q.}~\bibnamefont {Zhao}},\ }\bibfield  {title} {\bibinfo {title} {Red and processed meat consumption and colorectal cancer risk: a systematic review and meta-analysis},\ }\href@noop {} {\bibfield  {journal} {\bibinfo  {journal} {Oncotarget}\ }\textbf {\bibinfo {volume} {8}},\ \bibinfo {pages} {83306} (\bibinfo {year} {2017})}\BibitemShut {NoStop}%
\bibitem [{\citenamefont {Slattery}\ \emph {et~al.}(1998)\citenamefont {Slattery}, \citenamefont {Boucher}, \citenamefont {Caan}, \citenamefont {Potter},\ and\ \citenamefont {Ma}}]{slattery1998eating}%
  \BibitemOpen
  \bibfield  {author} {\bibinfo {author} {\bibfnamefont {M.~L.}\ \bibnamefont {Slattery}}, \bibinfo {author} {\bibfnamefont {K.~M.}\ \bibnamefont {Boucher}}, \bibinfo {author} {\bibfnamefont {B.~J.}\ \bibnamefont {Caan}}, \bibinfo {author} {\bibfnamefont {J.~D.}\ \bibnamefont {Potter}},\ and\ \bibinfo {author} {\bibfnamefont {K.-N.}\ \bibnamefont {Ma}},\ }\bibfield  {title} {\bibinfo {title} {Eating patterns and risk of colon cancer},\ }\href@noop {} {\bibfield  {journal} {\bibinfo  {journal} {American journal of epidemiology}\ }\textbf {\bibinfo {volume} {148}},\ \bibinfo {pages} {4} (\bibinfo {year} {1998})}\BibitemShut {NoStop}%
\bibitem [{\citenamefont {Lynch}\ and\ \citenamefont {De~la Chapelle}(2003)}]{lynch2003hereditary}%
  \BibitemOpen
  \bibfield  {author} {\bibinfo {author} {\bibfnamefont {H.~T.}\ \bibnamefont {Lynch}}\ and\ \bibinfo {author} {\bibfnamefont {A.}~\bibnamefont {De~la Chapelle}},\ }\bibfield  {title} {\bibinfo {title} {Hereditary colorectal cancer},\ }\href@noop {} {\bibfield  {journal} {\bibinfo  {journal} {New England Journal of Medicine}\ }\textbf {\bibinfo {volume} {348}},\ \bibinfo {pages} {919} (\bibinfo {year} {2003})}\BibitemShut {NoStop}%
\bibitem [{\citenamefont {Ahadova}\ \emph {et~al.}(2018)\citenamefont {Ahadova}, \citenamefont {Gallon}, \citenamefont {Gebert}, \citenamefont {Ballhausen}, \citenamefont {Endris}, \citenamefont {Kirchner}, \citenamefont {Stenzinger}, \citenamefont {Burn}, \citenamefont {von Knebel~Doeberitz}, \citenamefont {Bl{\"a}ker} \emph {et~al.}}]{ahadova2018three}%
  \BibitemOpen
  \bibfield  {author} {\bibinfo {author} {\bibfnamefont {A.}~\bibnamefont {Ahadova}}, \bibinfo {author} {\bibfnamefont {R.}~\bibnamefont {Gallon}}, \bibinfo {author} {\bibfnamefont {J.}~\bibnamefont {Gebert}}, \bibinfo {author} {\bibfnamefont {A.}~\bibnamefont {Ballhausen}}, \bibinfo {author} {\bibfnamefont {V.}~\bibnamefont {Endris}}, \bibinfo {author} {\bibfnamefont {M.}~\bibnamefont {Kirchner}}, \bibinfo {author} {\bibfnamefont {A.}~\bibnamefont {Stenzinger}}, \bibinfo {author} {\bibfnamefont {J.}~\bibnamefont {Burn}}, \bibinfo {author} {\bibfnamefont {M.}~\bibnamefont {von Knebel~Doeberitz}}, \bibinfo {author} {\bibfnamefont {H.}~\bibnamefont {Bl{\"a}ker}}, \emph {et~al.},\ }\bibfield  {title} {\bibinfo {title} {Three molecular pathways model colorectal carcinogenesis in l ynch syndrome},\ }\href@noop {} {\bibfield  {journal} {\bibinfo  {journal} {International journal of cancer}\ }\textbf {\bibinfo {volume} {143}},\ \bibinfo {pages} {139} (\bibinfo {year} {2018})}\BibitemShut {NoStop}%
\bibitem [{\citenamefont {Haggar}\ and\ \citenamefont {Boushey}(2009)}]{haggar2009colorectal}%
  \BibitemOpen
  \bibfield  {author} {\bibinfo {author} {\bibfnamefont {F.~A.}\ \bibnamefont {Haggar}}\ and\ \bibinfo {author} {\bibfnamefont {R.~P.}\ \bibnamefont {Boushey}},\ }\bibfield  {title} {\bibinfo {title} {Colorectal cancer epidemiology: incidence, mortality, survival, and risk factors},\ }\href@noop {} {\bibfield  {journal} {\bibinfo  {journal} {Clinics in colon and rectal surgery}\ }\textbf {\bibinfo {volume} {22}},\ \bibinfo {pages} {191} (\bibinfo {year} {2009})}\BibitemShut {NoStop}%
\bibitem [{\citenamefont {Simon}(1984)}]{simon1984fecal}%
  \BibitemOpen
  \bibfield  {author} {\bibinfo {author} {\bibfnamefont {J.~B.}\ \bibnamefont {Simon}},\ }\bibfield  {title} {\bibinfo {title} {Fecal occult blood screening for colorectal cancer},\ }\href@noop {} {\bibfield  {journal} {\bibinfo  {journal} {Gastroenterology}\ }\textbf {\bibinfo {volume} {86}},\ \bibinfo {pages} {820} (\bibinfo {year} {1984})},\ \bibinfo {note} {pMID: 6370782}\BibitemShut {NoStop}%
\bibitem [{\citenamefont {Deans}\ \emph {et~al.}(1994)\citenamefont {Deans}, \citenamefont {Krukowski},\ and\ \citenamefont {Irwin}}]{deans1994malignant}%
  \BibitemOpen
  \bibfield  {author} {\bibinfo {author} {\bibfnamefont {G.~T.}\ \bibnamefont {Deans}}, \bibinfo {author} {\bibfnamefont {Z.~H.}\ \bibnamefont {Krukowski}},\ and\ \bibinfo {author} {\bibfnamefont {S.~T.}\ \bibnamefont {Irwin}},\ }\bibfield  {title} {\bibinfo {title} {Malignant obstruction of the colorectal large bowel},\ }\href {https://doi.org/10.1002/bjs.1800810702} {\bibfield  {journal} {\bibinfo  {journal} {British Journal of Surgery}\ }\textbf {\bibinfo {volume} {81}},\ \bibinfo {pages} {894} (\bibinfo {year} {1994})}\BibitemShut {NoStop}%
\bibitem [{\citenamefont {Cunningham}\ \emph {et~al.}(2010)\citenamefont {Cunningham}, \citenamefont {Atkin},\ and\ \citenamefont {Lenz}}]{cunningham2010colorectal}%
  \BibitemOpen
  \bibfield  {author} {\bibinfo {author} {\bibfnamefont {D.}~\bibnamefont {Cunningham}}, \bibinfo {author} {\bibfnamefont {W.~S.}\ \bibnamefont {Atkin}},\ and\ \bibinfo {author} {\bibfnamefont {H.-J. e.~a.}\ \bibnamefont {Lenz}},\ }\bibfield  {title} {\bibinfo {title} {Colorectal cancer},\ }\href {https://doi.org/10.1016/S0140-6736(10)60353-4} {\bibfield  {journal} {\bibinfo  {journal} {The Lancet}\ }\textbf {\bibinfo {volume} {375}},\ \bibinfo {pages} {1030} (\bibinfo {year} {2010})}\BibitemShut {NoStop}%
\bibitem [{\citenamefont {van~de Velde}(2011)}]{van2011staging}%
  \BibitemOpen
  \bibfield  {author} {\bibinfo {author} {\bibfnamefont {C.~J.}\ \bibnamefont {van~de Velde}},\ }\bibfield  {title} {\bibinfo {title} {Staging of colorectal cancer: Current standards and future perspectives},\ }\href@noop {} {\bibfield  {journal} {\bibinfo  {journal} {Annals of Surgery}\ }\textbf {\bibinfo {volume} {2011}},\ \bibinfo {pages} {1} (\bibinfo {year} {2011})},\ \bibinfo {note} {available at http://www.lippincott.com}\BibitemShut {NoStop}%
\bibitem [{\citenamefont {Sobin}\ and\ \citenamefont {Wittekind}(2010)}]{sobin2010tnm}%
  \BibitemOpen
  \bibfield  {author} {\bibinfo {author} {\bibfnamefont {L.~H.}\ \bibnamefont {Sobin}}\ and\ \bibinfo {author} {\bibfnamefont {C.}~\bibnamefont {Wittekind}},\ }\href@noop {} {\emph {\bibinfo {title} {TNM Classification of Malignant Tumours}}},\ \bibinfo {edition} {7th}\ ed.\ (\bibinfo  {publisher} {Wiley-Liss},\ \bibinfo {address} {New York},\ \bibinfo {year} {2010})\BibitemShut {NoStop}%
\bibitem [{\citenamefont {Wittekind}\ and\ \citenamefont {Compton}(2009)}]{wittekind2009tnm}%
  \BibitemOpen
  \bibfield  {author} {\bibinfo {author} {\bibfnamefont {C.}~\bibnamefont {Wittekind}}\ and\ \bibinfo {author} {\bibfnamefont {e.~a.}\ \bibnamefont {Compton}, \bibfnamefont {Carolyn~C.}},\ }\bibfield  {title} {\bibinfo {title} {Tnm residual tumor classification revisited},\ }\href {https://doi.org/10.1002/cncr.10477} {\bibfield  {journal} {\bibinfo  {journal} {Cancer}\ }\textbf {\bibinfo {volume} {94}},\ \bibinfo {pages} {2511} (\bibinfo {year} {2009})}\BibitemShut {NoStop}%
\bibitem [{\citenamefont {van~de Velde}\ and\ \citenamefont {Boelens}(2010)}]{van2010multidisciplinary}%
  \BibitemOpen
  \bibfield  {author} {\bibinfo {author} {\bibfnamefont {C.~J.}\ \bibnamefont {van~de Velde}}\ and\ \bibinfo {author} {\bibfnamefont {e.~a.}\ \bibnamefont {Boelens}, \bibfnamefont {Petra~G.}},\ }\bibfield  {title} {\bibinfo {title} {Multidisciplinary management of colorectal cancer},\ }\href {https://doi.org/10.1038/nrclinonc.2010.112} {\bibfield  {journal} {\bibinfo  {journal} {Nature Reviews Clinical Oncology}\ }\textbf {\bibinfo {volume} {7}},\ \bibinfo {pages} {468} (\bibinfo {year} {2010})}\BibitemShut {NoStop}%
\bibitem [{\citenamefont {Sauer}\ and\ \citenamefont {Becker}(2004)}]{sauer2004preoperative}%
  \BibitemOpen
  \bibfield  {author} {\bibinfo {author} {\bibfnamefont {R.}~\bibnamefont {Sauer}}\ and\ \bibinfo {author} {\bibfnamefont {H.~e.~a.}\ \bibnamefont {Becker}},\ }\bibfield  {title} {\bibinfo {title} {Preoperative versus postoperative chemoradiotherapy for rectal cancer},\ }\href {https://doi.org/10.1056/NEJMoa040694} {\bibfield  {journal} {\bibinfo  {journal} {New England Journal of Medicine}\ }\textbf {\bibinfo {volume} {351}},\ \bibinfo {pages} {1731} (\bibinfo {year} {2004})}\BibitemShut {NoStop}%
\bibitem [{\citenamefont {Kjeldsen}\ \emph {et~al.}(1997)\citenamefont {Kjeldsen}, \citenamefont {Kronborg},\ and\ \citenamefont {Fenger}}]{kjeldsen1997recurrence}%
  \BibitemOpen
  \bibfield  {author} {\bibinfo {author} {\bibfnamefont {B.~J.}\ \bibnamefont {Kjeldsen}}, \bibinfo {author} {\bibfnamefont {O.}~\bibnamefont {Kronborg}},\ and\ \bibinfo {author} {\bibfnamefont {C.~e.~a.}\ \bibnamefont {Fenger}},\ }\bibfield  {title} {\bibinfo {title} {Recurrence of colorectal cancer and survival after resection},\ }\href@noop {} {\bibfield  {journal} {\bibinfo  {journal} {Danish Medical Journal}\ }\textbf {\bibinfo {volume} {44}},\ \bibinfo {pages} {65} (\bibinfo {year} {1997})},\ \bibinfo {note} {pMID: 9062763}\BibitemShut {NoStop}%
\bibitem [{\citenamefont {Pascual}\ \emph {et~al.}(2016)\citenamefont {Pascual}, \citenamefont {Salvans},\ and\ \citenamefont {Pera}}]{pascual2016laparoscopic}%
  \BibitemOpen
  \bibfield  {author} {\bibinfo {author} {\bibfnamefont {M.}~\bibnamefont {Pascual}}, \bibinfo {author} {\bibfnamefont {S.}~\bibnamefont {Salvans}},\ and\ \bibinfo {author} {\bibfnamefont {M.}~\bibnamefont {Pera}},\ }\bibfield  {title} {\bibinfo {title} {Laparoscopic colorectal surgery: current status and implementation of the latest technological innovations},\ }\href@noop {} {\bibfield  {journal} {\bibinfo  {journal} {World journal of gastroenterology}\ }\textbf {\bibinfo {volume} {22}},\ \bibinfo {pages} {704} (\bibinfo {year} {2016})}\BibitemShut {NoStop}%
\bibitem [{\citenamefont {Lacy}\ \emph {et~al.}(2002)\citenamefont {Lacy}, \citenamefont {García-Valdecasas},\ and\ \citenamefont {Delgado}}]{lacy2002laparoscopy}%
  \BibitemOpen
  \bibfield  {author} {\bibinfo {author} {\bibfnamefont {A.~M.}\ \bibnamefont {Lacy}}, \bibinfo {author} {\bibfnamefont {J.~C.}\ \bibnamefont {García-Valdecasas}},\ and\ \bibinfo {author} {\bibfnamefont {e.~a.}\ \bibnamefont {Delgado}, \bibfnamefont {Salvadora}},\ }\bibfield  {title} {\bibinfo {title} {Laparoscopy-assisted colectomy versus open colectomy for treatment of non-metastatic colon cancer: A randomised trial},\ }\href {https://doi.org/10.1016/S0140-6736(02)09290-5} {\bibfield  {journal} {\bibinfo  {journal} {The Lancet}\ }\textbf {\bibinfo {volume} {359}},\ \bibinfo {pages} {2224} (\bibinfo {year} {2002})}\BibitemShut {NoStop}%
\bibitem [{\citenamefont {Guillou}\ \emph {et~al.}(2005)\citenamefont {Guillou}, \citenamefont {Quirke},\ and\ \citenamefont {Thorpe}}]{guillou2005short}%
  \BibitemOpen
  \bibfield  {author} {\bibinfo {author} {\bibfnamefont {P.~J.}\ \bibnamefont {Guillou}}, \bibinfo {author} {\bibfnamefont {P.}~\bibnamefont {Quirke}},\ and\ \bibinfo {author} {\bibfnamefont {H.~e.~a.}\ \bibnamefont {Thorpe}},\ }\bibfield  {title} {\bibinfo {title} {Short-term endpoints of conventional versus laparoscopic-assisted surgery in patients with colorectal cancer (mrc clasicc trial): Multicentre, randomised controlled trial},\ }\href {https://doi.org/10.1016/S0140-6736(05)66545-2} {\bibfield  {journal} {\bibinfo  {journal} {The Lancet}\ }\textbf {\bibinfo {volume} {365}},\ \bibinfo {pages} {1718} (\bibinfo {year} {2005})}\BibitemShut {NoStop}%
\bibitem [{\citenamefont {Kaplan}\ and\ \citenamefont {Meier}(1958)}]{kaplan1958nonparametric}%
  \BibitemOpen
  \bibfield  {author} {\bibinfo {author} {\bibfnamefont {E.~L.}\ \bibnamefont {Kaplan}}\ and\ \bibinfo {author} {\bibfnamefont {P.}~\bibnamefont {Meier}},\ }\bibfield  {title} {\bibinfo {title} {Nonparametric estimation from incomplete observations},\ }\href {https://doi.org/10.1080/01621459.1958.10501452} {\bibfield  {journal} {\bibinfo  {journal} {Journal of the American Statistical Association}\ }\textbf {\bibinfo {volume} {53}},\ \bibinfo {pages} {457} (\bibinfo {year} {1958})}\BibitemShut {NoStop}%
\bibitem [{\citenamefont {Mantel}(1966)}]{mantel1966evaluation}%
  \BibitemOpen
  \bibfield  {author} {\bibinfo {author} {\bibfnamefont {N.}~\bibnamefont {Mantel}},\ }\bibfield  {title} {\bibinfo {title} {Evaluation of survival data and two new rank order statistics arising in its consideration},\ }\href@noop {} {\bibfield  {journal} {\bibinfo  {journal} {Cancer Chemotherapy Reports}\ }\textbf {\bibinfo {volume} {50}},\ \bibinfo {pages} {163} (\bibinfo {year} {1966})},\ \bibinfo {note} {pMID: 5910392}\BibitemShut {NoStop}%
\bibitem [{\citenamefont {Rosenbaum}\ and\ \citenamefont {Rubin}(1984)}]{rosenbaum1984reducing}%
  \BibitemOpen
  \bibfield  {author} {\bibinfo {author} {\bibfnamefont {P.~R.}\ \bibnamefont {Rosenbaum}}\ and\ \bibinfo {author} {\bibfnamefont {D.~B.}\ \bibnamefont {Rubin}},\ }\bibfield  {title} {\bibinfo {title} {Reducing bias in observational studies using subclassification on the propensity score},\ }\href {https://doi.org/10.2307/2288398} {\bibfield  {journal} {\bibinfo  {journal} {Journal of the American Statistical Association}\ }\textbf {\bibinfo {volume} {79}},\ \bibinfo {pages} {516} (\bibinfo {year} {1984})}\BibitemShut {NoStop}%
\bibitem [{\citenamefont {Xie}\ and\ \citenamefont {Liu}(2005)}]{xie2005adjusted}%
  \BibitemOpen
  \bibfield  {author} {\bibinfo {author} {\bibfnamefont {J.}~\bibnamefont {Xie}}\ and\ \bibinfo {author} {\bibfnamefont {C.}~\bibnamefont {Liu}},\ }\bibfield  {title} {\bibinfo {title} {Adjusted kaplan-meier estimator and log-rank test with inverse probability of treatment weighting for survival data},\ }\href {https://doi.org/10.1002/sim.2174} {\bibfield  {journal} {\bibinfo  {journal} {Statistics in Medicine}\ }\textbf {\bibinfo {volume} {24}},\ \bibinfo {pages} {3089} (\bibinfo {year} {2005})}\BibitemShut {NoStop}%
\bibitem [{\citenamefont {Austin}(2008)}]{austin2008assessing}%
  \BibitemOpen
  \bibfield  {author} {\bibinfo {author} {\bibfnamefont {P.~C.}\ \bibnamefont {Austin}},\ }\bibfield  {title} {\bibinfo {title} {Assessing balance in measured baseline covariates when using many-to-one matching on the propensity-score},\ }\href@noop {} {\bibfield  {journal} {\bibinfo  {journal} {Pharmacoepidemiology and drug safety}\ }\textbf {\bibinfo {volume} {17}},\ \bibinfo {pages} {1218} (\bibinfo {year} {2008})}\BibitemShut {NoStop}%
\bibitem [{\citenamefont {Preskill}(2018)}]{preskill2018quantum}%
  \BibitemOpen
  \bibfield  {author} {\bibinfo {author} {\bibfnamefont {J.}~\bibnamefont {Preskill}},\ }\bibfield  {title} {\bibinfo {title} {Quantum computing in the nisq era and beyond},\ }\href@noop {} {\bibfield  {journal} {\bibinfo  {journal} {Quantum}\ }\textbf {\bibinfo {volume} {2}},\ \bibinfo {pages} {79} (\bibinfo {year} {2018})}\BibitemShut {NoStop}%
\bibitem [{\citenamefont {Nov{\'a}k}\ \emph {et~al.}(2025{\natexlab{a}})\citenamefont {Nov{\'a}k}, \citenamefont {Zelinka}, \citenamefont {P{\v{r}}ibylov{\'a}}, \citenamefont {Mart{\'\i}nek},\ and\ \citenamefont {Ben{\v{c}}urik}}]{novak2025predicting}%
  \BibitemOpen
  \bibfield  {author} {\bibinfo {author} {\bibfnamefont {V.}~\bibnamefont {Nov{\'a}k}}, \bibinfo {author} {\bibfnamefont {I.}~\bibnamefont {Zelinka}}, \bibinfo {author} {\bibfnamefont {L.}~\bibnamefont {P{\v{r}}ibylov{\'a}}}, \bibinfo {author} {\bibfnamefont {L.}~\bibnamefont {Mart{\'\i}nek}},\ and\ \bibinfo {author} {\bibfnamefont {V.}~\bibnamefont {Ben{\v{c}}urik}},\ }\bibfield  {title} {\bibinfo {title} {Predicting post-surgical complications with quantum neural networks: A clinical study on anastomotic leak},\ }\href@noop {} {\bibfield  {journal} {\bibinfo  {journal} {arXiv preprint arXiv:2506.01708}\ } (\bibinfo {year} {2025}{\natexlab{a}})}\BibitemShut {NoStop}%
\bibitem [{\citenamefont {Zhou}\ and\ \citenamefont {Zhang}(2022)}]{zhou2022noise}%
  \BibitemOpen
  \bibfield  {author} {\bibinfo {author} {\bibfnamefont {Y.}~\bibnamefont {Zhou}}\ and\ \bibinfo {author} {\bibfnamefont {P.}~\bibnamefont {Zhang}},\ }\bibfield  {title} {\bibinfo {title} {Noise-resilient quantum machine learning for stability assessment of power systems},\ }\href@noop {} {\bibfield  {journal} {\bibinfo  {journal} {IEEE Transactions on Power Systems}\ }\textbf {\bibinfo {volume} {38}},\ \bibinfo {pages} {475} (\bibinfo {year} {2022})}\BibitemShut {NoStop}%
\bibitem [{\citenamefont {Schuld}\ \emph {et~al.}(2020)\citenamefont {Schuld}, \citenamefont {Bocharov}, \citenamefont {Svore},\ and\ \citenamefont {Wiebe}}]{schuld2020circuit}%
  \BibitemOpen
  \bibfield  {author} {\bibinfo {author} {\bibfnamefont {M.}~\bibnamefont {Schuld}}, \bibinfo {author} {\bibfnamefont {A.}~\bibnamefont {Bocharov}}, \bibinfo {author} {\bibfnamefont {K.~M.}\ \bibnamefont {Svore}},\ and\ \bibinfo {author} {\bibfnamefont {N.}~\bibnamefont {Wiebe}},\ }\bibfield  {title} {\bibinfo {title} {Circuit-centric quantum classifiers},\ }\href@noop {} {\bibfield  {journal} {\bibinfo  {journal} {Physical Review A}\ }\textbf {\bibinfo {volume} {101}},\ \bibinfo {pages} {032308} (\bibinfo {year} {2020})}\BibitemShut {NoStop}%
\bibitem [{\citenamefont {Li}\ \emph {et~al.}(2019)\citenamefont {Li}, \citenamefont {Ding},\ and\ \citenamefont {Xie}}]{li2019tackling}%
  \BibitemOpen
  \bibfield  {author} {\bibinfo {author} {\bibfnamefont {G.}~\bibnamefont {Li}}, \bibinfo {author} {\bibfnamefont {Y.}~\bibnamefont {Ding}},\ and\ \bibinfo {author} {\bibfnamefont {Y.}~\bibnamefont {Xie}},\ }\bibfield  {title} {\bibinfo {title} {Tackling the qubit mapping problem for nisq-era quantum devices},\ }in\ \href@noop {} {\emph {\bibinfo {booktitle} {Proceedings of the twenty-fourth international conference on architectural support for programming languages and operating systems}}}\ (\bibinfo {year} {2019})\ pp.\ \bibinfo {pages} {1001--1014}\BibitemShut {NoStop}%
\bibitem [{\citenamefont {Lloyd}\ \emph {et~al.}(2014)\citenamefont {Lloyd}, \citenamefont {Mohseni},\ and\ \citenamefont {Rebentrost}}]{lloyd2014quantum}%
  \BibitemOpen
  \bibfield  {author} {\bibinfo {author} {\bibfnamefont {S.}~\bibnamefont {Lloyd}}, \bibinfo {author} {\bibfnamefont {M.}~\bibnamefont {Mohseni}},\ and\ \bibinfo {author} {\bibfnamefont {P.}~\bibnamefont {Rebentrost}},\ }\bibfield  {title} {\bibinfo {title} {Quantum algorithms for supervised and unsupervised machine learning},\ }\href@noop {} {\bibfield  {journal} {\bibinfo  {journal} {arXiv preprint arXiv:1307.0411}\ } (\bibinfo {year} {2014})},\ \bibinfo {note} {available at https://arxiv.org/abs/1307.0411}\BibitemShut {NoStop}%
\bibitem [{\citenamefont {Bonet-Monroig}\ \emph {et~al.}(2023)\citenamefont {Bonet-Monroig}, \citenamefont {Wang}, \citenamefont {Vermetten}, \citenamefont {Senjean}, \citenamefont {Moussa}, \citenamefont {B{\"a}ck}, \citenamefont {Dunjko},\ and\ \citenamefont {O'Brien}}]{bonet2023performance}%
  \BibitemOpen
  \bibfield  {author} {\bibinfo {author} {\bibfnamefont {X.}~\bibnamefont {Bonet-Monroig}}, \bibinfo {author} {\bibfnamefont {H.}~\bibnamefont {Wang}}, \bibinfo {author} {\bibfnamefont {D.}~\bibnamefont {Vermetten}}, \bibinfo {author} {\bibfnamefont {B.}~\bibnamefont {Senjean}}, \bibinfo {author} {\bibfnamefont {C.}~\bibnamefont {Moussa}}, \bibinfo {author} {\bibfnamefont {T.}~\bibnamefont {B{\"a}ck}}, \bibinfo {author} {\bibfnamefont {V.}~\bibnamefont {Dunjko}},\ and\ \bibinfo {author} {\bibfnamefont {T.~E.}\ \bibnamefont {O'Brien}},\ }\bibfield  {title} {\bibinfo {title} {Performance comparison of optimization methods on variational quantum algorithms},\ }\href@noop {} {\bibfield  {journal} {\bibinfo  {journal} {Physical Review A}\ }\textbf {\bibinfo {volume} {107}},\ \bibinfo {pages} {032407} (\bibinfo {year} {2023})}\BibitemShut {NoStop}%
\bibitem [{\citenamefont {Ill{\'e}sov{\'a}}\ \emph {et~al.}(2025)\citenamefont {Ill{\'e}sov{\'a}}, \citenamefont {Nov{\'a}k}, \citenamefont {Bezd{\v{e}}k}, \citenamefont {Beseda},\ and\ \citenamefont {Possel}}]{illesova2025numerical}%
  \BibitemOpen
  \bibfield  {author} {\bibinfo {author} {\bibfnamefont {S.}~\bibnamefont {Ill{\'e}sov{\'a}}}, \bibinfo {author} {\bibfnamefont {V.}~\bibnamefont {Nov{\'a}k}}, \bibinfo {author} {\bibfnamefont {T.}~\bibnamefont {Bezd{\v{e}}k}}, \bibinfo {author} {\bibfnamefont {M.}~\bibnamefont {Beseda}},\ and\ \bibinfo {author} {\bibfnamefont {C.}~\bibnamefont {Possel}},\ }\bibfield  {title} {\bibinfo {title} {Numerical optimization strategies for the variational hamiltonian ansatz in noisy quantum environments},\ }\href@noop {} {\bibfield  {journal} {\bibinfo  {journal} {arXiv preprint arXiv:2505.22398}\ } (\bibinfo {year} {2025})}\BibitemShut {NoStop}%
\bibitem [{\citenamefont {Nov{\'a}k}\ \emph {et~al.}(2025{\natexlab{b}})\citenamefont {Nov{\'a}k}, \citenamefont {Zelinka},\ and\ \citenamefont {Sn{\'a}{\v{s}}el}}]{novak2025optimization}%
  \BibitemOpen
  \bibfield  {author} {\bibinfo {author} {\bibfnamefont {V.}~\bibnamefont {Nov{\'a}k}}, \bibinfo {author} {\bibfnamefont {I.}~\bibnamefont {Zelinka}},\ and\ \bibinfo {author} {\bibfnamefont {V.}~\bibnamefont {Sn{\'a}{\v{s}}el}},\ }\bibfield  {title} {\bibinfo {title} {Optimization strategies for variational quantum algorithms in noisy landscapes},\ }\href@noop {} {\bibfield  {journal} {\bibinfo  {journal} {arXiv preprint arXiv:2506.01715}\ } (\bibinfo {year} {2025}{\natexlab{b}})}\BibitemShut {NoStop}%
\bibitem [{\citenamefont {Cerezo}\ \emph {et~al.}(2021)\citenamefont {Cerezo}, \citenamefont {Arrasmith},\ and\ \citenamefont {Babbush}}]{cerezo2021variational}%
  \BibitemOpen
  \bibfield  {author} {\bibinfo {author} {\bibfnamefont {M.}~\bibnamefont {Cerezo}}, \bibinfo {author} {\bibfnamefont {A.}~\bibnamefont {Arrasmith}},\ and\ \bibinfo {author} {\bibfnamefont {R.~e.~a.}\ \bibnamefont {Babbush}},\ }\bibfield  {title} {\bibinfo {title} {Variational quantum algorithms},\ }\href {https://doi.org/10.1038/s42254-021-00348-9} {\bibfield  {journal} {\bibinfo  {journal} {Nature Reviews Physics}\ }\textbf {\bibinfo {volume} {3}},\ \bibinfo {pages} {625} (\bibinfo {year} {2021})}\BibitemShut {NoStop}%
\bibitem [{\citenamefont {Javadi-Abhari}\ \emph {et~al.}(2024)\citenamefont {Javadi-Abhari}, \citenamefont {Treinish}, \citenamefont {Krsulich}, \citenamefont {Wood}, \citenamefont {Lishman}, \citenamefont {Gacon}, \citenamefont {Martiel}, \citenamefont {Nation}, \citenamefont {Bishop}, \citenamefont {Cross}, \citenamefont {Johnson},\ and\ \citenamefont {Gambetta}}]{qiskit}%
  \BibitemOpen
  \bibfield  {author} {\bibinfo {author} {\bibfnamefont {A.}~\bibnamefont {Javadi-Abhari}}, \bibinfo {author} {\bibfnamefont {M.}~\bibnamefont {Treinish}}, \bibinfo {author} {\bibfnamefont {K.}~\bibnamefont {Krsulich}}, \bibinfo {author} {\bibfnamefont {C.~J.}\ \bibnamefont {Wood}}, \bibinfo {author} {\bibfnamefont {J.}~\bibnamefont {Lishman}}, \bibinfo {author} {\bibfnamefont {J.}~\bibnamefont {Gacon}}, \bibinfo {author} {\bibfnamefont {S.}~\bibnamefont {Martiel}}, \bibinfo {author} {\bibfnamefont {P.~D.}\ \bibnamefont {Nation}}, \bibinfo {author} {\bibfnamefont {L.~S.}\ \bibnamefont {Bishop}}, \bibinfo {author} {\bibfnamefont {A.~W.}\ \bibnamefont {Cross}}, \bibinfo {author} {\bibfnamefont {B.~R.}\ \bibnamefont {Johnson}},\ and\ \bibinfo {author} {\bibfnamefont {J.~M.}\ \bibnamefont {Gambetta}},\ }\href {https://doi.org/10.48550/arXiv.2405.08810} {\bibinfo {title} {Quantum computing with {Q}iskit}} (\bibinfo {year} {2024}),\ \Eprint {https://arxiv.org/abs/2405.08810} {arXiv:2405.08810 [quant-ph]}
  \BibitemShut {NoStop}%
\bibitem [{\citenamefont {Kreplin}\ \emph {et~al.}(2025)\citenamefont {Kreplin}, \citenamefont {Willmann}, \citenamefont {Schnabel}, \citenamefont {Rapp}, \citenamefont {Hagel{\"u}ken},\ and\ \citenamefont {Roth}}]{kreplin2025squlearn}%
  \BibitemOpen
  \bibfield  {author} {\bibinfo {author} {\bibfnamefont {D.~A.}\ \bibnamefont {Kreplin}}, \bibinfo {author} {\bibfnamefont {M.}~\bibnamefont {Willmann}}, \bibinfo {author} {\bibfnamefont {J.}~\bibnamefont {Schnabel}}, \bibinfo {author} {\bibfnamefont {F.}~\bibnamefont {Rapp}}, \bibinfo {author} {\bibfnamefont {M.}~\bibnamefont {Hagel{\"u}ken}},\ and\ \bibinfo {author} {\bibfnamefont {M.}~\bibnamefont {Roth}},\ }\bibfield  {title} {\bibinfo {title} {squlearn: A python library for quantum machine learning},\ }\href@noop {} {\bibfield  {journal} {\bibinfo  {journal} {IEEE Software}\ } (\bibinfo {year} {2025})}\BibitemShut {NoStop}%
\bibitem [{\citenamefont {Hansen}\ \emph {et~al.}(2019)\citenamefont {Hansen}, \citenamefont {Akimoto},\ and\ \citenamefont {Baudis}}]{hansen2019pycma}%
  \BibitemOpen
  \bibfield  {author} {\bibinfo {author} {\bibfnamefont {N.}~\bibnamefont {Hansen}}, \bibinfo {author} {\bibfnamefont {Y.}~\bibnamefont {Akimoto}},\ and\ \bibinfo {author} {\bibfnamefont {P.}~\bibnamefont {Baudis}},\ }\href {https://doi.org/10.5281/zenodo.2559634} {\bibinfo {title} {{CMA-ES/pycma} on {G}ithub}},\ \bibinfo {howpublished} {Zenodo, DOI:10.5281/zenodo.2559634} (\bibinfo {year} {2019})\BibitemShut {NoStop}%
\end{thebibliography}%

\end{document}